\newcommand\ckmfitter{{CKMfitter}}
\newcommand{\simgt}{\,\hbox{\lower0.6ex\hbox{$\sim$}\llap{\raise0.6ex\hbox{$>$}}}\,}
\newcommand{\simlt}{\,\hbox{\lower0.6ex\hbox{$\sim$}\llap{\raise0.6ex\hbox{$<$}}}\,} 
\documentclass[
  aps,
  prd,
  reprint,
  showpacs,
  groupedaddress,
  amsmath,
  amssymb,
  floatfix,
  preprintnumbers
]{revtex4-1}
\usepackage{graphicx,epsfig,dcolumn,multirow}
\newcolumntype{d}[1]{D{.}{.}{#1}}

\RequirePackage{xspace}
\usepackage{relsize}
\def\babar{\mbox{\slshape B\kern-0.1em{\smaller A}\kern-0.1em B\kern-0.1em{\smaller A\kern-0.2em R}}\xspace}

\begin{document}

\title{Disentangling weak and strong interactions in $B\to K^*(\to K\pi)\pi$ Dalitz-plot analyses}

\author{
J.~Charles$^{\,a}$,
S.~Descotes-Genon$^{\,b}$,
J.~Ocariz$^{\,c,d}$,
A.~P\'erez~P\'erez$^{\,e}$ \\
for the \ckmfitter\  Group
}

\vspace{0.1cm}

\affiliation{
                {$^{a}$CNRS, Aix Marseille Univ, Universit\'e de Toulon,
                     CPT, Marseille, France\\
                  {e-mail: charles@cpt.univ-mrs.fr}
                } \\
                {$^{b}$Laboratoire de Physique Th\'eorique (UMR8627)\\
                  CNRS, Univ.  Paris-Sud, Universit\'e Paris-Saclay, F-91405 Orsay Cedex, France \\
                  {e-mail: sebastien.descotes-genon@th.u-psud.fr}
                }\\
                {$^{c}$ Sorbonne Universit\'es, UPMC Univ. Paris 06, UMR 7585, LPNHE, F-75005, Paris, France
                }\\
                {$^{d}$ Universit\'e Paris Diderot, LPNHE UMR 7585, Sorbonne Paris Cit\'e, F-75252 Paris, France\\
                  {e-mail: ocariz@in2p3.fr}
                }\\
                {$^{e}$Universit\'e de Strasbourg, \\
                  CNRS, IPHC UMR 7178, F-67000 Strasbourg, France \\
                  {e-mail: luis\_alejandro.perez\_perez@iphc.cnrs.fr}
                }\\
}

\date{\today}

\preprint{LPT-Orsay-17-08}

\begin{abstract}
Dalitz-plot analyses of $B\rightarrow K\pi\pi$ decays provide direct access to decay 
amplitudes, and thereby weak and strong phases can be disentangled by resolving 
the interference patterns in phase space between intermediate resonant states. 
A phenomenological isospin analysis of  $B\rightarrow K^*(\rightarrow K\pi)\pi$  decay amplitudes is presented exploiting available amplitude analyses performed at the {\babar}, Belle and LHCb experiments.
A first application consists in constraining the CKM parameters thanks to an external hadronic input. A method, proposed some time ago by two different groups and relying on a bound on the electroweak penguin contribution, is shown to lack the desired robustness and accuracy, and we propose a more
alluring alternative using a bound on the annihilation contribution. A second application consists in extracting information on hadronic amplitudes assuming the values of the CKM parameters from a global fit to quark flavour data. The current data yields several solutions, which do not fully support the hierarchy of hadronic amplitudes usually expected from theoretical arguments (colour suppression, suppression of electroweak penguins), as illustrated from computations within QCD factorisation. Some prospects concerning the impact of future measurements at LHCb and Belle II are also presented. Results are obtained with the \ckmfitter\ analysis package, featuring the frequentist statistical approach and using the Rfit scheme to handle theoretical uncertainties. 
\end{abstract}

\maketitle

\section{Introduction}\label{sec:Introduction}

Non-leptonic $B$ decays have been extensively studied at the $B$-factories {\babar} and Belle~\cite{Bevan:2014iga}, as well at the LHCb experiment~\cite{Bediaga:2012py}. 
Within the Standard Model (SM) some of these modes provide valuable information on the Cabibbo-Kobayashi-Maskawa (CKM) matrix and the structure of  $CP$ violation~\cite{Cabibbo:1963yz,Kobayashi:1973fv}, entangled with ha\-dronic amplitudes describing processes either at the tree level or the loop level (the so-called 
penguin contributions). Depending on the transition considered, one may or may not get rid of hadronic contributions which are notoriously difficult to assess. 
For instance, in $b\rightarrow c\bar{c}s$ processes, the CKM phase in the dominant tree amplitude is the same  as that of the Cabibbo-suppressed penguin one, so the 
only relevant weak phase is the $B_d$-mixing phase $2\beta$ (up to a very high accuracy) and it can be extracted from a $CP$ asymmetry  out of which QCD contributions 
drop to a very high accuracy. For charmless $B$ decays, the two leading amplitudes often carry different CKM and strong phases, and thus the  extraction of CKM 
couplings can be more challenging. In some cases, for instance the determination of $\alpha$ from $B\to\pi\pi$~\cite{Olivier}, one can use flavour symmetries such as isospin in order 
to extract all hadronic contributions from experimental measurements, while constraining CKM parameters. This has  provided many useful constraints for the global 
analysis of the CKM matrix within the Standard Model and the accurate determination of its parameters~\cite{Charles:2004jd,Charles:2015gya,CKMfitterwebsite,Koppenburg:2017mad}, 
as well as inputs for some models of New Physics~\cite{Deschamps:2009rh,Lenz:2010gu,Lenz:2012az,Charles:2013aka}.

The constraints obtained from some of the non-leptonic two-body $B$ decays can be contrasted with the unclear situation of the theoretical computations for these processes. 
Several methods (QCD factorisation~\cite{Beneke:1999br,Beneke:2000ry,Beneke:2003zv,Beneke:2006hg}, perturbative QCD approach~\cite{Li:2001ay,Li:2002mi,Li:2003yj,Ali:2007ff,Li:2014rwa,Wang:2016rlo}, 
Soft-Collinear Effective Theory~\cite{Bauer:2002aj,Beneke:2003pa,Bauer:2004tj,Bauer:2005kd,Becher:2014oda}) were devised more than a decade ago to compute hadronic contributions 
for non-leptonic decays. However, some of their aspects remain debated at the conceptual 
level~\cite{DescotesGenon:2001hm,Ciuchini:2001gv,Beneke:2004bn,Manohar:2006nz,Li:2009wba,Feng:2009rp,Beneke:2009az,Becher:2011dz,Beneke:2015wfa}, and they struggle to reproduce 
some data on $B$ decays into two mesons, especially $\pi^0\pi^0$, $\rho^0\rho^0$, $K\pi$, $\phi K^*$, $\rho K^*$~\cite{Beneke:2015wfa}. Considering the progress 
performed meanwhile in the determination of the CKM matrix, it is clear that by now, most of these non-leptonic modes provide more a test of our understanding of hadronic 
process rather than competitive constraints on the values of the CKM parameters, even though it can be interesting to consider them from one point of view or the other.

Our analysis is focused on the study of $B\rightarrow K^*(\rightarrow K\pi)\pi$ decay amplitudes, with the help of isospin symmetry. Among the various 
$b\rightarrow u\bar{u}s$ processes, the choice of $B\rightarrow K^*\pi$ system is motivated by the fact that an amplitude (Dalitz-plot)  analysis of the three-body 
final state $K\pi\pi$ provides access to several interference phases among different intermediate $K^*\pi$ states. The information provided by these physical 
observables highlights the potential of the $B\rightarrow K^*\pi$ system $(VP)$ compared with $B\rightarrow K\pi$ $(PP)$ where only branching ratios and $CP$ 
asymmetries are accessible. Similarly,  the $B\rightarrow K^*\pi$ system leads to the final $K\pi\pi$ state with a richer pattern of interferences and thus 
a larger set of observables than other pseudoscalar-vector states, like, say, $B\to K\rho$ (indeed, $K\pi\pi$ exhibits $K^*$ resonances from either of the two 
combinations of $K\pi$ pairs, whereas the $\rho$ meson comes from the only $\pi\pi$ pair available).  In addition,  the study of these modes provides experimental 
information on the dynamics of pseudoscalar-vector modes, which is less known and more challenging from the theoretical point of view. Finally, this system has 
been studied extensively at the {\babar}~\cite{Aubert:2008bj,Aubert:2009me,BABAR:2011ae,Lees:2015uun} and  Belle~\cite{Garmash:2006bj,Dalseno:2008wwa} experiments, 
and a large set of observables is readily available. 

Let us mention that other approaches, going beyond isospin symmetry, have been proposed to  study this system. For instance, one can use
$SU(3)$ symmetry and $SU(3)$-related channels in addition to the ones that we consider in this paper~\cite{Bhattacharya:2013boa,Bhattacharya:2015uua}. Another proposal is the construction of the fully SU(3)-symmetric amplitude~\cite{Bhattacharya:2014eca} to which the spin-one intermediate resonances that we consider here do not contribute.

The rest of this article is organised in the following way. In Sec.~\ref{sec:Dalitz}, we discuss the observables provided by the analysis of the 
$K\pi\pi$ Dalitz plot analysis. In Sec.~\ref{sec:Isospin}, we recall how isospin symmetry is used to reduce the set of hadronic amplitudes and their 
connection with diagram topologies. In Sec.~\ref{sec:CKM}, we discuss two methods to exploit these decays in order to extract information on the CKM matrix, 
making some assumptions about the size of specific contributions (either electroweak penguins or annihilation). In Sec.~\ref{sec:Hadronic}, we take the 
opposite point of view. Taking into account our current knowledge of the CKM matrix from global analysis, we set constraints on the hadronic amplitudes 
used to describe these decays, and we make a brief comparison with theoretical estimates based on QCD factorisation. In Sec.~\ref{sec:prospect}, we perform a brief prospective study, 
determining how the improved measurements expected from LHCb and Belle II may modify the determination of the hadronic amplitudes before concluding. In the 
Appendices, we discuss various technical aspects concerning the inputs and the fits presented in the paper.

\section{Dalitz-plot amplitudes}\label{sec:Dalitz}

Charmless hadronic $B$ decays are a particularly rich source of experimental information~\cite{Bevan:2014iga,Bediaga:2012py}. 
For $B$ decays into three light mesons (pions and kaons), the kinematics of the three-body final state 
can be completely determined experimentally, thus allowing for a complete characterisation of the Dalitz-plot (DP) phase space. In addition to quasi-two-body event-counting 
observables, the interference phases between pairs of resonances can also be accessed, and $CP$-odd (weak) phases can be disentangled from $CP$-even (strong) ones. Let us 
however stress that the extraction of the experimental information relies heavily on the so-called isobar approximation, widely used in experimental analyses because of its simplicity, and in spite of its known shortcomings~\cite{Amato:2016xjv}.

The $B\rightarrow K\pi\pi$  system is particularly interesting, as the decay amplitudes from intermediate $B\rightarrow PV$ resonances ($K^\star(892)$ and $\rho(770)$) 
receive sizable contributions from both tree-level and loop diagrams, and interfere directly in the common phase-space regions (namely the ``corners'' of the DP). The 
presence of additional resonant intermediate states further constrain the interference patterns and help resolving potential  phase ambiguities. In the case of 
$B^0\rightarrow K^+\pi^-\pi^0$ and $B^+\rightarrow K^0_S\pi^+\pi^0$, two different $K^\star(892)$ states contribute to  the decay amplitude,  and their interference 
phases can be directly measured. For $B^0\rightarrow K^0_S\pi^+\pi^-$, the time-dependent evolution of the decay amplitudes for $B^0$ and $\overline{B^0}$ provides 
(indirect) access to the relative phase between the $B^0\rightarrow K^{\star+}\pi^-$ and $\overline{B^0}\rightarrow K^{\star-}\pi^+$ amplitudes.

In the isobar approximation~\cite{Amato:2016xjv}, the total 
decay amplitude for a given mode is a sum of intermediate resonant contributions, and each of these is a complex function of phase-space: ${\cal A}(DP)= \sum_i A_iF_i(DP)$, 
where the sum rolls over all the intermediate resonances providing sizable contributions, the $F_i$ functions are the ``lineshapes'' of each resonance, and the isobar parameters 
$A_i$ are complex coefficients indicating the strength of each intermediate amplitude. The corresponding relation is 
$\overline{{\cal A}}(DP)=\sum_i \overline{A_i}~\overline{F_i}(DP)$ for $CP$-conjugate amplitudes.

Any convention-independent function of isobar parameters is a physical observable. For instance, for a given resonance ``$i$'', its direct $CP$ asymmetry $A_{CP}$ 
is expressed as
\begin{equation}
A_{CP}^i = \frac{|\overline{A_i}|^2-|A_i|^2}{|\overline{A_i}|^2+|A_i|^2} ,
\end{equation}
and its partial fit fraction $FF^i$ is 
\begin{equation}
FF^i = \frac{(|A_i|^2+|\overline{A_i}|^2) \int_{DP} |F_i(DP)|^2 d(DP)}
                 {\sum_{jk} (A_j A^*_k + \overline{A_j}~\overline{A^*_k}) \int_{DP} F_j(DP) F^*_k(DP) d(DP)}.
\end{equation}
To obtain the partial branching fraction ${\cal B}^i$, the fit fraction has to be multiplied by the total branching fraction 
of the final state (e.g., $B^0\to K^0_S\pi^+\pi^-$),
\begin{equation}
{\cal B}^i = {\it FF}^i \times {\cal B}_{incl} .
\end{equation}
A phase difference $\varphi_{ij}$ between two resonances ``$i$'' and ``$j$'' contributing to the same
total decay amplitude (i.e., between resonances in the same DP) is 
\begin{equation}\label{eq:phasediff1}
\varphi^{ij} = \arg(A_i/A_j) ,\qquad \overline{\varphi}_{ij} = \arg\left(\overline{A_i}/\overline{A_j}\right)\,,
\end{equation}
and a phase difference between the two $CP$-conjugate amplitudes for resonance ``$i$'' is
\begin{equation}\label{eq:phasediff2}
\Delta\varphi^{i} = \arg\left(\frac{q}{p}\frac{\overline{A_i}}{A_i}\right)\,,
\end{equation}
where $q/p$ is the $B^0-\overline{B^0}$ oscillation parameter.

For $B\rightarrow K^\star\pi$ modes,  there are in total 13 physical observables. These can be classified as four branching fractions, four direct $CP$ asymmetries and 
five phase differences:
\begin{itemize} 
\item The $CP$-averaged ${\cal B}^{+-}=BR(B^0\rightarrow K^{\star+}\pi^{-})$ branching fraction and its corresponding $CP$ asymmetry $A_{CP}^{+-}$. These observables can 
be measured independently in the $B^0\rightarrow K^0_S\pi^+\pi^-$ and $B^0\rightarrow K^+\pi^-\pi^0$ Dalitz planes.

\item The $CP$-averaged ${\cal B}^{00}=BR(B^0\rightarrow K^{\star 0}\pi^{0})$ branching fraction and its corresponding $CP$ asymmetry $A_{CP}^{00}$. These observables can 
be accessed both in the $B^0\rightarrow K^+\pi^-\pi^0$ and $B^0\rightarrow K^0_S\pi^0\pi^0$ Dalitz planes.

\item The $CP$-averaged ${\cal B}^{+0}=BR(B^+\rightarrow K^{\star+}\pi^{0})$ branching fraction and its corresponding $CP$ asymmetry $A_{CP}^{+0}$. These observables can 
be measured both in the $B^+\rightarrow K^0_S\pi^+\pi^0$ and $B^+\rightarrow K^+\pi^0\pi^0$ Dalitz planes.

\item The $CP$-averaged ${\cal B}^{0+}=BR(B^+\rightarrow K^{\star 0}\pi^{+})$ branching fraction and its corresponding $CP$ asymmetry $A_{CP}^{0+}$. They can 
be measured 
both in the $B^+\rightarrow K^+\pi^+\pi^-$ and $B^+\rightarrow K^0_S\pi^0\pi^+$ Dalitz planes.

\item The phase difference $\varphi^{00,+-}$ between $B^0\rightarrow K^{\star+}\pi^{-}$ and $B^0\rightarrow K^{\star 0}\pi^{0}$, and its corresponding $CP$ conjugate 
$\overline\varphi^{{00},-+}$. They can be measured in the $B^0\rightarrow K^+\pi^-\pi^0$ Dalitz plane and in its $CP$ conjugate DP $\overline{B^0}\rightarrow K^-\pi^+\pi^0$, 
respectively.

\item The phase difference $\varphi^{+0,0+}$ between $B^+\rightarrow K^{\star+}\pi^{0}$ and $B^+\rightarrow K^{\star 0}\pi^{+}$, and its corresponding $CP$ conjugate 
$\overline\varphi^{{-0},0-}$. They can be measured in the $B^+\rightarrow K^0_S\pi^+\pi^0$ Dalitz plane and in its $CP$ conjugate DP  $B^-\rightarrow K^0_S\pi^-\pi^0$, 
respectively.

\item The phase difference $\Delta\varphi^{+-}$ between $B^0\rightarrow K^{\star+}\pi^-$ and its $CP$ conjugate $\overline{B^0}\rightarrow K^{\star -}\pi^+$. This phase 
difference can only be measured in a time-dependent analysis of the $K^0_S\pi^+\pi^-$ DP. As $K^{\star +}\pi^-$ is only accessible for $B^0$ and $K^{\star -}\pi^+$ to 
$\overline{B^0}$ only, the $B^0\rightarrow K^{\star+}\pi^-$ and $\overline{B^0}\rightarrow K^{\star -}\pi^+$ amplitudes do not interfere directly (they contribute to 
different DPs). But they do interfere with intermediate resonant amplitudes that are accessible to both $B^0$ and $\overline{B^0}$, like $\rho^0(770)K^0_S$ or $f_0(980)K^0_S$, 
and thus the time-dependent oscillation is sensitive to the combined phases from mixing and decay amplitudes.
\end{itemize}

\subsection{Real-valued physical observables} \label{sec:Dalitz_new_Observbles}

The set of physical observables described in the previous paragraph (branching fractions, $CP$ asymmetries and phase differences) has the advantage of providing 
straightforward physical interpretations. From a technical point of view though, the phase differences suffer from the drawback of their definition with a $2\pi$ periodicity. This feature becomes an issue when the experimental uncertainties on the phases are large and the correlations between observables are significant, 
since there is no straightforward way to properly implement their covariance into a fit algorithm. Moreover the uncertainties on the phases are related to the 
moduli of the corresponding amplitudes, leading to problems when the latter are not known precisely and can reach values compatible with zero. As a solution to 
this issue, a set of real-valued Cartesian physical observables is defined, in which the $CP$ asymmetries and phase differences are expressed in terms of the real and imaginary 
parts of ratios of isobar amplitudes scaled by the ratios of the corresponding branching fractions and $CP$ asymmetries. The new observables are functions of branching 
fractions, $CP$ asymmetries and phase differences, and are thus physical observables. The new set of observables, similar to the $U$ and $I$ observables defined 
in $B\to \rho\pi$~\cite{Olivier}, are expressed as the real and imaginary parts 
of ratios of amplitudes as follows,

\begin{eqnarray}
{\mathcal Re}\left(A_i/A_j\right)                       = \sqrt{\frac{{\cal B}^i}{{\cal B}^j}\frac{A_{CP}^i - 1}{A_{CP}^j - 1}} \cos(\varphi_{ij})\,, \\
{\mathcal Im}\left(A_i/A_j\right)                       = \sqrt{\frac{{\cal B}^i}{{\cal B}^j}\frac{A_{CP}^i - 1}{A_{CP}^j - 1}} \sin(\varphi_{ij})\,, \\
{\mathcal Re}\left(\overline{A}_i/\overline{A}_j\right) = \sqrt{\frac{{\cal B}^i}{{\cal B}^j}\frac{A_{CP}^i + 1}{A_{CP}^j + 1}} \cos(\overline{\varphi}_{ij})\,, \\
{\mathcal Im}\left(\overline{A}_i/\overline{A}_j\right) = \sqrt{\frac{{\cal B}^i}{{\cal B}^j}\frac{A_{CP}^i + 1}{A_{CP}^j + 1}} \sin(\overline{\varphi}_{ij})\,.
\end{eqnarray}

We see that some observables are not defined in the case $A_{CP}^j=\pm 1$, as could be expected from the following argument.
Let us suppose that $A_{CP}^j=+1$ for the $j$-th resonance, i.e., we have the amplitude $A_j=0$: the quantities ${\mathcal Re}(A_i/A_j)$ and
${\mathcal Im}(A_i/A_j)$ are not defined, but neither is the phase difference between $A_i$ and $A_j$. Therefore, in both parametrisations (real and imaginary part of ratios, 
or branching ratios, $CP$ asymmetries and phase differences), the singular case $A_{CP}^j=\pm 1$ leads to some undefined observables. Let us add that this case does not occur in practice for our analysis.

For each $B\rightarrow K\pi\pi$ mode considered in this paper, the real and imaginary parts of amplitude ratios used as inputs are the following:

\begin{eqnarray}
&&  B^0\rightarrow K^0_S\pi^+\pi^- :  \label{eq:KsPiPi_inputs}     \\ && \nonumber \qquad
    {\mathcal B}(K^{*+}\pi^-)
         \ ; \ \\&&\nonumber \qquad
         {\mathcal Re}\left[ \frac{q}{p} \frac{\overline{A}(K^{*-}\pi^+)}{A(K^{*+}\pi^-)} \right] 
         \ ; \
         {\mathcal Im}\left[ \frac{q}{p} \frac{\overline{A}(K^{*-}\pi^+)}{A(K^{*+}\pi^-)} \right]\ , \
\\
&& B^0\rightarrow K^+\pi^-\pi^0 :  \label{eq:KPiPi0_inputs}  \\ && \nonumber \qquad
    \left\{   \begin{array}{l}\displaystyle  {\mathcal B}(K^{*0}\pi^0) 
         \ ; \ \qquad
         \left| \frac{\overline{A}(K^{*-}\pi^+)}{A(K^{*+}\pi^-)} \right| \ ; \   \\
         \displaystyle 
       {\mathcal Re}\left[ \frac{A(K^{*0}\pi^0)}{A(K^{*+}\pi^-)} \right]
         \ ; \
         {\mathcal Im}\left[ \frac{A(K^{*0}\pi^0)}{A(K^{*+}\pi^-)} \right] 
         \ ; \ \\
         \displaystyle 
         {\mathcal Re}\left[ \frac{\overline{A}(\overline{K}^{*0}\pi^0)}{\overline{A}(K^{*-}\pi^+)} \right]
         \ ; \
         {\mathcal Im}\left[ \frac{\overline{A}(\overline{K}^{*0}\pi^0)}{\overline{A}(K^{*-}\pi^+)} \right]\ , \
         \end{array}\right.
 \\
&& B^+\rightarrow K^+\pi^-\pi^+ :  \label{eq:KPiPi_inputs}\\&& \nonumber\qquad
         {\mathcal B}(K^{*0}\pi^+) 
         \ ; \ 
         \left| \frac{\overline{A}(\overline{K}^{*0}\pi^-)}{A(K^{*0}\pi^+)} \right|\ , \\
&& B^+\rightarrow K^0_S\pi^+\pi^0 : \label{eq:K0PiPi0_inputs}\\&& \nonumber\qquad
    \left\{   \begin{array}{l}\displaystyle 
         {\mathcal B}(K^{*+}\pi^0) 
         \ ; \
         \left| \frac{\overline{A}(K^{*-}\pi^0)}{A(K^{*+}\pi^0)} \right| \ ; \   \\
         \displaystyle 
         {\mathcal Re}\left[ \frac{A(K^{*+}\pi^0)}{A(K^{*0}\pi^+)} \right]
         \ ; \
         {\mathcal Im}\left[ \frac{A(K^{*+}\pi^0)}{A(K^{*0}\pi^+)} \right] 
         \ ; \ \\ \displaystyle
         {\mathcal Re}\left[ \frac{\overline{A}(K^{*-}\pi^0)}{\overline{A}(\overline{K}^{*0}\pi^-)} \right]
         \ ; \
         {\mathcal Im}\left[ \frac{\overline{A}(K^{*-}\pi^0)}{\overline{A}(\overline{K}^{*0}\pi^-)} \right]\ .

                \end{array}\right.
\end{eqnarray}

This choice of inputs is motivated by the fact that amplitude analyses are sensitive to ratios of isobar amplitudes. The sensitivity to phase differences leads to a  
sensitivity to the real and imaginary part of these ratios. It has to be said that the set of inputs listed previously is just one of the possible sets of independent 
observables that can be extracted from this set of amplitude analyses. In order to combine {\babar} and Belle results, it is straightforward to express the 
experimental results in the above format, and then combine them as is done for independent measurements. Furthermore, experimental information from other analyses 
which are not amplitude and/or time-dependent, i.e., which are only sensitive to ${\mathcal B}$ and $A_{CP}$, can be also added in a straightforward fashion.

In order to properly use the experimental information in the above format it will be necessary to use the full covariance matrix, both statistical and systematic, 
of the isobar amplitudes. This will allow us to properly propagate the uncertainties as well as the correlations of the experimental inputs to the ones exploited in 
the phenomenological fit.
 
\section{Isospin analysis of $B\rightarrow K^*\pi$ decays}\label{sec:Isospin}

The isospin formalism used in this work is described in detail in Ref.~\cite{PerezPerez:2008gna}. Only the main ingredients are summarised below.

Without any loss of generality, exploiting the unitarity of the CKM matrix, the $B^0\rightarrow K^{*+}\pi^-$ decay amplitude $A^{+-}$ can be parametrised as
\begin{equation}\label{eq:BtoKstarPlusPiMinusAmplitude}
A^{+-} = V_{ub}^*V_{us} T^{+-} + V_{tb}^*V_{ts} P^{+-} ,
\end{equation}
with similar expressions for the $CP$-conjugate amplitude $\bar{A}^{-+}$ (the CKM factors appearing as complex conjugates), and for the remaining three amplitudes $A^{ij}=A(B^{i+j}\rightarrow K^{*i}\pi^j)$, corresponding to the $(i,j)=(0,+)$, $(+,0)$, $(00)$ modes. The tree and penguin contributions are now defined through their CKM factors rather than their diagrammatic 
structure: they can include contributions from additional $c$-quark penguin diagrams due to the re-expression of  $V_{cb}^*V_{cs}$ in Eq.~(\ref{eq:BtoKstarPlusPiMinusAmplitude}). 
In the following, $T^{ij}$ and $P^{ij}$ will be called hadronic amplitudes.

Note that the relative CKM matrix elements in Eq.~(\ref{eq:BtoKstarPlusPiMinusAmplitude}) significantly enhance the penguin contributions with respect to the tree ones, providing an 
improved sensitivity to the former. The isospin invariance imposes a quadrilateral relation among these four decay amplitudes, derived in Ref.~\cite{Nir:1991cu} for $B\to K\pi$, 
but equivalently applicable in the $K^*\pi$ case:
\begin{equation}\label{eq:isospinRelations}
A^{0+} + \sqrt{2} A^{+0} = A^{+-} + \sqrt{2} A^{00},
\end{equation}
and a similar expression for the $CP$-conjugate amplitudes. These can be used to rewrite the decay amplitudes in the ``canonical'' parametrisation,
\begin{eqnarray}\label{eq:canonicalParametrisation}
\begin{array}{cclclc}
A^{+-} & = & V_{us}V_{ub}^*T^{+-} & + & V_{ts}V_{tb}^*P^{+-} & ,
\\
A^{0+} & = & V_{us}V_{ub}^*N^{0+} & + & V_{ts}V_{tb}^*(-P^{+-}+P_{\rm EW}^{\rm C}) & , 
\\
\sqrt{2}A^{+0} & = & V_{us}V_{ub}^*T^{+0} & + & V_{ts}V_{tb}^*P^{+0} & ,
\\
\sqrt{2}A^{00} & = & V_{us}V_{ub}^*T^{00}_{\rm C} & + & V_{ts}V_{tb}^*(-P^{+-}+P_{\rm EW}) & ,
\end{array}
\end{eqnarray}
with
\begin{eqnarray}
T^{+0}&=&T^{+-}+T_{\rm C}^{00}-N^{0+}\,,\\
P^{+0}&=&P^{+-}+P_{\rm EW}-P_{\rm EW}^{\rm C}\,.
\end{eqnarray}
This parametrisation is frequently used in the literature with various slightly different conventions, and is expected to hold up to a very high accuracy (see 
Refs.~\cite{Gronau:2005pq,Botella:2006zi} for isospin-breaking contributions to $B\to\pi\pi$ decays). The notation is chosen to illustrate the main diagram topologies 
contributing to the decay amplitude under consideration. $N^{0+}$ makes reference to the fact that the contribution to $B^+\rightarrow K^{*0}\pi^+$ with a $V_{us}V_{ub}^*$ term
corresponds to an annihilation/exchange topology; $T^{00}_{\rm C}$ denotes the colour-suppressed  $B^0\rightarrow K^{*0}\pi^0$ tree amplitude; the EW subscript 
in the $P_{\rm EW}$ and $P_{\rm EW}^{\rm C}$ terms refers to the $\Delta I=1$ electroweak penguin contributions to the decay amplitudes. We can also introduce the $\Delta I=3/2$ 
combination $T_{3/2}=T^{+-}+T^{00}_{\rm C}$.

One naively expects that colour-suppressed contributions will indeed be suppressed compared to their colour-allowed partner, and that electroweak penguins and annihilation 
contributions will be much smaller than tree and QCD penguins. These expectations can be expressed quantitatively using theoretical approaches like QCD 
factorisation~\cite{Beneke:1999br,Beneke:2000ry,Beneke:2003zv,Beneke:2006hg}. Some of these assumptions have been challenged by the experimental data gathered, in particular the mechanism of 
colour suppression in $B\to\pi\pi$ and the smallness of the annihilation part for $B\to K\pi$~\cite{Beneke:2006mk,Bell:2009fm,Bell:2015koa,Beneke:2015wfa,Li:2014rwa,Olivier}.

The complete set of $B\rightarrow K^*\pi$ decay amplitudes, constrained by the isospin relations described in Eq.~(\ref{eq:isospinRelations}) are fully described by 13 parameters, 
which can be classified as 11 hadronic and 2 CKM parameters following Eq.~(\ref{eq:canonicalParametrisation}). A unique feature of the $B\rightarrow K^*\pi$ system is that this 
number of unknowns matches the total number of physical observables discussed in Sec.~\ref{sec:Dalitz}. One could thus expect that all parameters (hadronic and CKM) could be fixed 
from the data. However, it turns out that the weak and strong phases can be redefined in such a way as to absorb in the CKM parameters any constraints on the hadronic ones. This 
property, known as {\it reparametrisation invariance}, is derived in detail in Refs.~\cite{Botella:2005ks,PerezPerez:2008gna} and we recall its essential aspects here. The decay 
amplitude of a $B$ meson into a final state can be written as:
\begin{eqnarray}\label{eq:Af1}
A_f&=&m_1 e^{i\phi_1}e^{i\delta_1}+m_2 e^{i\phi_2}e^{i\delta_2} \ , \\
\bar{A}_{\bar{f}}&=&m_1 e^{-i\phi_1}e^{i\delta_1}+m_2 e^{-i\phi_2}e^{i\delta_2} \ , \label{eq:Abarf1}
\end{eqnarray}
where $\phi_i$ are $CP$-odd (weak) phases, $\delta_i$ are $CP$-even (strong) phases, and $m$ are real magnitudes. Any additional term $M_3e^{i\phi_3}e^{i\delta_3}$ can be expressed as 
a linear combination of $e^{i\phi_1}$ and $e^{i\phi_2}$ (with the appropriate properties under $CP$ violation), leading to the fact that the decay amplitudes can be written in terms 
of any other pair of weak phases $\{\varphi_1, \varphi_2\}$ as long as $\varphi_1\neq \varphi_2$ (mod $\pi$):
\begin{eqnarray}\label{eq:Af2}
A_f&=&M_1 e^{i\varphi_1}e^{i\Delta_1}+M_2 e^{i\varphi_2}e^{i\Delta_2} \ , \\
\bar{A}_{\bar{f}}&=&M_1 e^{-i\varphi_1}e^{i\Delta_1}+M_2 e^{-i\varphi_2}e^{i\Delta_2} \ , \label{eq:Abarf2}
\end{eqnarray}
with
\begin{eqnarray}
M_1e^{i\Delta_1} &=&[m_1e^{i\delta _1}\sin(\phi_1-\varphi_2)+m_2e^{i\delta_2}\sin(\phi_2-\varphi_2)]
\nonumber\\ &&\qquad\qquad/\sin(\varphi_2-\varphi_1)  \ ,  \label{eq:m1}\\
M_2e^{i\Delta_2} &=&[m_1e^{i\delta _1}\sin(\phi_1-\varphi_1)+m_2e^{i\delta_2}\sin(\phi_2-\varphi_1)] 
\nonumber\\ &&\qquad\qquad/\sin(\varphi_2-\varphi_1)\ . \label{eq:m2}
\end{eqnarray}

This change in the set of weak basis does not have any physical implications, hence the name of re-para\-meterisation invariance. We can now take two different sets of weak phases 
$\{\phi_1, \phi_2\}$ and $\{\varphi_1, \varphi_2\}$ with $\phi_1=\varphi_1$ but $\phi_2\neq\varphi_2$. If an algorithm existed to extract $\phi_2$ as a function of physical 
observables related to these decay amplitudes, the similarity of Eqs.~(\ref{eq:Af1})-(\ref{eq:Abarf1}) and Eqs.~(\ref{eq:Af2})-(\ref{eq:Abarf2}) indicate that $\varphi_2$ 
would be extracted exactly using the same function with the same measurements as input, leading to $\varphi_2=\phi_2$, in contradiction with the original statement that we 
are free to express the physical observables using an arbitrary choice for the weak basis.

We have thus to abandon the idea of an algorithm allowing one to extract both CKM and hadronic parameters from a set of physical observables.  The weak phases in the 
parameterisation of the decay amplitudes  cannot be extracted  without additional hadronic hypothesis. This discussion holds if the two weak phases used to describe the decay amplitudes are different (modulo $\phi$). The argument does not apply when only one weak 
phase can be used to describe the decay amplitude: setting one of the amplitudes to zero, say $m_2=0$, breaks reparametrisation invariance, as can be seen easily in Eqs.~(\ref{eq:m1})-(\ref{eq:m2}).
In such cases, weak phases can be extracted from experiment, e.g., the extraction of $\alpha$ from $B\to \pi\pi$, the extraction of $\beta$ from $J/\psi K_S$ or 
$\gamma$ from $B\to DK$. In each case, an amplitude is assumed to vanish, either approximately (extraction of $\alpha$ and $\beta$) or exactly (extraction of $\gamma$)~\cite{Olivier,Bevan:2014iga,Bediaga:2012py}. 
 
In view of this limitation, two main strategies can be considered for the system considered here: either implementing additional constraints on some hadronic parameters in order to extract the CKM phases 
using the $B \rightarrow K^*\pi$ observables, or fix the CKM parameters to their known values from a global fit and use the $B \rightarrow K^*\pi$ observables to extract 
information on the hadronic contributions to the decay amplitudes. Both approaches are described below.

\section{Constraints on CKM phases}\label{sec:CKM}

We illustrate the first strategy using two specific examples. The first example is similar in spirit to the Gronau-London method for extracting the CKM angle 
$\alpha$~\cite{Gronau:1990ka}, which relies on neglecting the contributions of electroweak penguins to the $B\rightarrow\pi\pi$ decay amplitudes. The second example 
assumes that upper bounds on annihilation/exchange contributions can be estimated from external information.

\subsection{The CPS/GPSZ method: setting a bound on electroweak penguins}\label{subsec:CPS}

In $B\rightarrow\pi\pi$ decays, the electroweak penguin contribution can be related to the tree amplitude in a model-independent way using Fierz transformations of 
the relevant current-current operators in the effective Hamiltonian for $B\to \pi\pi$ decays~\cite{Buras:1998rb,Neubert:1998pt,Neubert:1998jq,Charles:2004jd}. One can predict the 
ratio $R=P_{\rm EW}/T_{3/2}\simeq -3/2 (C_9+C_{10})/(C_1+C_2)=(1.35\pm 0.12) \%$ only in terms of short-distance Wilson Coefficients, since long-distance 
hadronic matrix elements drop from the ratio (neglecting the operators $O_7$ and $O_8$ due to their small Wilson coefficients compared to $O_9$ and $O_{10}$). This 
leads to the prediction that there is no strong phase difference between $P_{\rm EW}$ and $T_{3/2}$ so that electroweak penguins do not generate a charge asymmetry in 
$B^+\to \pi^+\pi^0$ if this picture holds: this prediction is in agreement with the present experimental average of the corresponding asymmetry. Moreover, this 
assumption is crucial to ensure the usefulness of the Gronau-London method to extract the CKM angle $\alpha$ from an isospin analysis of $B\rightarrow\pi\pi$ decay 
amplitudes~\cite{Charles:2004jd,Olivier}: setting the electroweak penguin to zero in the Gronau-London breaks the reparametrisation invariance described in Sec.~\ref{sec:Isospin} 
and opens the possibility of extracting weak phases.

One may want to follow a similar approach and use some knowledge or assumptions on the electroweak penguin in the case of $B\to K\pi$ or $B\to K^*\pi$ in order to constrain 
the CKM factors. This approach is sometimes referred to as the CPS/GPSZ method~\cite{Ciuchini:2006kv,Gronau:2006qn}. Indeed, as shown in Eq.~(\ref{eq:canonicalParametrisation}), 
the penguins in $A^{00}$ and $A^{+-}$ differ only by the $P_{\rm EW}$ term. By neglecting its contribution to $A^{00}$, these two decay amplitudes can be combined so 
that their (now identical) penguin terms can be eliminated,
\begin{eqnarray}
A^0 = A^{+-} + \sqrt{2}A^{00} = V_{us}V_{ub}^*(T^{+-}+T^{00}_{\rm C}),
\end{eqnarray}
and then, together with its $CP$-conjugate amplitude $\bar{A}^0$,  a convention-independent amplitude ratio $R^0$ can be defined as
\begin{eqnarray}\label{eq:Eq4}
R^0 = \frac{q}{p}\frac{\bar{A}^0}{A^0} = e^{-2i\beta}e^{-2i\gamma} = e^{2i\alpha}.
\end{eqnarray}
The $A^0$ amplitude can be extracted using the decay chains $B^0\rightarrow K^{*+}(\rightarrow K^+\pi^0)\pi^-$ and $B^0\rightarrow K^{*0}(\rightarrow K^+\pi^-)\pi^0$ contributing 
to the same $B^0\rightarrow K^+\pi^-\pi^0$ Dalitz plot, so that both the partial decay rates and their interference phase can be measured in an amplitude analysis. Similarly, 
$\bar{A}^0$ can be extracted from the $CP$-conjugate $\bar{B}^0\rightarrow K^-\pi^+\pi^0$ DP using the same procedure. Then, the phase difference between $A^{+-}$ and $\bar{A}^{-+}$ 
can be extracted from the $B^0\rightarrow K^0_{\rm S}\pi^+\pi^-$ DP, considering the $B^0\rightarrow K^{*+}(\rightarrow K^0\pi^+)\pi^-$ decay chain, and its $CP$-conjugate 
$\bar{B}^0\rightarrow K^{*-}(\rightarrow \bar{K}^0\pi^-)\pi^+$, which do interfere through mixing. Let us stress that this method is  a measurement of $\alpha$ rather than a 
measurement of $\gamma$, in contrast with the claims in Refs.~\cite{Ciuchini:2006kv,Gronau:2006qn}.

However, the method used to bound $P_{\rm EW}$ for the $\pi\pi$ system cannot be used directly in the $K^*\pi$ case. In the $\pi\pi$ case, $SU(2)$ symmetry guarantees that the matrix 
element with the combination of operators $O_1-O_2$ vanishes, so that it does not enter tree amplitudes. A similar argument would hold for $SU(3)$ symmetry in the case of the $K\pi$ 
system, but it does not for the vector-pseudoscalar $K^*\pi$ system. It is thus not possible to cancel hadronic matrix elements when considering $P_{\rm EW}/T_{3/2}$, which becomes a 
complex quantity suffering from (potentially large) hadronic uncertainties~\cite{Gronau:2003yf,Ciuchini:2006kv}. The size of the electroweak penguin (relative to the tree contributions), is parametrised as
\begin{equation}
\frac{P_{\rm EW}}{T_{3/2}} = R \frac{1-r_{\rm VP}}{1+r_{\rm VP}} ,
\label{eq:PEWfromCPS}
\end{equation}
where $R\simeq (1.35\pm 0.12)\%$ is the value obtained in the $SU(3)$ limit for $B\to \pi K$ (and identical to the one obtained from $B\to\pi\pi$ using the arguments in 
Refs.~\cite{Buras:1998rb,Neubert:1998pt,Neubert:1998jq}), and $r_{\rm VP}$ is a complex parameter measuring the deviation of $P/T_{3/2}$ from this value corresponding to
\begin{equation}
r_{\rm VP}=\frac{\langle K^*\pi(I=3/2)|Q_1-Q_2|B\rangle}{\langle K^*\pi(I=3/2)|Q_1+Q_2|B\rangle}\,.
\end{equation}
Estimates on factorisation and/or $SU(3)$ flavour relations suggest $|r_{\rm VP}|\leq 0.05$~\cite{Ciuchini:2006kv,Gronau:2006qn}. However it is clear that both approximations 
can easily be broken, suggesting a more conservative upper bound $|r_{\rm VP}|\leq 0.30$.

The presence of these hadronic uncertainties have important consequences for the method. Indeed, it turns out that including a non-vanishing $P_{\rm EW}$ completely disturbs the extraction of $\alpha$. The electroweak penguin can provide a ${\cal O}(1)$ contribution to $CP$-violating effects in 
charmless $b\rightarrow s$ processes, as its CKM coupling amplifies its contribution to the decay amplitude: $P_{\rm EW}$ is multiplied by a large CKM factor $V_{ts}V_{tb}^*=O(\lambda^2)$ 
compared to the tree-level amplitudes multiplied by a CKM factor $V_{us}V_{ub}^*=O(\lambda^4)$. Therefore, unless $P_{\rm EW}$ is particularly suppressed due to some specific hadronic dynamics, 
its presence modifies the CKM constraint obtained following this method in a very significant way.

It would be difficult to illustrate this point using the current data, due to the experimental uncertainties described in the next sections. We choose thus to discuss this problem using 
a reference scenario described in Tab.~\ref{tab:IdealCase}, where the hadronic amplitudes have been assigned arbitrary (but realistic) values and they are used to derive a 
complete set of experimental inputs with arbitrary (and much more precise than currently available) uncertainties. As shown in App.~\ref{App:Exp_inputs} (cf. Tab.~\ref{tab:IdealCase}), 
the current world averages for branching ratios and $CP$ asymmetries in $B^0\rightarrow K^{*+}\pi^-$ and $B^0\rightarrow K^{*0}\pi^0$  agree broadly with these values, which also reproduce the expected hierarchies among hadronic amplitudes,
if we set the CKM parameters to their current values from our global fit~\cite{Charles:2004jd,Charles:2015gya,CKMfitterwebsite}.
 We choose a penguin parameter $P^{+-}$ with a magnitude 
$28$ times smaller than the tree parameter $T^{+-}$, and a phase fixed at $-7^\circ$. The electroweak $P_{\rm EW}$ parameter has a value $66$ times smaller in magnitude 
than the tree parameter $T^{+-}$, and its phase is arbitrarily fixed to $+15^\circ$ in order to get a good agreement with the current central values. Our results do not depend 
significantly on this phase, and a similar outcome occurs if we choose sets with a vanishing phase for $P_{\rm EW}$ (though the agreement with the current data will be less good).

\begin{figure}[t]
\begin{center}
\includegraphics[width=7cm]{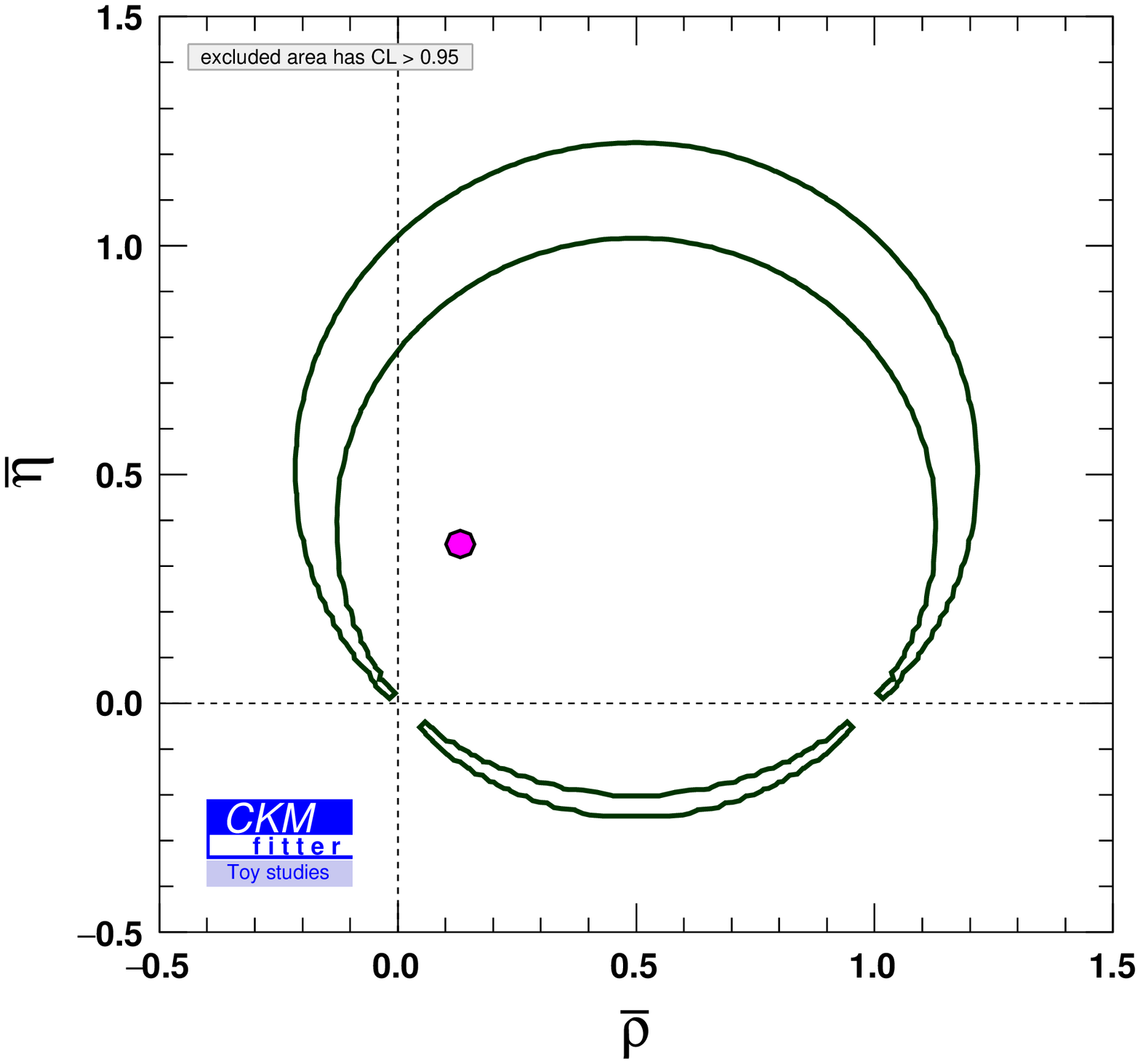}
\includegraphics[width=7cm]{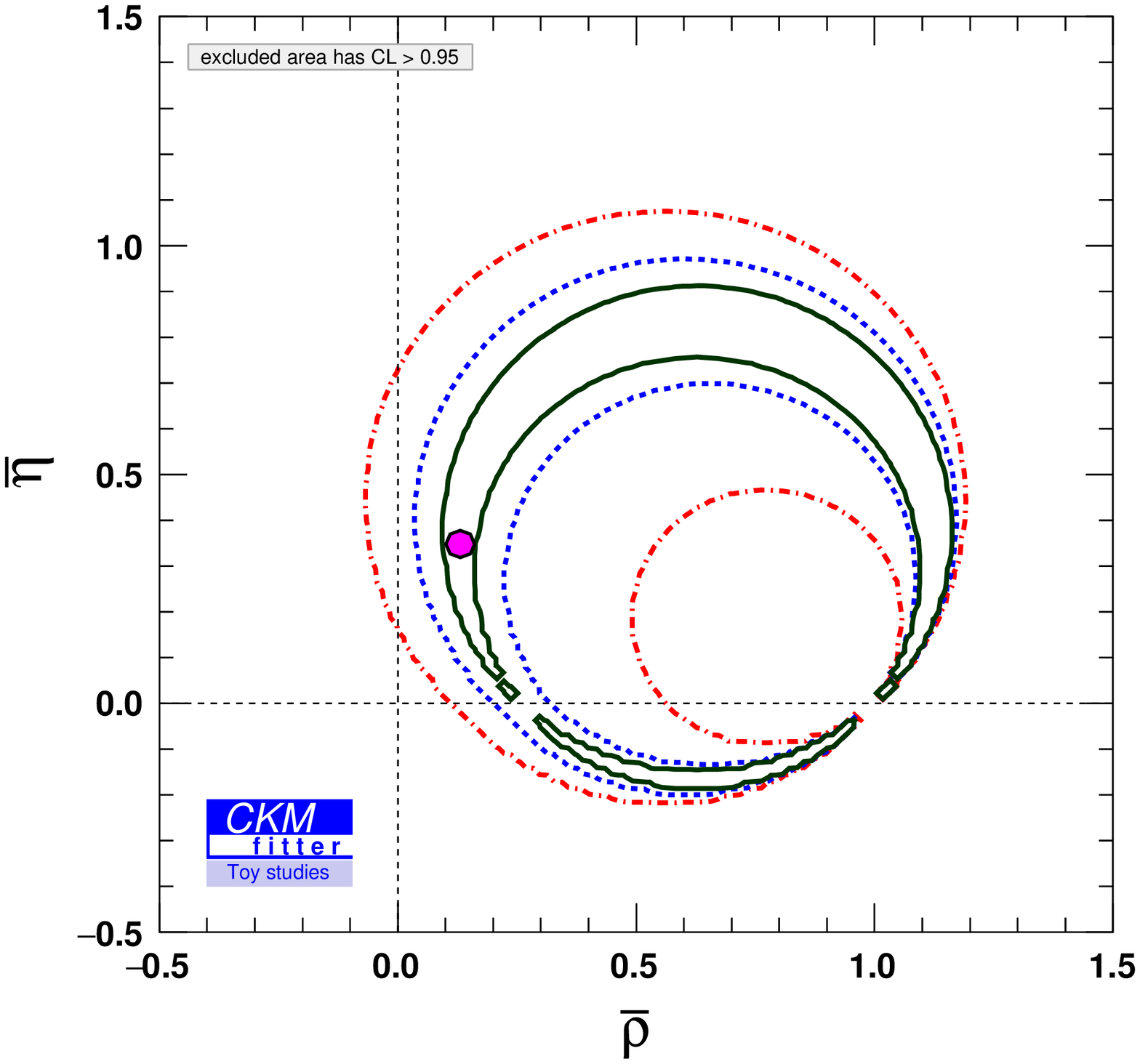}
\caption{
Constraints in the $\bar{\rho}-\bar{\eta}$ plane from the amplitude ratio $R^0$ method,
using the arbitrary but realistic numerical values for the input parameters, detailed
in the text. In the top panel, the $P_{\rm EW}$ hadronic parameter is set to zero. 
In the bottom panel, the $P_{\rm EW}$ hadronic parameter is set to its true generation value with 
different theoretical errors on $R$ and $r_{VP}$ parameters (defined in Eq.~(\ref{eq:PEWfromCPS})), 
either zero (green solid-line contour), 10\% and 5\% (blue dashed-line contour), and 10\% and 30\% 
(red solid-dashed-line contour). The parameters $\bar{\rho}$ and $\bar{\eta}$ are fixed to 
their current values from the global CKM fit~\cite{Charles:2004jd,Charles:2015gya,CKMfitterwebsite}, 
indicated by the magenta point.
}
\label{fig:RhoEta_CPS}
\end{center}
\end{figure}

We use the values of the observables derived with this set of hadronic parameters, and we perform a CPS/ GPSZ analysis to 
extract a constraint on the CKM parameters. Fig.~\ref{fig:RhoEta_CPS} 
shows the constraints derived in the $\bar{\rho}-\bar{\eta}$ plane. If we assume $P_{\rm EW}=0$ (upper panel), the extracted constraint is equivalent to a constraint on the CKM angle $\alpha$, 
as expected from Eq.~(\ref{eq:Eq4}). However, the confidence regions in the $\bar{\rho}-\bar{\eta}$ plane are very strongly biased, and the true value of the parameters are far from 
belonging to the 95\% confidence regions. On the other hand, if we fix $P_{\rm EW}$ to its true value (with a magnitude of $0.038$), the bias is removed but the constraint deviates from a pure $\alpha$-like 
shape (for instance, it does not include the origin point $\bar\rho=\bar\eta=0$). We notice that the uncertainties on $R$ and, more significantly, $r_{VP}$, have an important impact on the 
precision of the constraint on $(\bar\rho,\bar\eta)$.

This simple illustration with our reference scenario shows that the CPS/GPSZ method is limited both in robustness and accuracy due to the assumption on a negligible $P_{\rm EW}$: a small non-vanishing value 
breaks the relation between the phase of $R^0$ and the CKM angle $\alpha$, and therefore, even a small uncertainty on the $P_{\rm EW}$ value would translate into large biases on the CKM 
constraints. It shows that this method would require a very accurate understanding of hadronic amplitudes in order to extract a meaningful constraint on the unitarity triangle, and the 
presence of non-vanishing electroweak penguins dilutes the potential of this method significantly.

\begin{figure}[t]
\begin{center}
\includegraphics[width=7cm]{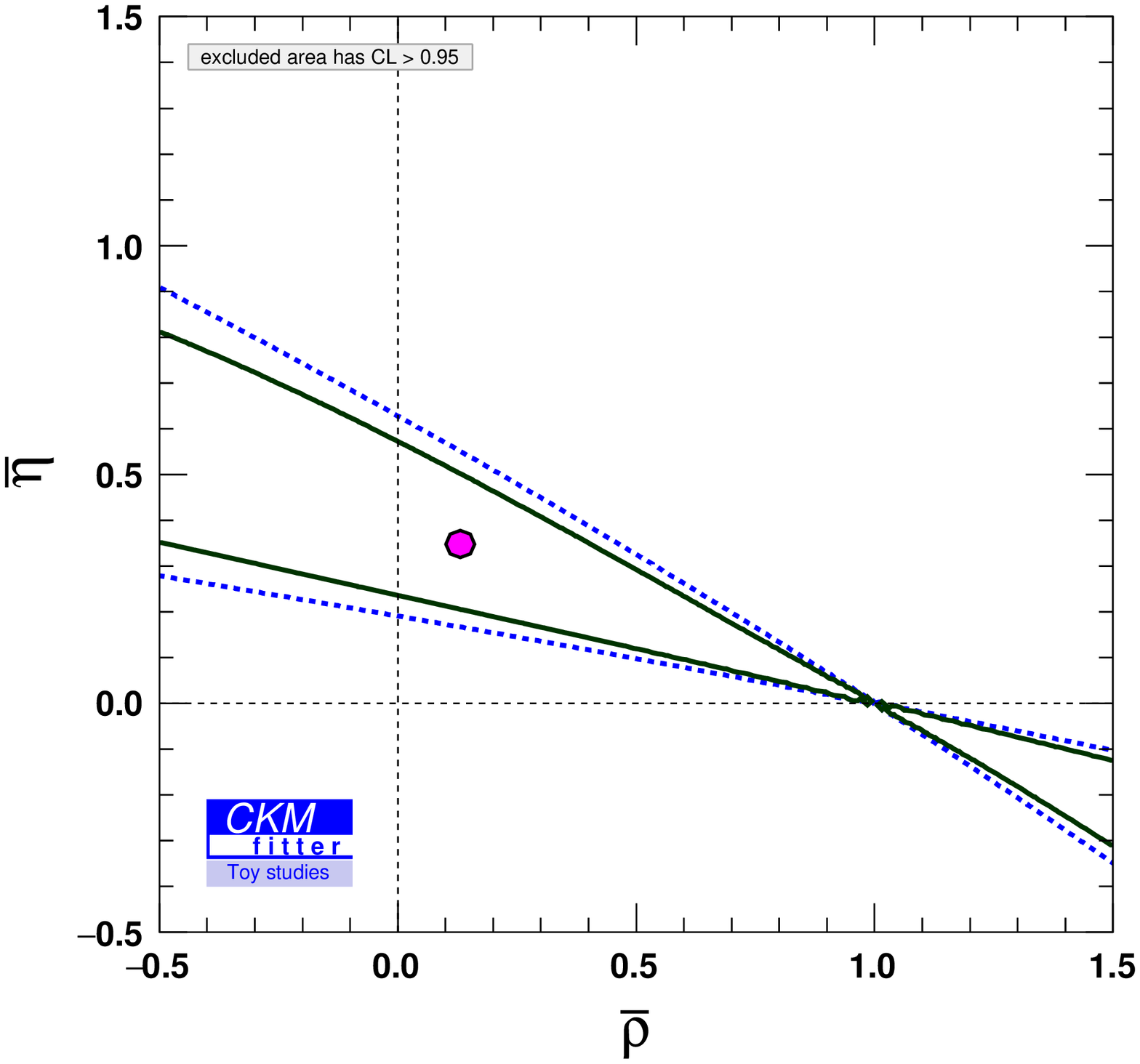}
\includegraphics[width=8cm]{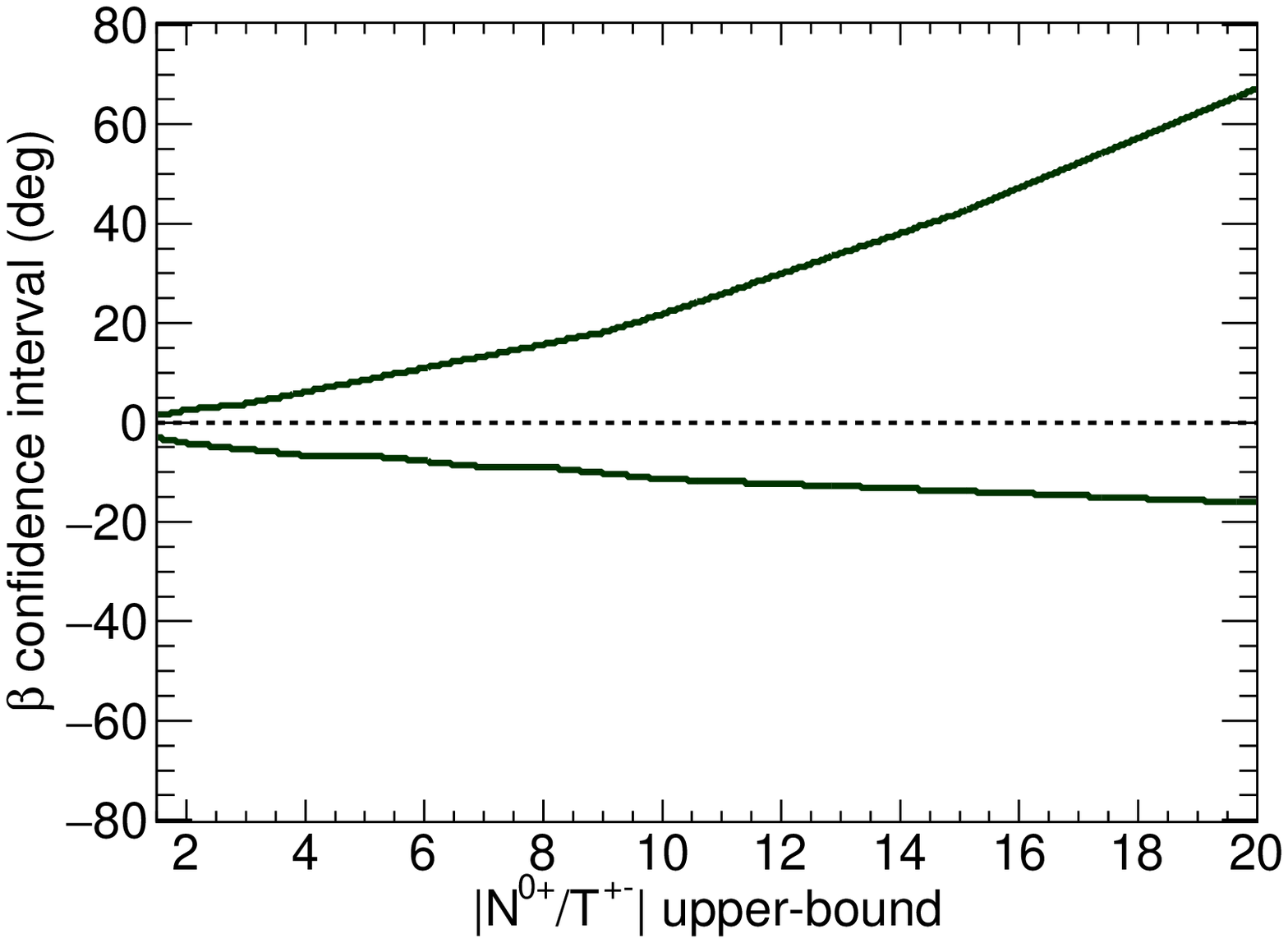}
\caption{
Top: constraints in the $\bar{\rho}-\bar{\eta}$ plane from the annihilation/exchange method, using the arbitrary but realistic numerical values for the input parameters detailed in the text. 
The green solid-line contour is the constraint obtained by fixing the $N^{0+}$ hadronic parameter to its generation value; the blue dotted-line contour is the constraint obtained by setting 
an upper bound on the $\left|N^{0+}/T^{+-}\right|$ ratio at twice its generation value. The parameters $\bar{\rho}$ and $\bar{\eta}$ are fixed to their current values from the global 
CKM fit~\cite{Charles:2004jd,Charles:2015gya,CKMfitterwebsite}, indicated by the magenta point.
Bottom: size of the $\beta - \beta_{\rm gen}$ 68\% confidence interval vs the upper-bound on $|N^{0+}/T^{+-}|$ in units of its generation value.
}
\label{fig:RhoEta_All}
\end{center}
\end{figure}

\subsection{Setting bounds on annihilation/exchange contributions}\label{subsec:N}

As discussed in the previous paragraphs, the penguin contributions for $B\rightarrow K^*\pi$ decays are strongly CKM-enhanced, impacting the CPS/GPSZ method based on neglecting a penguin amplitude 
$P_{\rm EW}$. This method exhibits a strong sensitivity to small changes or uncertainties in values assigned to the electroweak penguin contribution. An alternative and safer approach 
consists in constraining a tree amplitude, with a CKM-suppressed contribution. Among the various hadronic amplitudes introduced, it seems appropriate to choose the annihilation amplitude 
$N^{0+}$, which is expected to be smaller than $T^{+-}$, and which could even be smaller than the colour-suppressed $T^{00}_{\rm C}$. Unfortunately, no direct, clean constraints on $N^{0+}$ can 
be extracted  from data and from the theoretical point of view, $N^{0+}$ is dominated by incalculable non-factorisable contributions in QCD 
factorisation~\cite{Beneke:1999br,Beneke:2000ry,Beneke:2003zv,Beneke:2006hg}. On the other hand, indirect upper bounds on $N^{0+}$ 
may be inferred from either the $B^+\to K^{*0} \pi^+$ decay rate 
or from the $U$-spin related mode $B^+\to K^{*0}K^+$.

This method, like the previous one, hinges on a specific assumption on hadronic amplitudes. Fixing $N^{0+}$ breaks the reparametrisation invariance in Sec.~\ref{sec:Isospin}, and thus provides a way of measuring weak phases.
We can compare the two approaches by using the same reference scenario as in Sec.~\ref{subsec:CPS}, i.e., the values gathered in Tab.~\ref{tab:IdealCase}. We have an annihilation parameter $N^{0+}$ with a magnitude 
$18$ times smaller than the tree parameter $T^{+-}$, and a phase fixed at $108^\circ$. All $B\rightarrow K^*\pi$ physical observables are used  as inputs. 
This time, all hadronic parameters are free to vary in the fits, except for the annihilation/exchange parameter $N^{0+}$, which is subject to two different hypotheses: either 
its value is fixed to  its generation value, or the ratio $\left|N^{0+}/T^{+-}\right|$ is constrained in a range (up to twice its generation value).

The resulting constraints on the $\bar{\rho}-\bar{\eta}$ are shown on the upper plot of Fig.~\ref{fig:RhoEta_All}. We stress that in this fit, the value of $N^{0+}$ is bound, but the other 
amplitudes (including $P_{\rm EW}$) are left free to vary. Using a loose bound on  $\left|N^{0+}/T^{+-}\right|$ yields a less tight constraint, but in contrast 
with the CPS/GPSZ method, the CKM generation value is here included. One may notice that the resulting  constraint  is similar to the one corresponding to the CKM angle $\beta$. This can be understood 
in the following way. Let us assume that we neglect the contribution from $N^{0+}$. We obtain the 
following amplitude to be considered
\begin{equation}
A'=A^{0+}=V_{ts}V_{tb}^*(-P^{+-}+P_{\rm EW}^{\rm C}),
\end{equation}
and then, together with its $CP$-conjugate amplitude $\bar{A}'$,  a convention-independent amplitude ratio $R'$ can be defined as
\begin{equation}
R' = \frac{q}{p}\frac{\bar{A}'}{A} = e^{-2i\beta}\,,
\end{equation}
in agreement with the convention used to fix the phase of the $B$-meson state. This justifies the $\beta$-like shape of the constraint obtained when fixing the value of the annihilation parameter.
The presence of the oscillation phase $q/p$ here, starting from a decay of a charged $B$, may seem surprising. However, one should keep in mind that the measurement of $B^+\to K^{*0}\pi^+$ 
and its $CP$-conjugate amplitude are not sufficient to determine the relative phase between $A'$ and $\bar{A}'$: this requires one to reconstruct the whole quadrilateral equation 
Eq.~(\ref{eq:isospinRelations}), where the phases are provided by interferences between mixing and decay amplitudes in $B_0$ and $\bar{B}_0$ decays. In other words, the phase observables 
obtained from the Dalitz plot are always of the form Eq.~(\ref{eq:phasediff1})-(\ref{eq:phasediff2}): their combination can only lead to a ratio of $CP$-conjugate amplitudes multiplied by 
the oscillation parameter $q/p$.

The lower plot of Fig.~\ref{fig:RhoEta_All} describes how the constraint on $\beta$ loosens around its true value when the range allowed for $\left|N^{0+}/T^{+-}\right|$ is increased compared to its initial 
value ($0.143$). We see that the method is stable and keeps on including the true value for $\beta$ even in the case of a mild constraint on $\left|N^{0+}/T^{+-}\right|$.

\section{Constraints on hadronic parameters using current data}\label{sec:Hadronic}

As already anticipated in Sec.~\ref{sec:Isospin}, a second strategy to exploit the data consists in assuming that the CKM matrix is
already well determined from 
the CKM global fit~\cite{Charles:2004jd,Charles:2015gya,CKMfitterwebsite}. The measurements of 
$B\rightarrow K^\star\pi$ observables (isobar parameters) can then be used to extract constraints on the hadronic 
parameters in Eq.~(\ref{eq:canonicalParametrisation}).

\subsection{Experimental inputs}\label{subsec:exp_inputs}

For this study, the complete set of available results from the \babar\ and Belle experiments is used. The level of detail for the publicly available results  varies  according to the decay 
mode in consideration. In most cases, at least one amplitude DP analysis of $B^0$ and $B^+$ decays is 
public~\cite{Amhis:2016xyh}, and at least one input from each physical observable is available.
In addition, the conventions used in the various DP analyses are usually different. Ideally, one would like to have access to the complete covariance matrix, including statistical and systematic 
uncertainties, for all isobar parameters, as done for instance in Ref.~\cite{Aubert:2009me}. Since such information is not always available, the published results are used in order to derive ad-hoc approximate covariance matrices, implementing all the available information (central values, total uncertainties, correlations among parameters). The inputs 
for this study are the following:
\begin{itemize}

\item Two three-dimensional covariance matrices, cf. Eq.~(\ref{eq:KsPiPi_inputs}), from the {\babar} time-dependent 
DP analysis of $B^0\rightarrow K^0_S\pi^+\pi^-$ in Ref.~\cite{Aubert:2009me},
and two three-dimensional covariance matrices from the Belle time-dependent
 DP analysis of $B^0\rightarrow K^0_S\pi^+\pi^-$ in Ref.~\cite{Dalseno:2008wwa}. 
Both the {\babar} and Belle  analyses found two quasi-degenerate solutions each, with very similar goodness-of-fit merits. 
The combination of these solutions is described in App.~\ref{App:comb_inputs}, and is taken as input for this study.

\item A five-dimensional covariance matrix, cf. Eq.~(\ref{eq:KPiPi0_inputs}), from 
the {\babar} $B^0\rightarrow K^+\pi^-\pi^0$ DP analysis~\cite{BABAR:2011ae}.

\item A two-dimensional covariance matrix, cf. Eq.~(\ref{eq:KPiPi_inputs}), from 
the {\babar} $B^+\rightarrow K^+\pi^+\pi^-$ DP analysis~\cite{Aubert:2008bj}, and
a two-dimensional covariance matrix from the 
Belle $B^+\rightarrow K^+\pi^+\pi^-$ DP analysis~\cite{Garmash:2006bj}.

\item A simplified uncorrelated four-dimensional input, cf. Eq.~(\ref{eq:K0PiPi0_inputs}), from 
the {\babar} $B^+\rightarrow K^0_S\pi^+\pi^0$ preliminary DP analysis~\cite{Lees:2015uun}.
\end{itemize}

Besides the inputs described previously, there are other experimental measurements on different three-body final states performed in 
the quasi-two-body approach, which provide measurements of
branching ratios and $CP$ asymmetries only. Such is the case of the \babar\ result on the $B^+\to K^+\pi^0\pi^0$ final state~\cite{Lees:2011aaa},  where the branching ratio and the $CP$ asymmetry of the 
$B^+\to K^{*}(892)^+\pi^0$ contribution are measured. In this study, these two measurements are treated as uncorrelated, and they are combined  with the inputs from the DP analyses mentioned previously.

These sets of experimental central values and covariance matrices are described in App.~\ref{App:Exp_inputs}, where the combinations of the results from {\babar} and Belle are also described.

Finally, we notice that the time-dependent asymmetry in $B\to K_S\pi^0\pi^0$ has been measured~\cite{Abe:2007xd,Aubert:2007ub}. As these are global analyses integrated over the whole DP, we cannot take these measurements into account. In principle a time-dependent isobar analysis of the $K_S\pi^0\pi^0$ DP could be performed and it could bring some independent information on $B\to K^{*0}\pi^0$ intermediate amplitudes. Since this more challenging analysis has not been done yet, we will not consider this channel for the time being.

\subsection{Selected results for $CP$ asymmetries and hadronic amplitudes}

Using the experimental inputs described in Sec.~\ref{subsec:exp_inputs}, a fit to the complete set of hadronic parameters 
is performed. We discuss the fit results focusing on three aspects: the most significant direct $CP$ asymmetries, 
the significance of electroweak penguins, and the relative hierarchies of hadronic contributions
to the tree amplitudes. 
As will be seen in the following, the fit results can be interpreted in terms of two sets of local minima, out of which one yields constraints on the hadronic parameters in better agreement with 
the expectations from CPS/GPSZ, the measured direct $CP$ asymmetries and the expected relative hierarchies of hadronic contributions.

\subsubsection{Direct $CP$ violation in $B^0\rightarrow K^{\star+}\pi^-$}

The $B^0\rightarrow K^{\star+}\pi^-$ amplitude can be accessed both in the $B^0\rightarrow K^0_{\rm S}\pi^+\pi^-$ and 
$B^0\rightarrow K^+\pi^-\pi^0$ Dalitz-plot analyses. The direct $CP$ asymmetry $A_{\rm CP}(B^0\rightarrow K^{\star+}\pi^-)$ has been measured 
by {\babar} in both modes~\cite{BABAR:2011ae,Aubert:2009me} and by Belle in  the  $B^0\rightarrow K^0_{\rm S}\pi^+\pi^-$ mode~\cite{Dalseno:2008wwa}.
All three measurements yield a negative value: incidentally, this matches also the sign of
the two-body $B^0\rightarrow K^+\pi^-$ $CP$ asymmetry, for which direct $CP$ violation is clearly established.

\begin{figure}[t]
\begin{center}
\includegraphics[width=8cm]{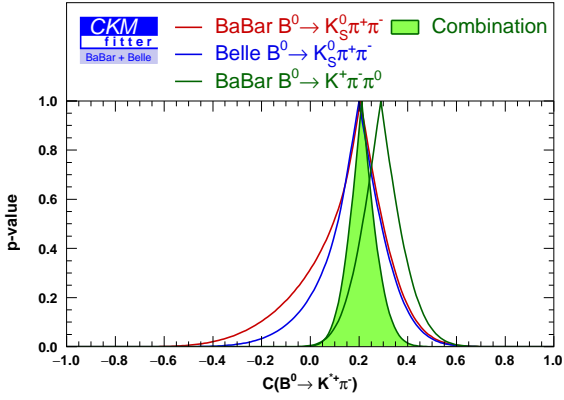}
\caption{
Constraint on the direct $CP$ asymmetry parameter $C(B^0\rightarrow K^{\star+}\pi^-) = -A_{\rm CP}(B^0\rightarrow K^{\star+}\pi^-)$ from {\babar} data on $B^0\to K^0_S\pi^+\pi^-$ (red curve), 
Belle data on $B^0\to K^0_S\pi^+\pi^-$ (blue curve), {\babar} data on $B^0\to K^+\pi^-\pi^0$ (green curve) and the combination of all these measurements (green shaded curve). 
The constraints are obtained using the observables described in the text.
}
\label{fig:ACP_KstpPim}
\end{center}
\end{figure}

Using the amplitude DP analysis results from these three measurements as inputs, the 
combined constraint on  $A_{\rm CP}(B^0\rightarrow K^{\star+}\pi^-)$ is shown in Fig.~\ref{fig:ACP_KstpPim}.
The combined value is 3.0~$\sigma$ away from zero, and the 68\% confidence interval on this $CP$ asymmetry is $0.21\pm 0.07$ approximately.
This result is to be compared with the $0.23\pm 0.06$ value provided by HFLAV~\cite{Amhis:2016xyh}. The difference is likely to come from the fact that HFLAV performs an average of the $CP$ asymmetries extracted from individual experiments, while this analysis uses isobar values as inputs which are averaged over the various experiments before being translated into values for the $CP$ parameters: since the relationships between these two sets of quantities are non-linear, the two steps (averaging over experiments and translating from one type of observables to another) yield the same central values only in the case of very small uncertainties. In the current situation, where sizeable uncertainties affect the determinations from individual experiments, it is not surprising that minor discrepancies arise between our approach and the HFLAV result.

\begin{figure}[t]
\begin{center}
\includegraphics[width=8cm]{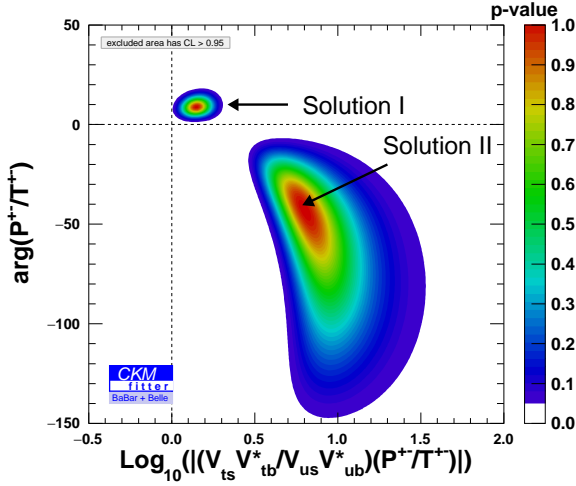}
\caption{
Two-dimensional constraint on the modulus and phase of the  $P^{+-}/T^{+-}$ ratio. 
For convenience, the modulus is multiplied by
the ratio of CKM factors appearing in the tree and penguin contributions to 
the $B^0\rightarrow K^{\star+}\pi^-$  decay amplitude.
}
\label{fig:PpmOTpm_BtoKstPi}
\end{center}
\end{figure}

As can be readily seen from Eq.~(\ref{eq:BtoKstarPlusPiMinusAmplitude}), a non-vanishing asymmetry in this mode requires a strong phase difference between
the tree $T^{+-}$ and penguin $P^{+-}$ hadronic parameters that is strictly different from zero. Fig.~\ref{fig:PpmOTpm_BtoKstPi} shows the two-dimensional constraint on the 
modulus and phase of the $P^{+-}/T^{+-}$ ratio. Two solutions with very similar $\chi^2$ are found, both incompatible
with a vanishing phase difference. The first solution corresponds to a small (but non-vanishing) positive strong phase, with similar
$\left|V_{ts}V_{tb}^\star P^{+-}\right|$ and $\left|V_{us}V_{ub}^\star T^{+-}\right|$ contributions to the total decay amplitude,
and is called Solution I in the following.
The other solution, denoted Solution II, 
corresponds to a larger, negative, strong phase, with a significantly larger penguin contribution. We notice that Solution I is closer to usual theoretical expectations concerning the relative size of penguin and tree contributions.

Let us stress that the presence of two solutions for $P^{+-}/T^{+-}$  is not related to the presence of ambiguities in the 
individual {\babar} and Belle measurements for $B^+\to K^+\pi^+\pi^-$ and $B^0\to K^0_S\pi^+\pi^-$, since we have performed 
their combinations in order to select a single solution for each process.
Therefore, the presence of two solutions in Fig.~\ref{fig:PpmOTpm_BtoKstPi} is a
global feature of our non-linear fit, arising from the  overall structure of the current combined measurements (central values and uncertainties) 
that we use as inputs.

\subsubsection{Direct $CP$ violation in $B^+\rightarrow K^{\star+}\pi^0$}

The $B^+\rightarrow K^{\star+}\pi^0$ amplitude can be accessed in a $B^+\rightarrow K^0_{\rm S}\pi^+\pi^0$ Dalitz-plot analysis, for which
only a preliminary result from {\babar} is available~\cite{Lees:2015uun}. A large, negative $CP$ asymmetry 
$A_{\rm CP}(B^+\rightarrow K^{\star+}\pi^0) = -0.52\pm 0.14\pm 0.04 ^{+0.04}_{-0.02}$  is reported there with a 3.4~$\sigma$ significance.
This $CP$ asymmetry has also been measured by \babar\ through a quasi-two-body analysis of the $B^+\rightarrow K^+\pi^0\pi^0$ final state~\cite{Lees:2011aaa}, 
obtaining $A_{\rm CP}(B^+\rightarrow K^{\star+}\pi^0) = -0.06\pm 0.24\pm 0.04$. The combination of these two measurement yields 
$A_{\rm CP}(B^+\rightarrow K^{\star+}\pi^0) = -0.39\pm 0.12\pm 0.03$, with a 3.2~$\sigma$ significance.

\begin{figure}[t]
\begin{center}
\includegraphics[width=8cm]{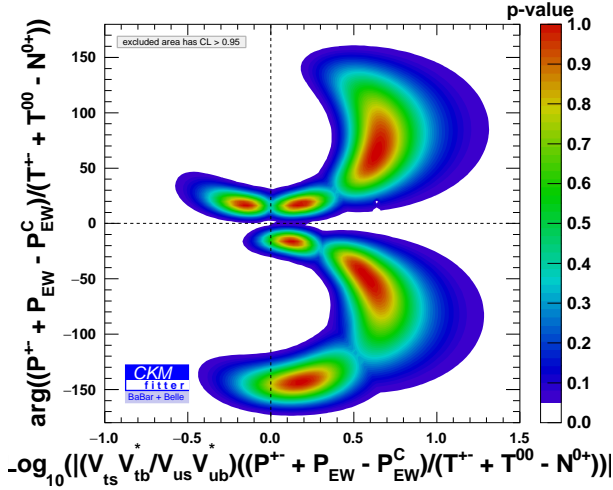}
\includegraphics[width=8cm]{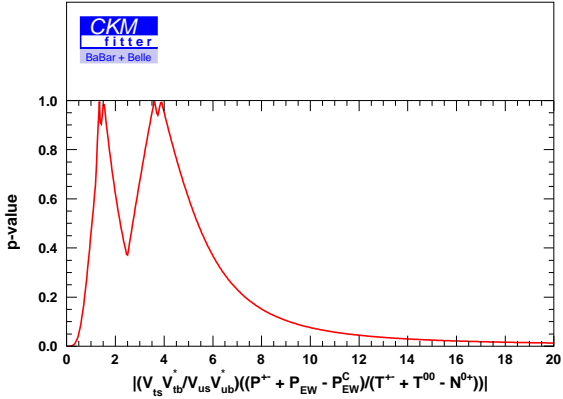}
\caption{
Top: two-dimensional constraint on the modulus and phase of the  $(P^{+-}+P_{\rm EW}-P_{\rm EW}^{\rm C})/(T^{+-}+T_{\rm C}^{00}-N^{0+})$ ratio. 
For convenience, the modulus is multiplied by the ratio of CKM factors appearing in the tree and penguin contributions to the $B^+\rightarrow K^{\star+}\pi^0$  
decay amplitude. Bottom: one-dimensional constraint on the modulus of the  $(P^{+-}+P_{\rm EW}-P_{\rm EW}^{\rm C})/(T^{+-}+T_{\rm C}^{00}-N^{0+})$ ratio.
}
\label{fig:POTp0_BtoKstPi}
\end{center}
\end{figure}

In contrast with the $B^0\rightarrow K^{\star+}\pi^-$ case, in the canonical parametrisation 
Eq.~(\ref{eq:canonicalParametrisation}), the decay amplitude 
for  $B^+\rightarrow K^{\star+}\pi^0$ includes several hadronic contributions both to the total tree and penguin
terms, namely
\begin{eqnarray}
 \sqrt{2}A^{+0} & = & V_{us}V_{ub}^*T^{+0} + V_{ts}V_{tb}^*P^{+0} \\ \nonumber
& = & V_{us}V_{ub}^*(T^{+-}+T_{\rm C}^{00}-N^{0+}) \\ \nonumber
& &+ V_{ts}V_{tb}^*(P^{+-}+P_{\rm EW}-P_{\rm EW}^{\rm C}) \ , 
\end{eqnarray}
and therefore no straightforward constraint on a single pair of hadronic parameters  can be extracted,
as several degenerate combinations can reproduce the observed value of the $CP$ asymmetry $A_{\rm CP}(B^+\rightarrow K^{\star+}\pi^0)$.
This is illustrated in Fig.~\ref{fig:POTp0_BtoKstPi},
where six different local minima are found in the fit, all with similar $\chi^2$ values. 
The three minima with positive strong phases correspond to Solution I, while the three minima with negative
strong phases correspond to Solution II.
The relative size of the total tree and penguin contributions is bound within a relatively narrow range: we get
$|P^{+0}/T^{+0}| \in (0.018,0.126)$ at $68\%$~C.L.

\begin{figure}[t]
\begin{center}
\includegraphics[width=8cm]{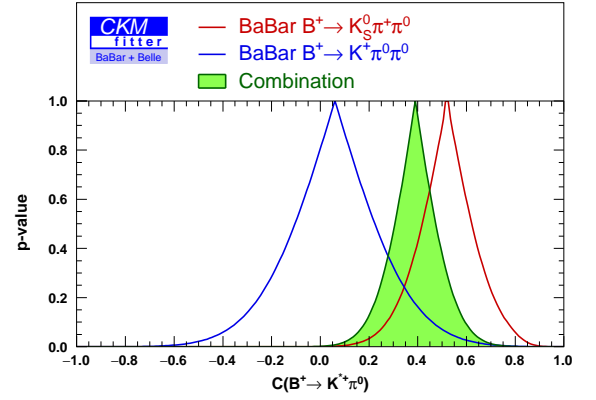}
\caption{
Constraint on the direct $CP$ asymmetry parameter $C(B^+\rightarrow K^{\star+}\pi^0) = -A_{\rm CP}(B^+\rightarrow K^{\star+}\pi^0)$ from {\babar} data on $B^+\to K^0_S\pi^+\pi^0$ (red curve), 
{\babar} data on $B^+\to K^+\pi^0\pi^0$ (blue curve) and the combination (green shaded curve). The constraints are obtained using the observables described in the text.
}
\label{fig:ACP_KstpPi0}
\end{center}
\end{figure}

\subsubsection{Hierarchy among penguins: electroweak penguins}\label{sec:hierarchypenguins}

\begin{figure}[t]
\begin{center}
\includegraphics[width=8cm]{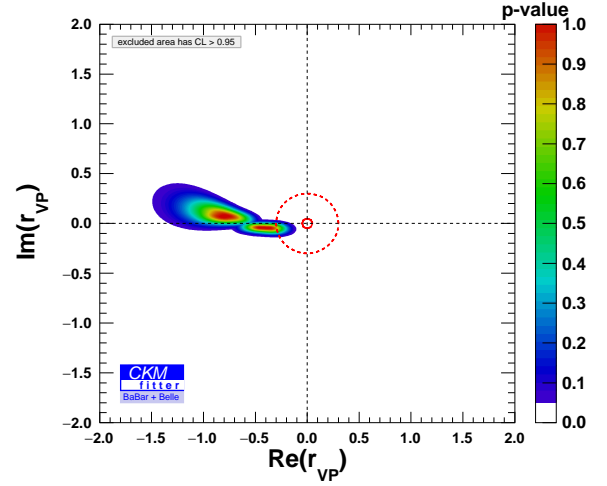}
\caption{
Two-dimensional constraint on real and imaginary parts on the $r_{VP}$ parameter defined in  Eq.~(\ref{eq:PEWfromCPS}). 
The area encircled with the solid (dashed) red line corresponds to the CPS/GPSZ prediction, with a $5\%$ ($30\%$) uncertainty 
on the $r_{\rm VP}$ parameter. 
\label{fig:PewOT3o2}}
\end{center}
\end{figure}

\begin{figure}[t]
\begin{center}
\includegraphics[width=8cm]{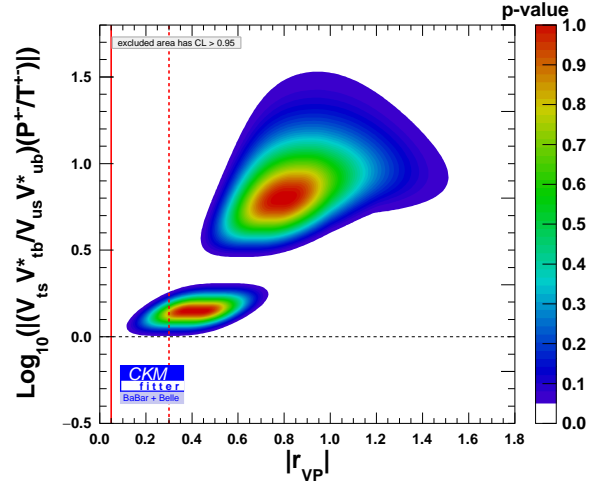}
\caption{
Two-dimensional constraint on $|r_{VP}|$ defined in  Eq.~(\ref{eq:PEWfromCPS}) and ${\rm Log}_{10}\left(|P^{+-}/T^{+-}|\right)$. 
The vertical solid (dashed) red line corresponds 
to the CPS/GPSZ prediction, with a $5\%$ ($30\%$) uncertainty.
\label{fig:POTvsrVP}
}
\end{center}
\end{figure}

In Sec.~\ref{subsec:CPS}, we described the CPS/GPSZ method designed to extract weak phases from $B\to\pi K$ assuming some control on the size of the electroweak penguin. According to this method, the electroweak penguin is expected to yield a small contribution to the decay amplitudes,
with no significant phase difference.
We are actually in a position to test this expectation by fitting the hadronic parameters using the {\babar} and Belle data as inputs. 
Fig.~\ref{fig:PewOT3o2} shows the
two-dimensional constraint on $r_{VP}$, in other words, the ratio $P_{\rm EW}/T_{3/2}$ ratio, showing two local minima.
The CPS/GPSZ prediction is also indicated in this figure. In Fig.~\ref{fig:POTvsrVP}, we provide
 the regions allowed for $|r_{VP}|$ and the modulus of the ratio $|P^{+-}/T^{+-}|$, exhibiting two favoured values, the smaller one 
being associated with Solution I and
 the larger one with Solution II. The latter one corresponds to a significantly large electroweak penguin amplitude and it is clearly incompatible with the CPS/GPSZ prediction by more than one order of magnitude.
A better agreement, yet still marginal, is found for the smaller minimum that corresponds to Solution I:
the central value for the ratio is about a factor of three larger than CPS/GPSZ, and a small, 
positive phase is preferred. For this minimum,
an inflation of the uncertainty on  $\left| r_{\rm VP}\right|$ up to $30\%$ would be needed to ensure proper agreement. 
In any case, it is clear that the data prefers a larger value of $|r_{\rm VP}|$ than the estimates originally proposed.

\begin{figure}[t]
\begin{center}
\includegraphics[width=8cm]{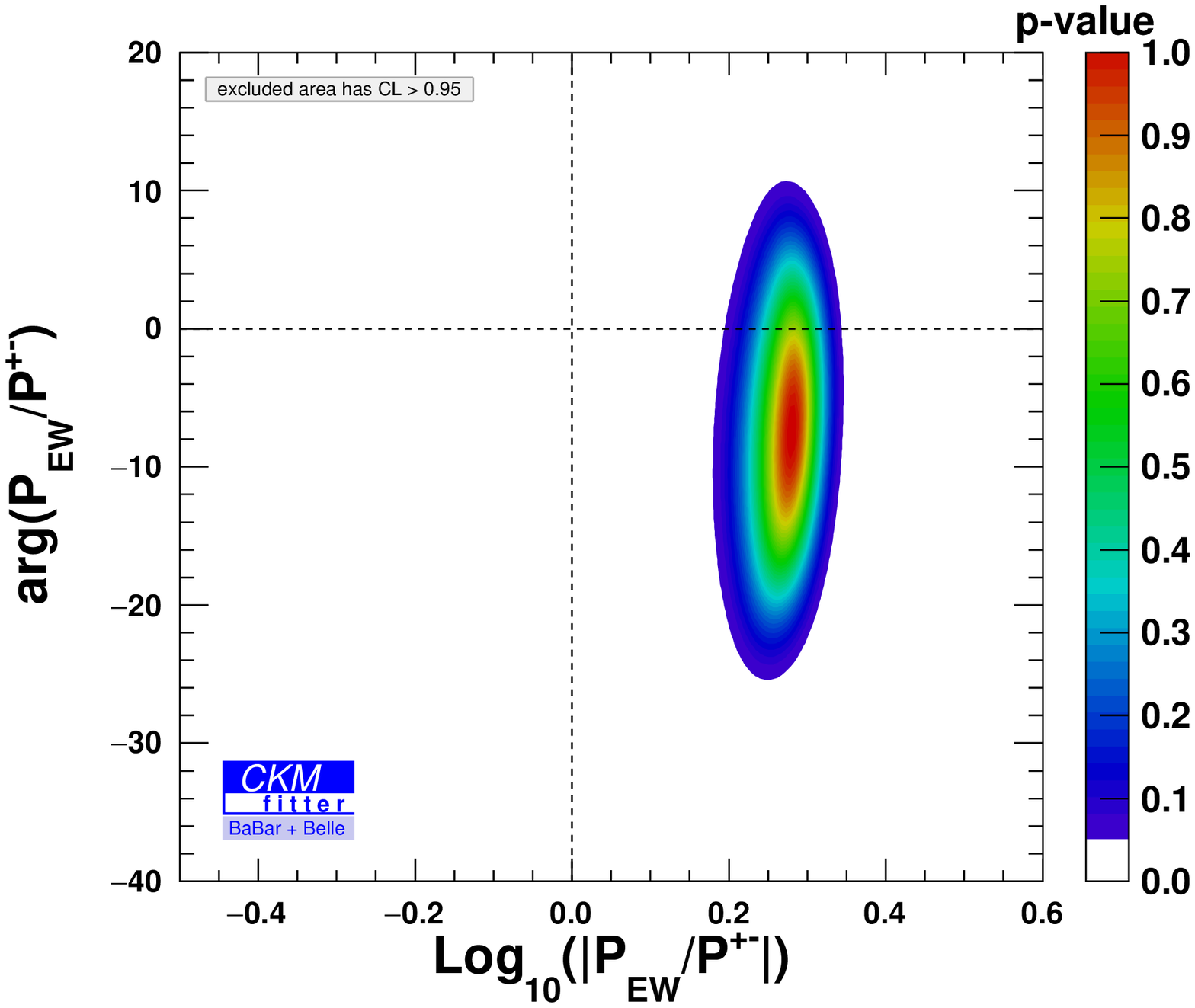}
\includegraphics[width=8cm]{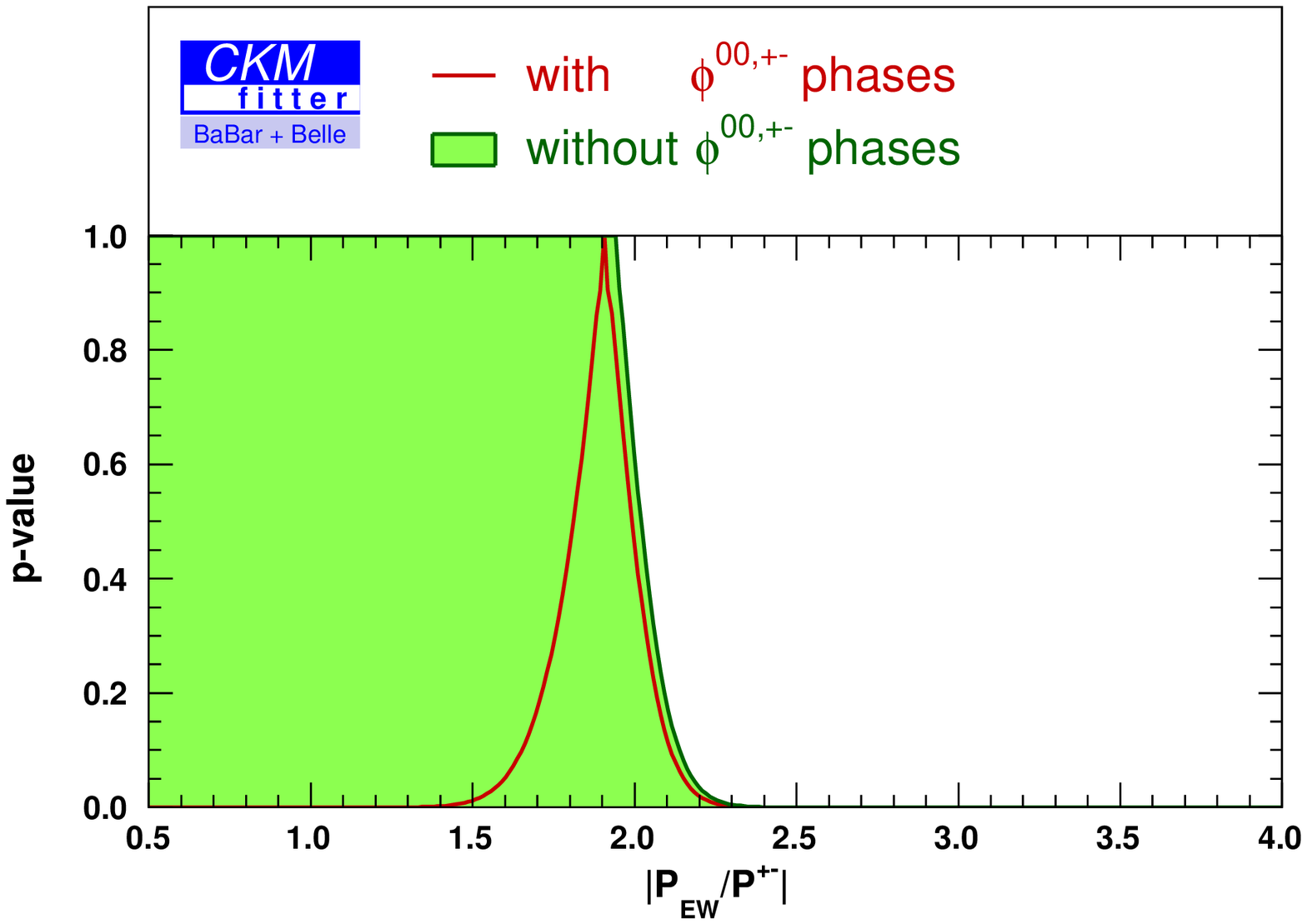}
\caption{
Top: two-dimensional constraint on the modulus and phase of the complex $P_{\rm EW}/P^{+-}$ ratio.
Bottom: constraint on the $\left|P_{\rm EW}/P^{+-}\right|$ ratio, using the complete set of experimental inputs (red curve),
and removing the {\babar} measurement of the $\varphi^{00,+-}$ phases from the $B^0\rightarrow K^+\pi^+\pi^0$ Dalitz-plot analysis (green shaded curve).
\label{fig:PewOP}}
\end{center}
\end{figure}

Moreover, the contribution from the electroweak penguin is found to be about twice larger than the main penguin 
contribution $P^{+-}$. This is illustrated
in Fig.~\ref{fig:PewOP}, where only one narrow solution is found in the $P_{\rm EW}/P^{+-}$ plane,
as both solutions I and II provide essentially the same constraint. The relative phase between these two parameters
is bound to the interval $(-25,+10)^\circ$ at $95\%$~C.L.
Additional tests allow us to demonstrate  that this strong constraint on the relative
$P_{\rm EW}/P^{+-}$ penguin contributions is predominantly driven by the $\varphi^{00,+-}$ phase differences 
measured in the {\babar}
Dalitz-plot analysis of $B^0\rightarrow K^+\pi^+\pi^0$ decays. The strong constraint on the $P_{\rm EW}/P^{+-}$ ratio
is turned into a mild upper bound when removing the $\varphi^{00,+-}$ phase differences
from the experimental inputs. The addition of these two observables as fit inputs increases the 
minimal $\chi^2$ by 7.7 units, which
corresponds to a 2.6~$\sigma$ discrepancy. Since the latter is 
 driven by a measurement from a single experiment, 
additional experimental results are needed to confirm such a large value for the electroweak penguin parameter.

\begin{figure}[t]
\begin{center}
\includegraphics[width=8cm]{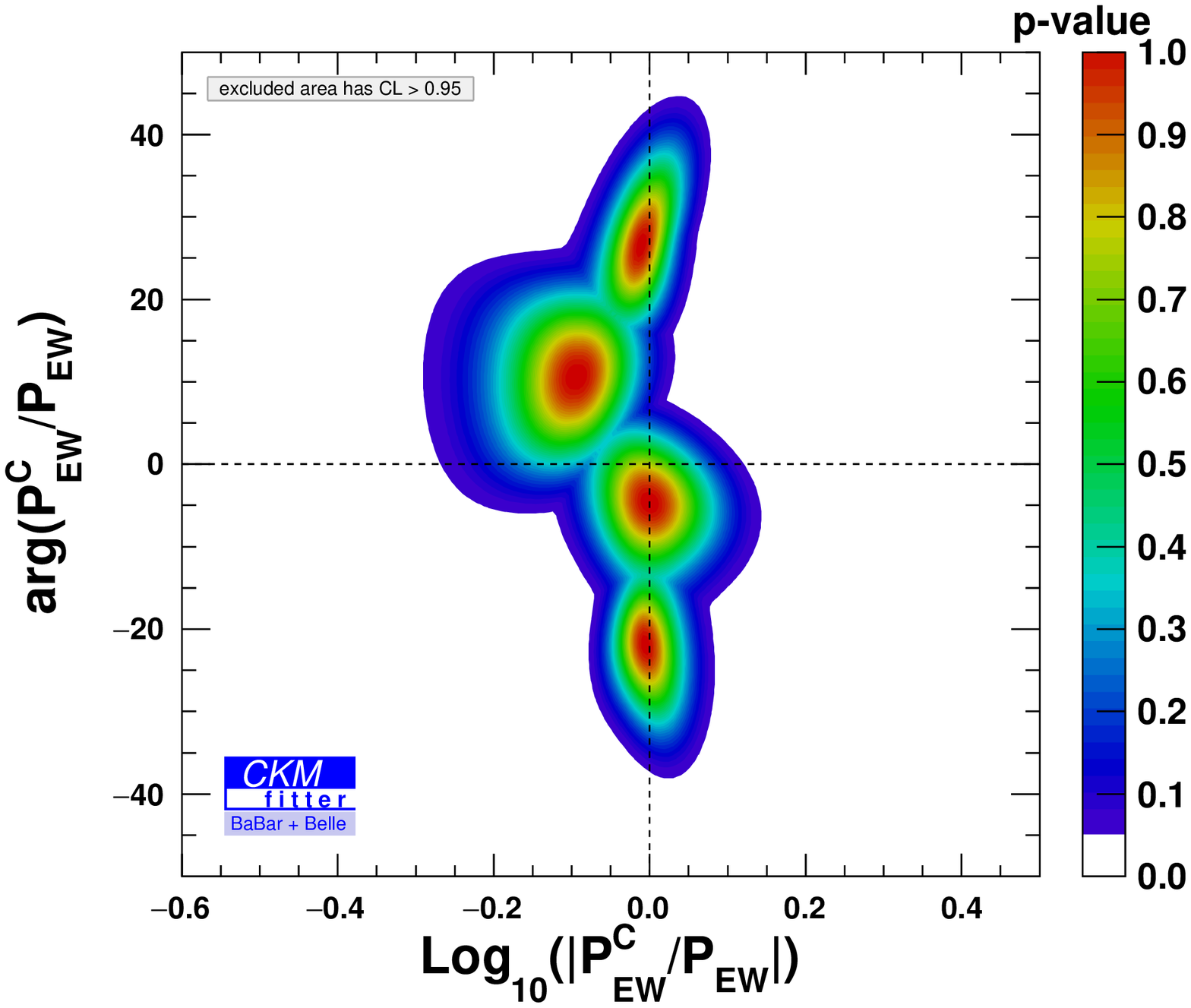}
\includegraphics[width=8cm]{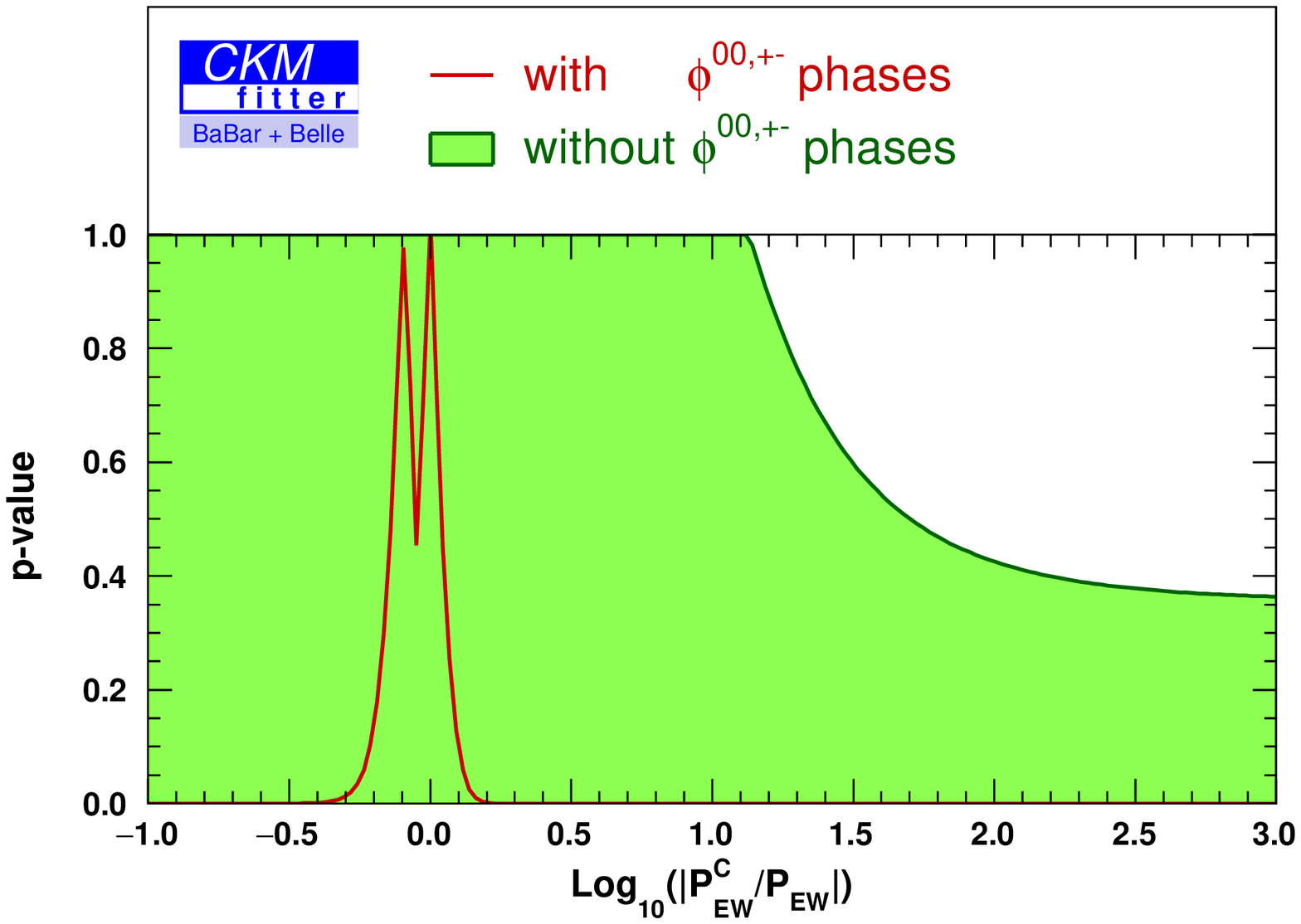}
\caption{
Top: two-dimensional constraint  on the modulus and phase of the $P_{\rm EW}^{\rm C}/P_{\rm EW}$ ratio. 
Bottom: one-dimensional constraint on ${\rm Log}_{10}\left(\left|P_{\rm EW}^{\rm C}/P_{\rm EW}\right|\right)$,
using the complete set of experimental inputs (red curve), and removing the {\babar} measurement of the $\varphi^{00,+-}$ 
phases from the $B^0\rightarrow K^+\pi^+\pi^0$ Dalitz-plot analysis (green shaded curve).
}
\label{fig:PewCOPew}
\end{center}
\end{figure}

In view of colour suppression, the  electroweak penguin $P_{\rm EW}^{\rm C}$ is expected to yield a smaller 
contribution than $P_{\rm EW}$ to the decay amplitudes. This hypothesis is
tested in Fig.~\ref{fig:PewCOPew}, which shows that current data favours a similar size for the 
two contributions, and a small relative phase  (up to $40^{\circ}$)
between the colour-allowed and the colour-suppressed electroweak penguins. Both Solutions I and II show the same
structure with four different local minima.

\subsubsection{Hierarchy among tree amplitudes: colour suppression and annihilation}

\begin{figure}[t]
\begin{center}
\includegraphics[width=7cm]{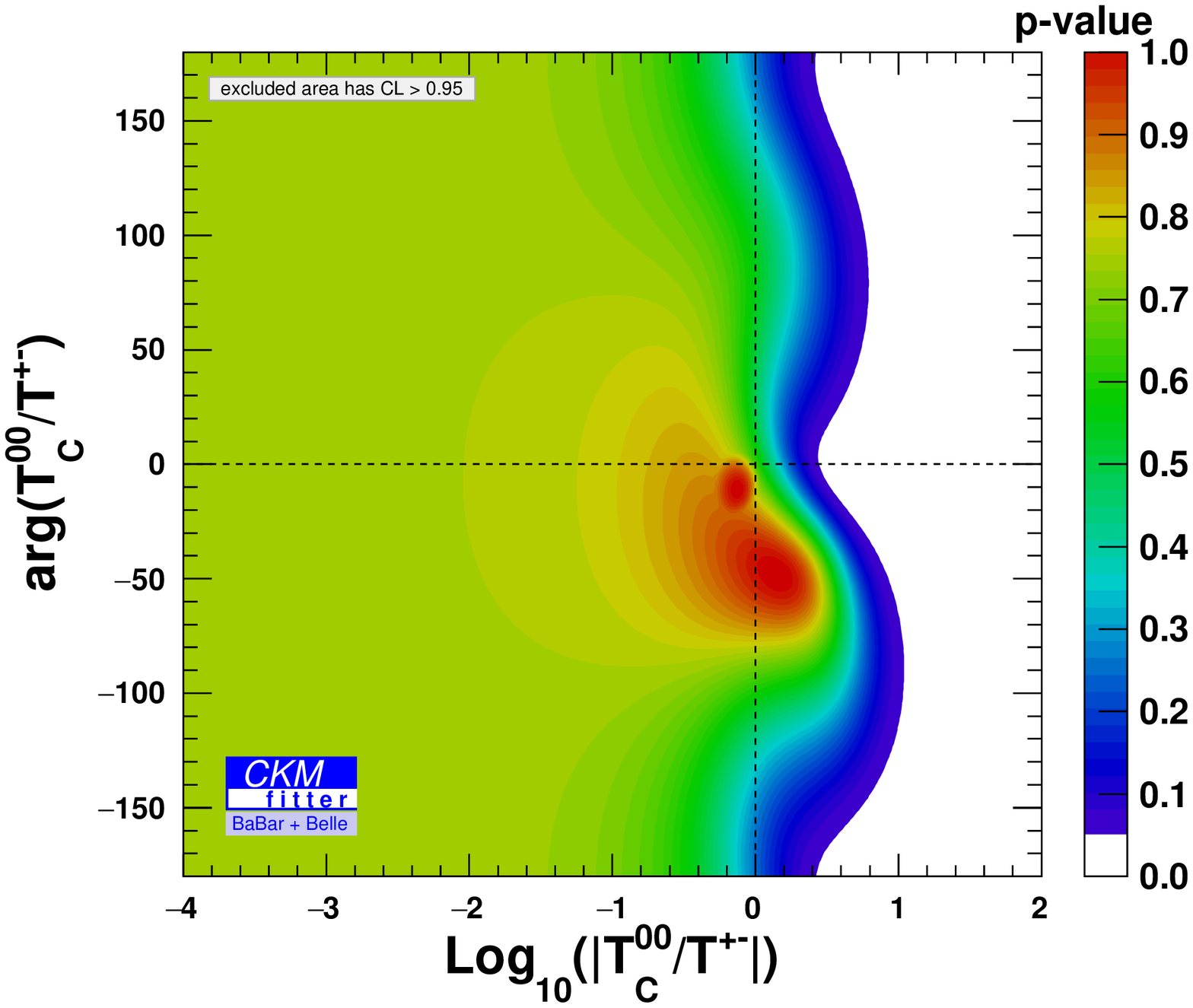}
\includegraphics[width=7cm]{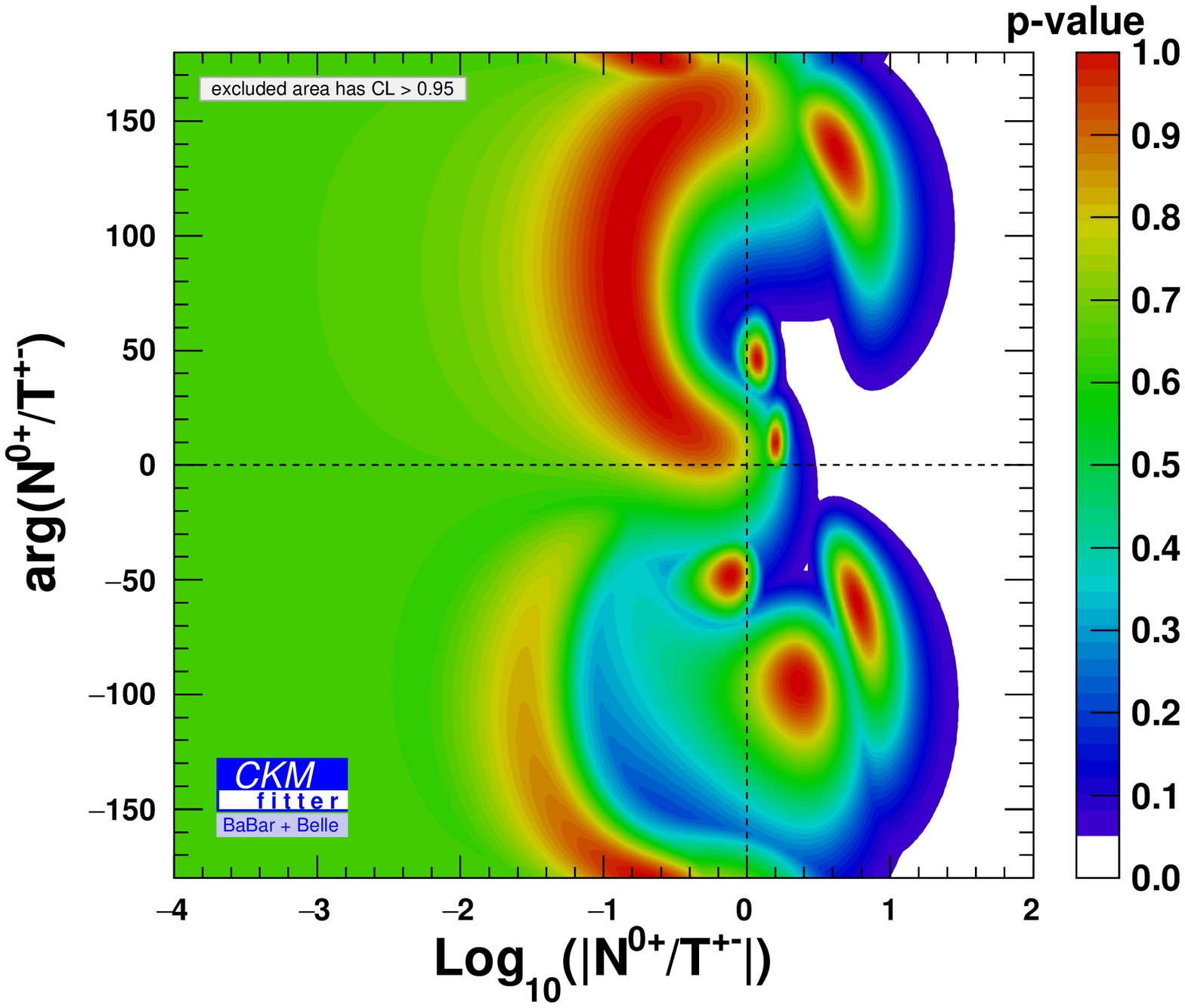}
\caption{
Two-dimensional constraint on the modulus and phase of the $T^{00}_{\rm C}/T^{+-}$ (top) and $N^{0+}/T^{+-}$ (bottom) ratios. 
}
\label{fig:T00OTpm_N0pOTpm}
\end{center}
\end{figure}

As already discussed,
the  hadronic parameter $T^{00}_{\rm C}$ is expected to be suppressed with respect to the main tree parameter $T^{+-}$. 
Also, the annihilation topology is expected to provide negligible contributions to the decay amplitudes.
These expectations can be compared with the extraction of these hadronic parameters from data in
Fig.~\ref{fig:T00OTpm_N0pOTpm}. 

For colour suppression, the current data provides no constraint on the relative phase
between the $T^{00}_{\rm C}$ and  $T^{+-}$ tree parameters,
and only a mild upper bound on the 
modulus can be inferred; the tighter constraint is provided by Solution I that 
excludes values of $|T^{00}_{\rm C}/T^{+-}|$ larger than $1.6$ at $95\%$~C.L. The constraint from Solution II is more than one order of
magnitude looser.

Similarly, for annihilation, Solution I provides slightly tighter constraints on its contribution to the total
tree amplitude with the bound $|N^{0+}/T^{+-}|<2.5$ at $95\%$~C.L., while the bound from Solution II  is much looser.

\subsection{Comparison with theoretical expectations}\label{QCDFcomparison}

\begin{figure*}[t]
\begin{center}
\includegraphics[width=6.5cm]{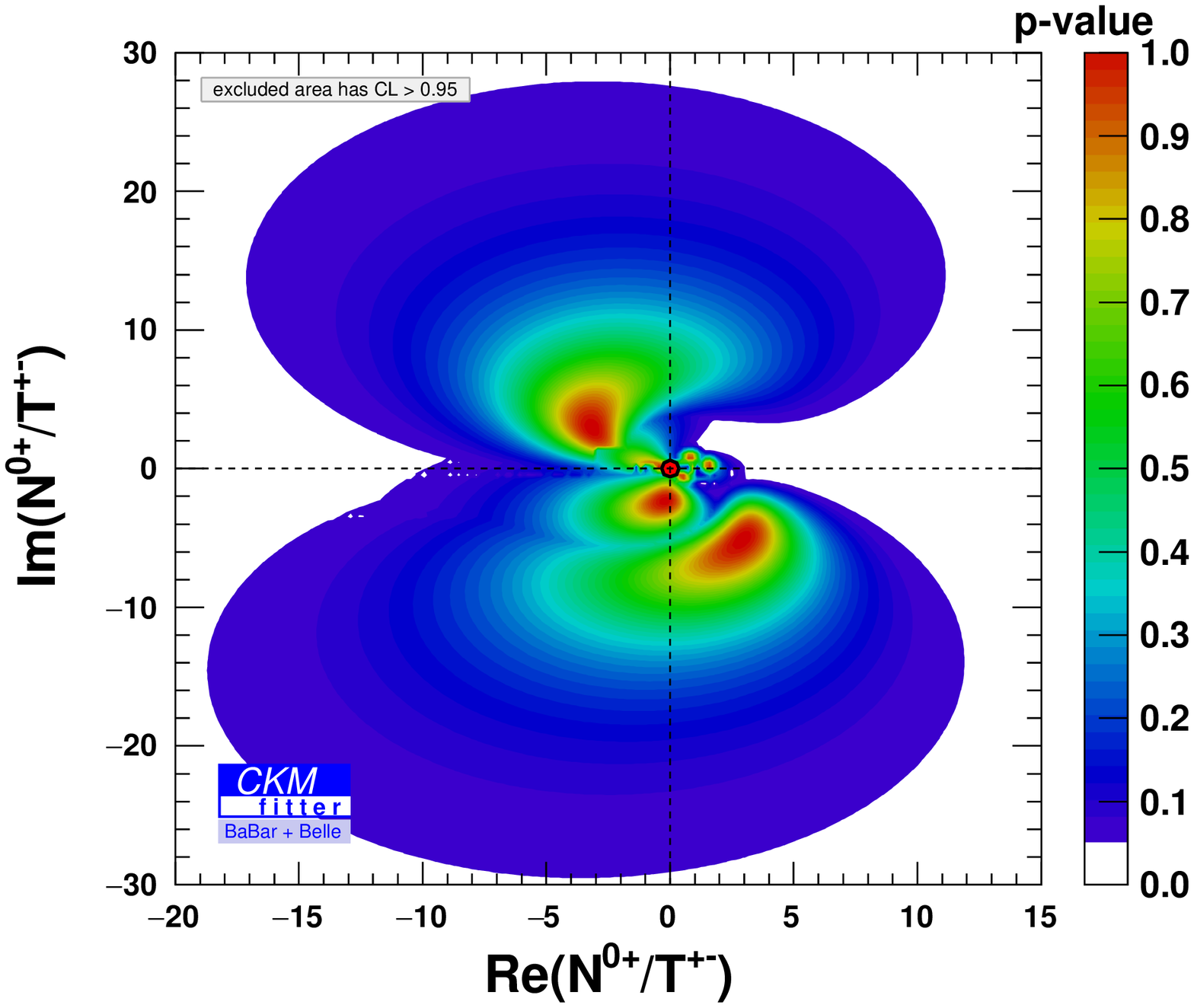}  \includegraphics[width=6.5cm]{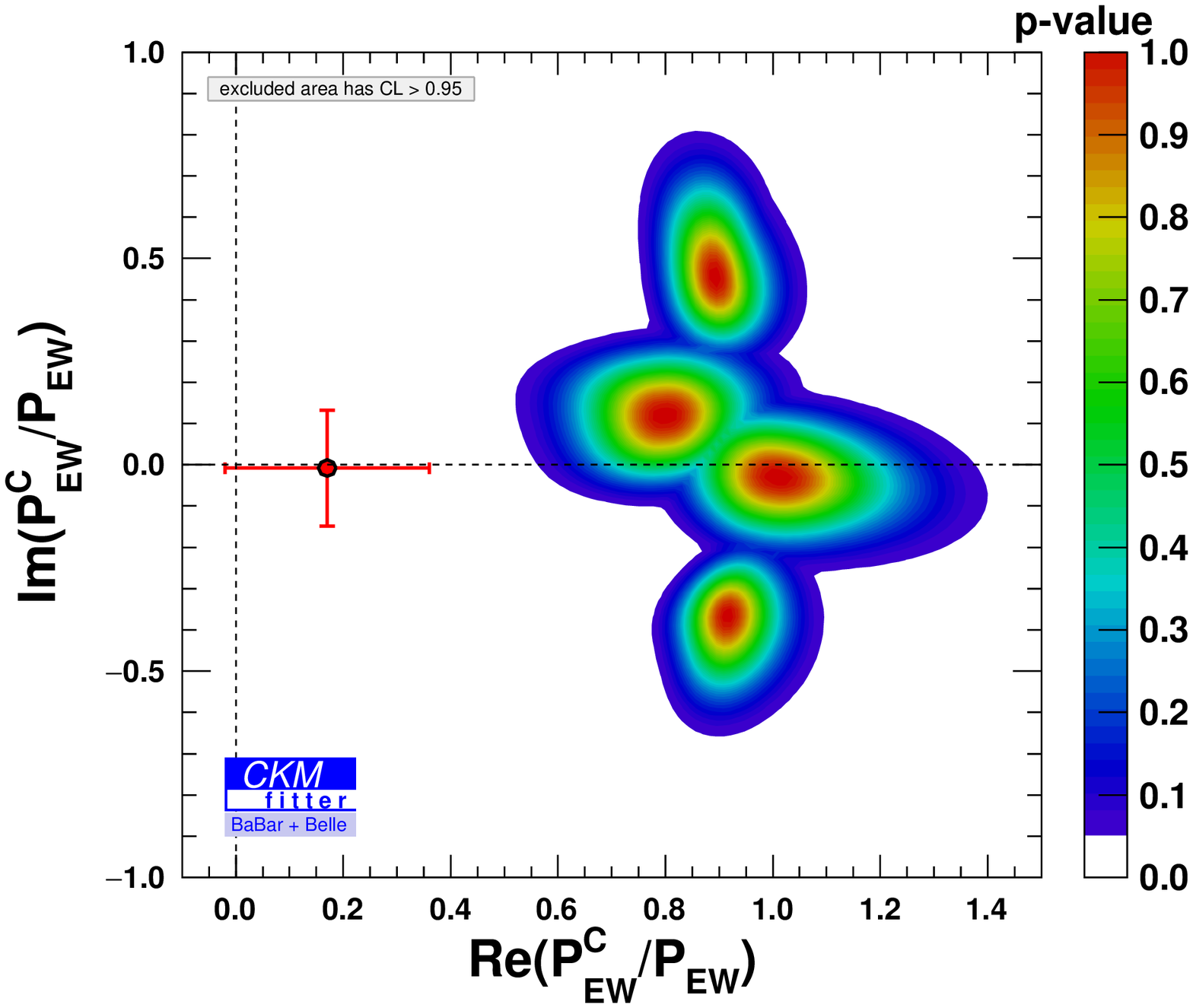}
\includegraphics[width=6.5cm]{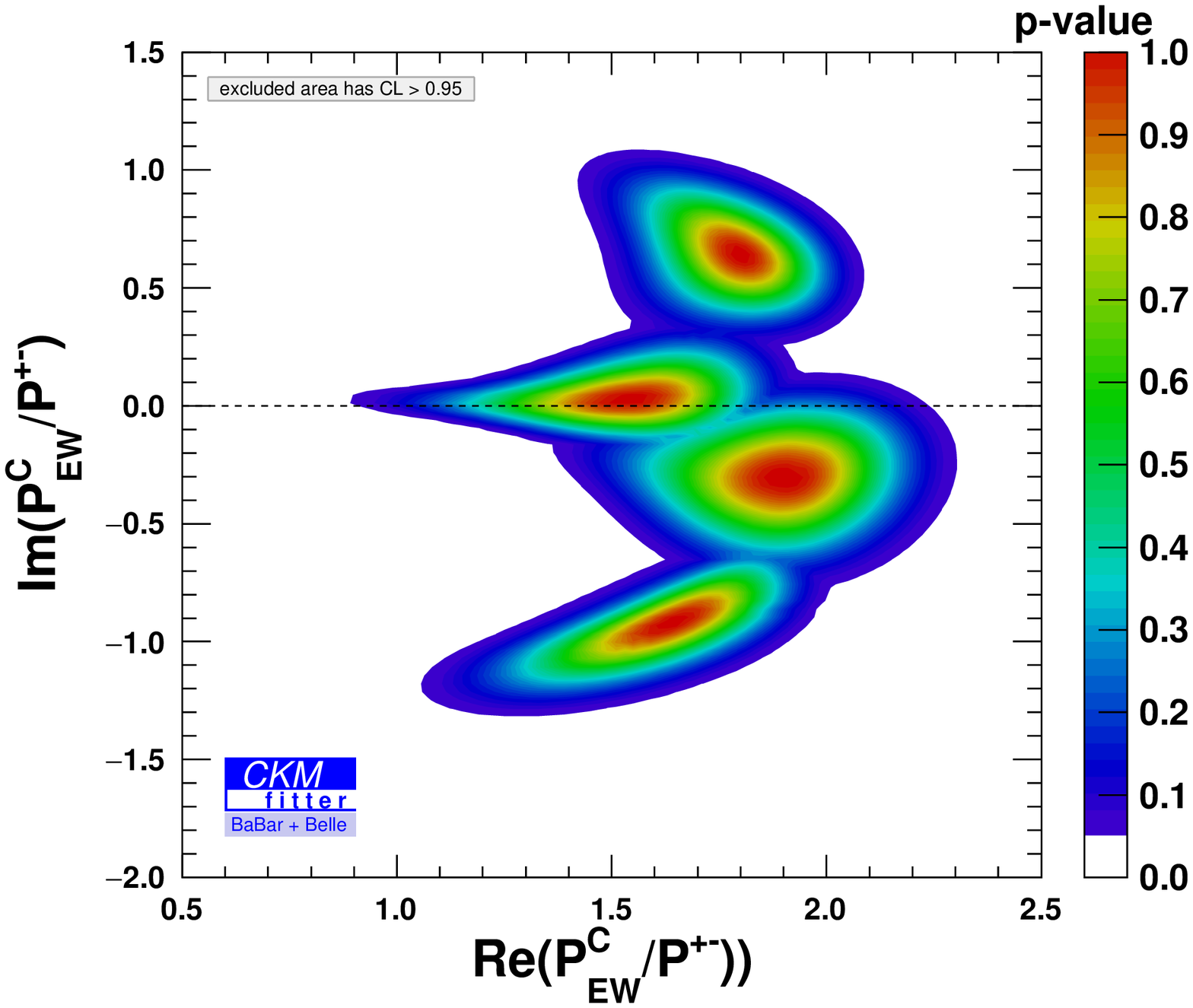} \includegraphics[width=6.5cm]{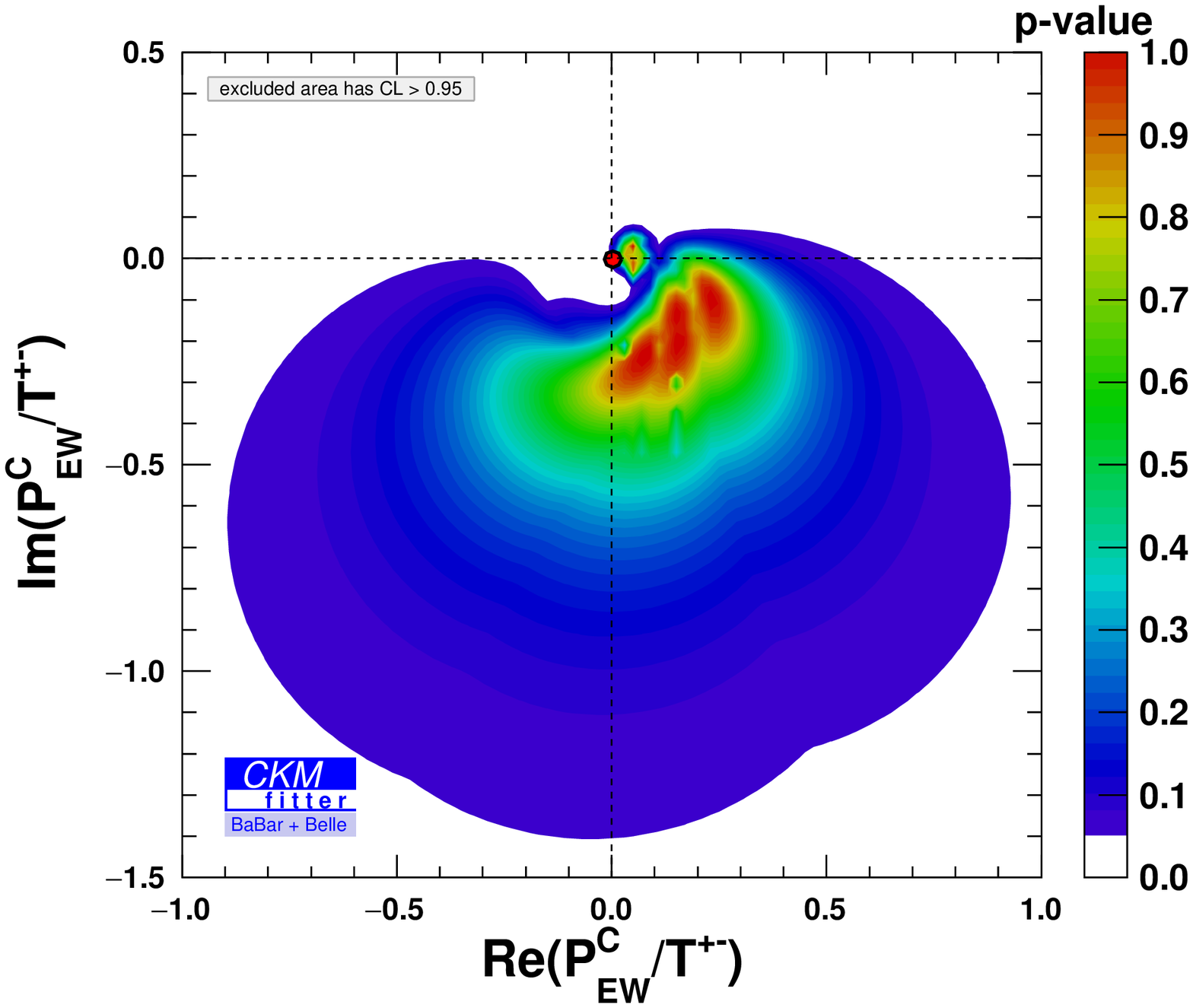}
\includegraphics[width=6.5cm]{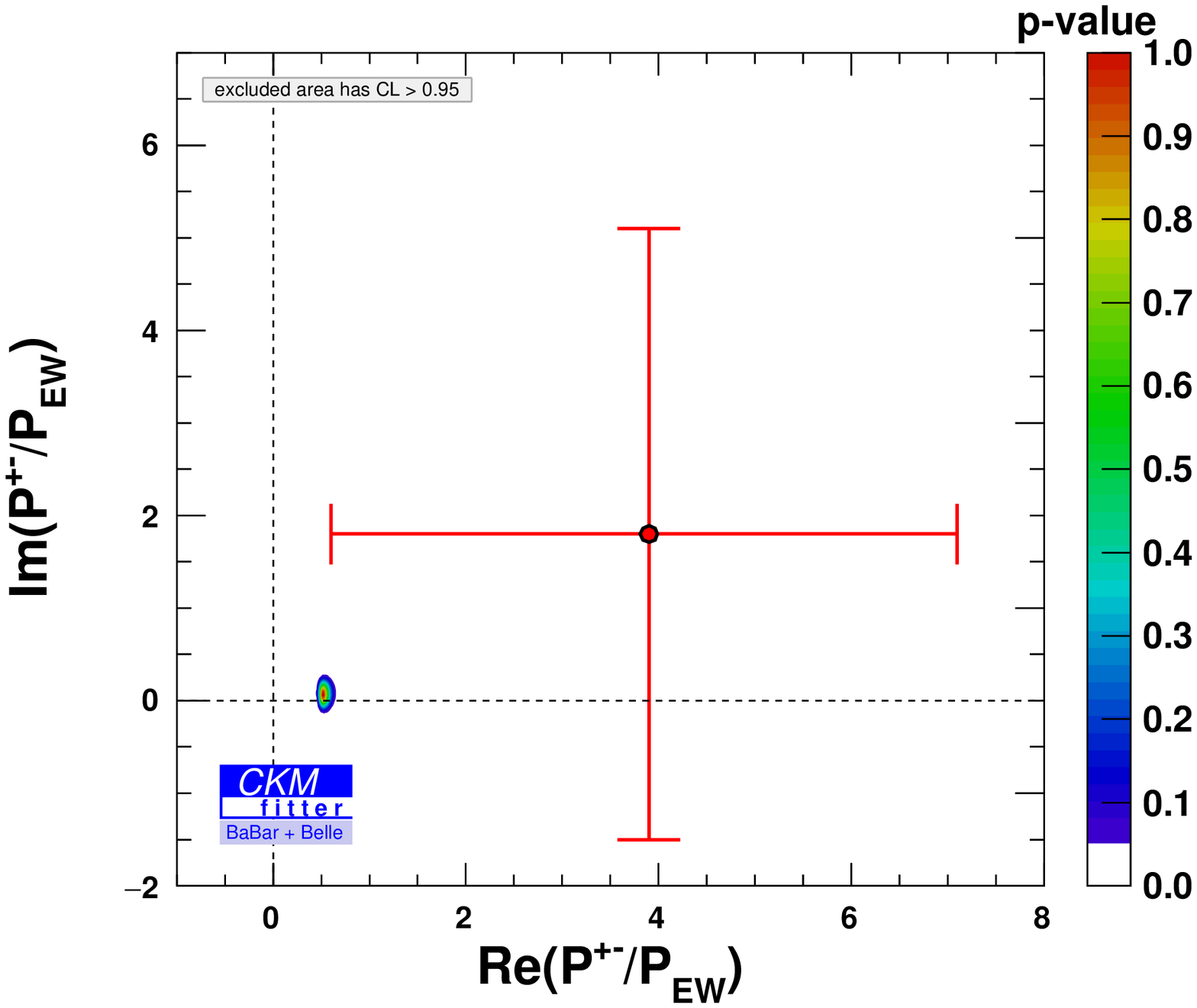}  \includegraphics[width=6.5cm]{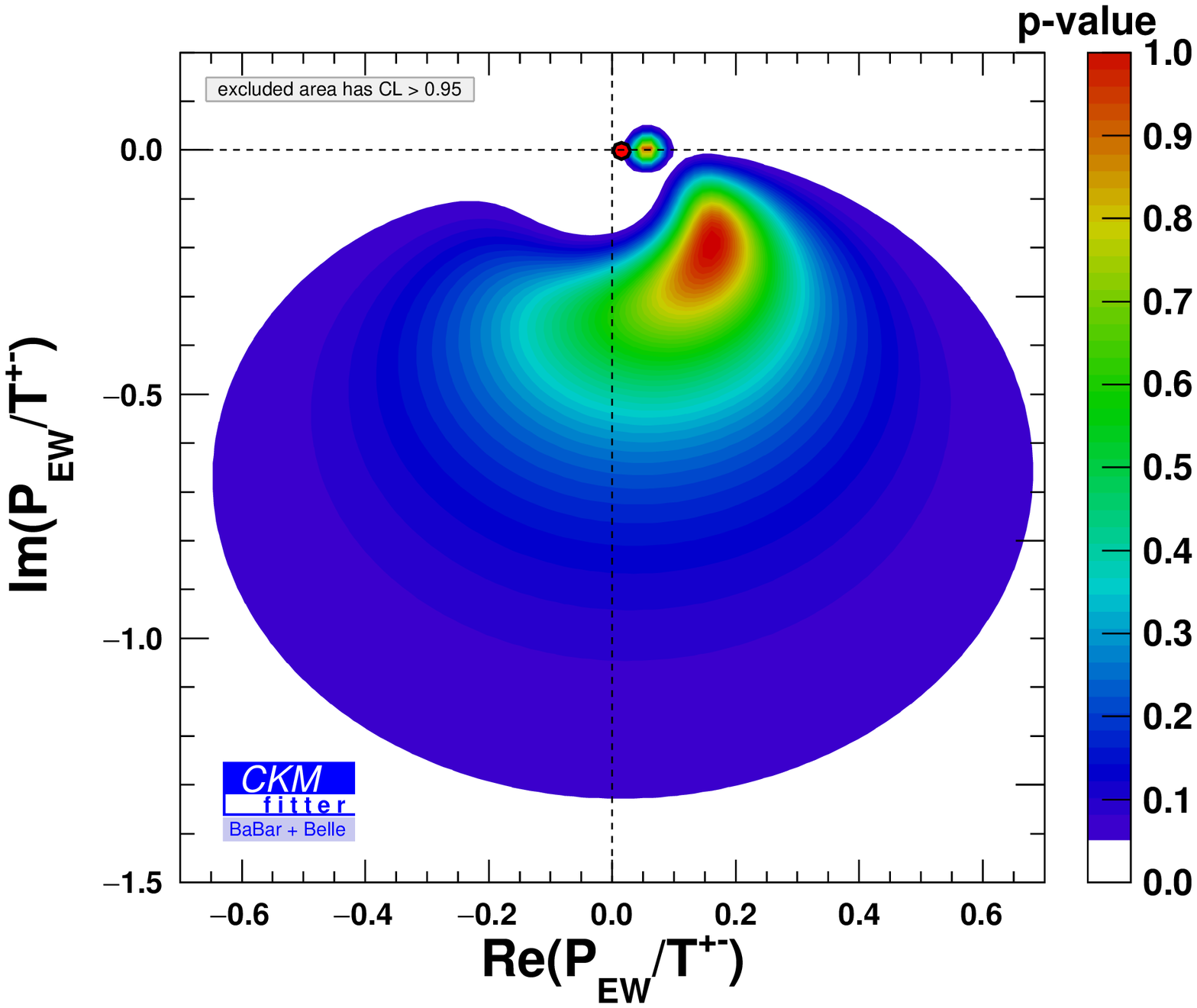}
\includegraphics[width=6.5cm]{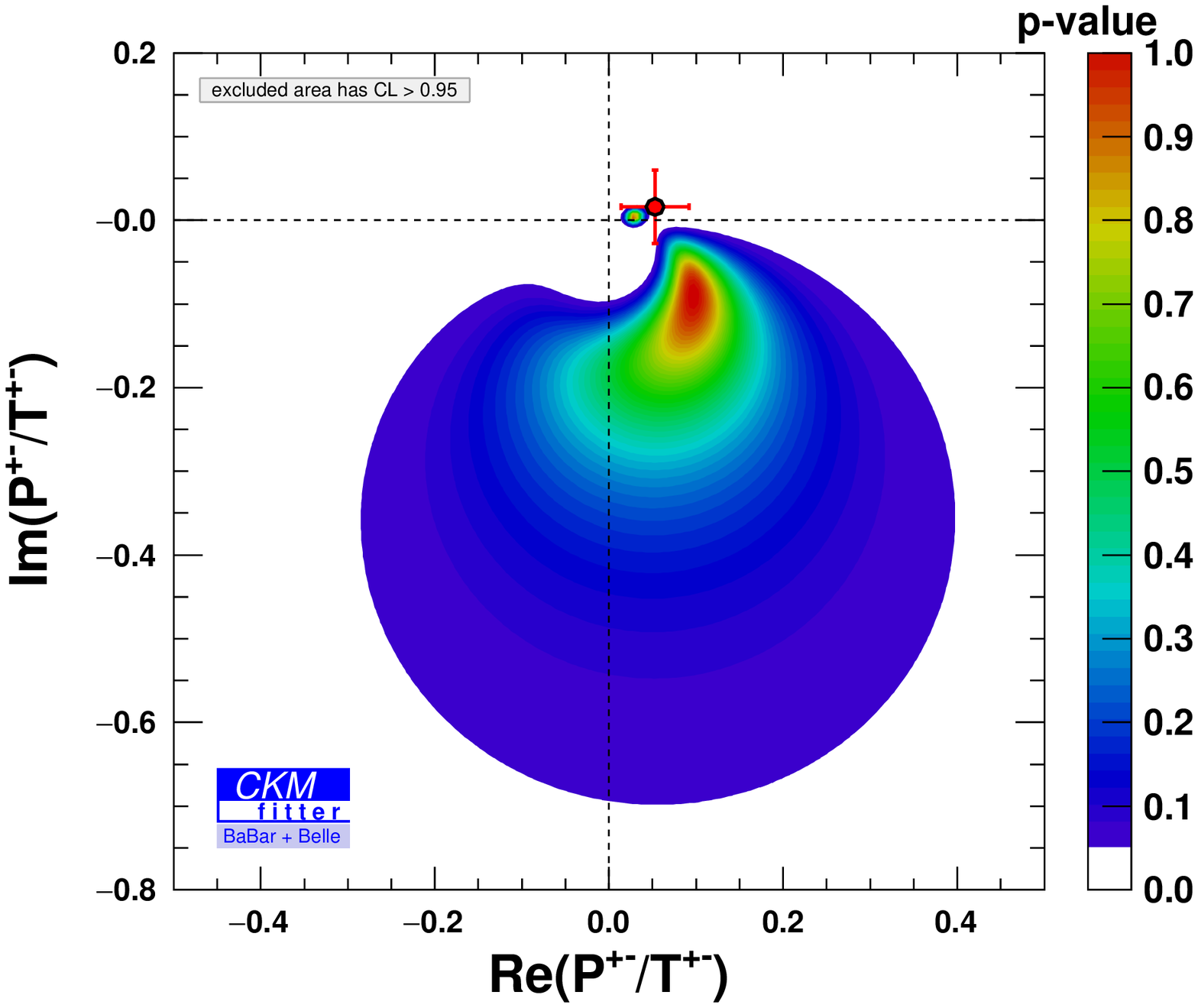}  \includegraphics[width=6.5cm]{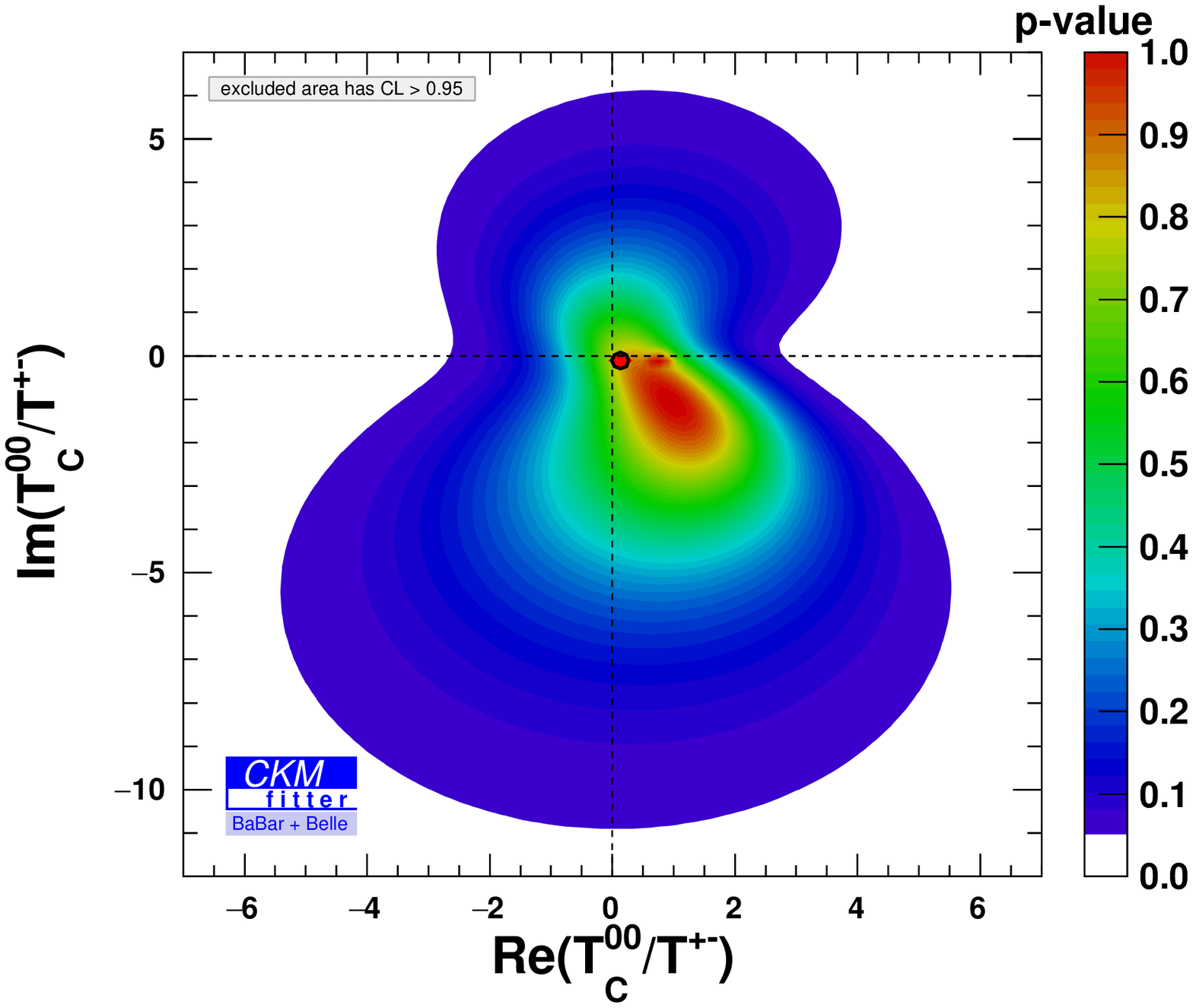}
\caption{Two-dimensional constraints on the real and imaginary parts of hadronic ratios, respectively from left to right and from top to bottom: 
$N^{0+}/T^{+-}$, $P_{\rm EW}^{\rm C}/P_{\rm EW}$, $P_{\rm EW}^{\rm C}/P^{+-}$, $P_{\rm EW}^{\rm C}/T^{+-}$, $P^{+-}/P_{\rm EW}$, $P_{\rm EW}/T^{+-}$, $P^{+-}/T^{+-}$ and $T^{00}_{\rm C}/T^{+-}$. 
The red crosses and dots indicate our predictions based on QCD factorisation. No prediction is given for the ratio ${P_{\rm EW}^{\rm C}}/{P^{+-}}$ due to numerical instabilities (see text).}
\label{fig:QCDFcomparison2D}
\end{center}
\end{figure*}

\begin{table*}[t]
 \begin{center}
\begin{tabular}{l|cc}
Quantity                                 &    Fit result &         QCDF \\ 
\hline
$\displaystyle {\mathcal Re}\frac{N^{0+}}{T^{+-}}$     &   $(-5.31, 4.73)$         &    $0.011 \pm 0.027$\\
$\displaystyle {\mathcal Im}\frac{N^{0+}}{T^{+-}}$     &   $(-9.59, 7.73)$         &    $0.003\pm 0.028$\\

$\displaystyle {\mathcal Re}\frac{P_{\rm EW}^{\rm C}}{P_{\rm EW}}$   &   $(0.69,1.14)$           &    $0.17\pm 0.19$\\
$\displaystyle {\mathcal Im}\frac{P_{\rm EW}^{\rm C}}{P_{\rm EW}}$   &   $(-0.48,-0.28)~\cup~(-0.13,0.22)~\cup$          &    $-0.08\pm0.14$\\
                                         &   $(0.34,0.60)$                  &         \\
                                         
$\displaystyle {\mathcal Re}\frac{P_{\rm EW}^{\rm C}}{P^{+-}}$   &   $(1.29,2.08)$           &    $-$\\
$\displaystyle {\mathcal Im}\frac{P_{\rm EW}^{\rm C}}{P^{+-}}$   &   $(-1.09,-0.75)~\cup~(-0.51,-0.10)~\cup$          &    $-$\\
                                         &   $(-0.08,0.16)~\cup~(0.47,0.83)$                  &         \\
                                         
$\displaystyle {\mathcal Re}\frac{P_{\rm EW}^{\rm C}}{T^{+-}}$   &   $(-0.12,0.34)$          &    $0.0027\pm 0.0031$\\
$\displaystyle {\mathcal Im}\frac{P_{\rm EW}^{\rm C}}{T^{+-}}$   &   $(-0.42,0.05)$          &    $-0.0015^{+0.0024}_{-0.0025}$\\

$\displaystyle {\mathcal Re}\frac{P^{+-}}{P_{\rm EW}}$     &   $(0.49,0.56)$           &    $3.9^{+3.2}_{-3.3}$\\
$\displaystyle {\mathcal Im}\frac{P^{+-}}{P_{\rm EW}}$     &   $(-0.03,0.16)$          &    $1.8\pm 3.3$\\

$\displaystyle {\mathcal Re}\frac{P_{\rm EW}}{T^{+-}}$     &   $(0.0, 0.25)$          &    $0.0154^{+0.0059}_{-0.0060}$\\
$\displaystyle {\mathcal Im}\frac{P_{\rm EW}}{T^{+-}}$     &   $(-0.40,-0.09)~\cup~(-0.02,0.02)$          &    $-0.0014^{+0.0023}_{-0.0022}$\\

$\displaystyle {\mathcal Re}\frac{P^{+-}}{T^{+-}}$     &   $( 0.023,0.140)$          &    $0.053\pm0.039$\\
$\displaystyle {\mathcal Im}\frac{P^{+-}}{T^{+-}}$     &   $(-0.20,-0.04)~\cup~(0.0, 0.01)$          &    $0.016\pm0.044$\\

$\displaystyle {\mathcal Re}\frac{T^{00}_{\rm C}}{T^{+-}}$   &   $(-0.26,2.24)$          &    $0.13\pm0.17$\\
$\displaystyle {\mathcal Im}\frac{T^{00}_{\rm C}}{T^{+-}}$   &   $(-3.28,0.74)$          &    $-0.11\pm0.15$
\end{tabular}
\caption{$68\%$ confidence intervals for the real and imaginary parts of hadronic ratios according to our 
fit and the corresponding predictions in our implementation of QCD factorisation (QCDF). No prediction is given for the ratio ${P_{\rm EW}^{\rm C}}/{P^{+-}}$ due to numerical instabilities (see text).}\label{tab:QCDFcomparison1D}
\end{center}
\end{table*}

We have extracted the values of the hadronic amplitudes from the data currently available. It may prove interesting to compare these results with theoretical expectations. 
For this exercise, we use QCD factorisation~\cite{Beneke:1999br,Beneke:2000ry,Beneke:2003zv,Beneke:2006hg} as a benchmark point, keeping in mind that other approaches (discussed in the introduction) are available. In order to keep the comparison simple and meaningful, we consider the real and imaginary part of several ratios of hadronic amplitudes.

We obtain our theoretical values in the following way. We follow Ref.~\cite{Beneke:2003zv} for the expressions within QCD factorisation, and we use the same model for the power-suppressed and infrared-divergent contributions   coming from hard scattering and weak annihilation: these contributions are formally $1/m_b$-suppressed but numerically non negligible, and play a crucial role in some of the amplitudes. On the other hand, we update the hadronic parameters in order to take into account more recent determinations of these quantities, see App.~\ref{app:QCDFinputs}. We use the Rfit scheme to handle theoretical uncertainties~\cite{Charles:2004jd,Charles:2015gya,CKMfitterwebsite,Charles:2016qtt} (in particular for the hadronic parameters and the $1/m_b$ power-suppressed contributions), and we compute only ratios of hadronic amplitudes using QCD factorisation.
We stress that we provide the estimates within QCD factorisation simply to compare the results of our experimental fit for the hadronic amplitudes with typical theoretical expectations concerning the same quantities. In particular we neglect Next-Next-to-Leading Order corrections that have been partially computed in Refs.~\cite{BHWS,Bell:2007tv,Bell:2009nk,Beneke:2009ek,Bell:2015koa}, and we do not attempt to perform a fully combined fit of the theoretical predictions with the experimental data, as the large uncertainties would make the interpretation difficult. 

Our results for the ratios of hadronic amplitudes are shown in Fig.~\ref{fig:QCDFcomparison2D} and in Tab.~\ref{tab:QCDFcomparison1D}. We notice that for most of the ratios, a good agreement is found. The global fit to the experimental data has often much larger uncertainties than theoretical predictions: with better data in the future, we may be able to perform very non trivial tests of the non-leptonic dynamics and the isobar approximation. The situation for $P_{\rm EW}^{\rm C}/P_{\rm EW}$ is slightly different, since the two determinations (experiment and theory) exhibit similar uncertainties and disagree with each other, providing an interesting test for QCD factorisation, which however goes beyond the scope of this study.

There are two cases where the theoretical output from QCD factorisation is significantly less precise than the constraints from the combined fit. For $P_{\rm EW}^C/P^{+-}$, both numerator and denominator can be (independently) very small in QCD factorisation, and numerical instabilities in this ratio prevent us from having a precise prediction. For $P^{+-}/P_{\rm EW}$, the impressively accurate experimental determination, as discussed in Sec.~\ref{sec:hierarchypenguins}, is predominantly driven by the $\varphi^{00,+-}$ phase differences  measured in the {\babar} Dalitz-plot analysis of $B^0\rightarrow K^+\pi^+\pi^0$ decays. Removing this input yields a much milder constraint on $P^{+-}/P_{\rm EW}$. On the other hand in QCD factorisation, the formally leading contributions to the $P^{+-}$ penguin amplitude are somewhat numerically suppressed, and compete with the model estimate of power corrections: due to the Rfit treatment used, the two contributions can either compensate each other almost exactly or add up 
coherently, leading to a $\sim\pm 100\%$ relative uncertainty, which 
is only in marginal agreement with the fit output. Thus we conclude that the $P^{+-}/P_{\rm EW}$ ratio is both particularly sensitive to the power corrections to QCD factorisation and experimentally well constrained, so that it can be used to provide an insight on non factorisable contributions, provided one assumes negligible effects from New Physics.

\section{Prospects for LHCb and Belle II} \label{sec:prospect}

\begin{figure*}[t]
\begin{center}
\includegraphics[width=7cm]{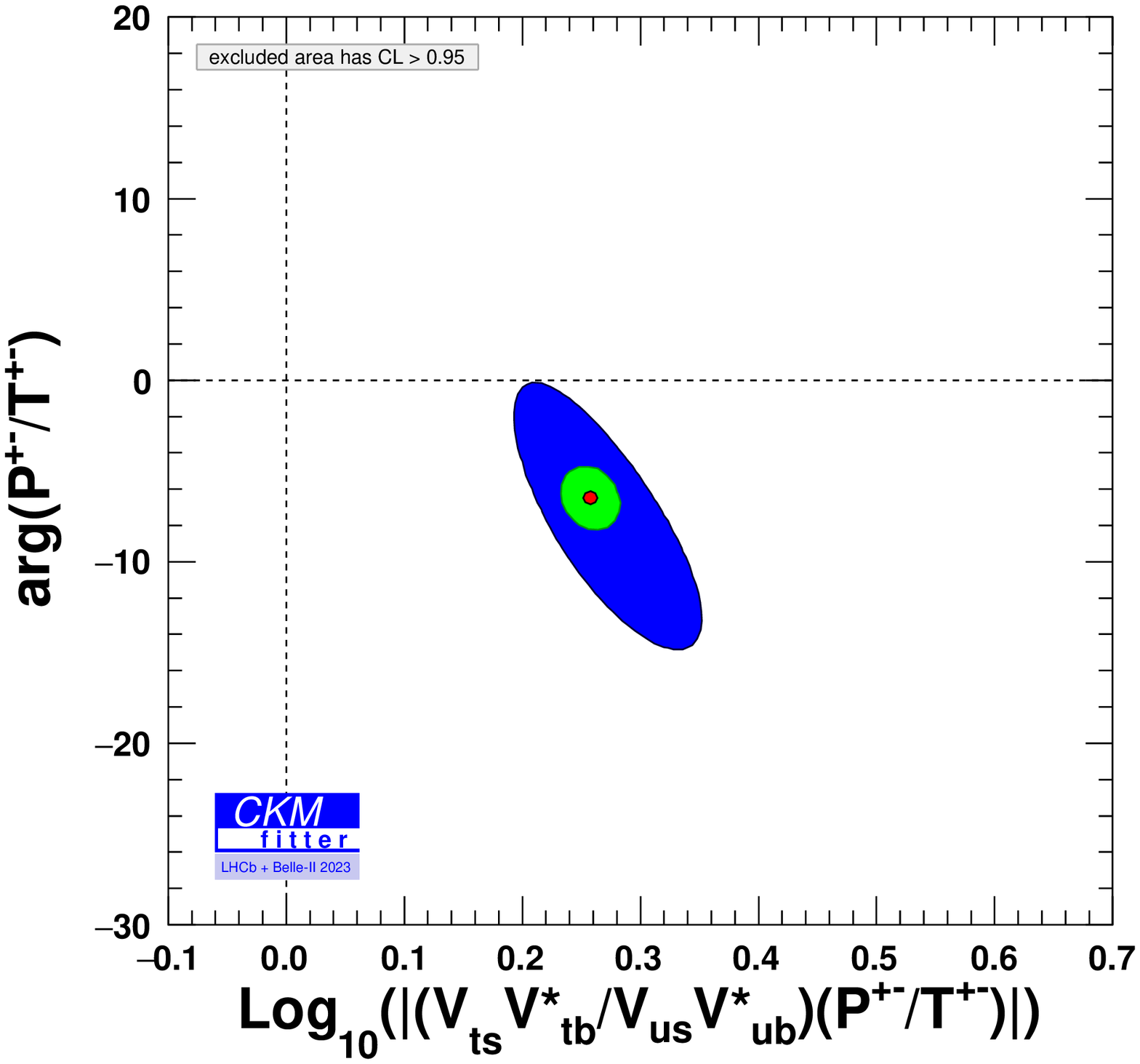}
\includegraphics[width=7cm]{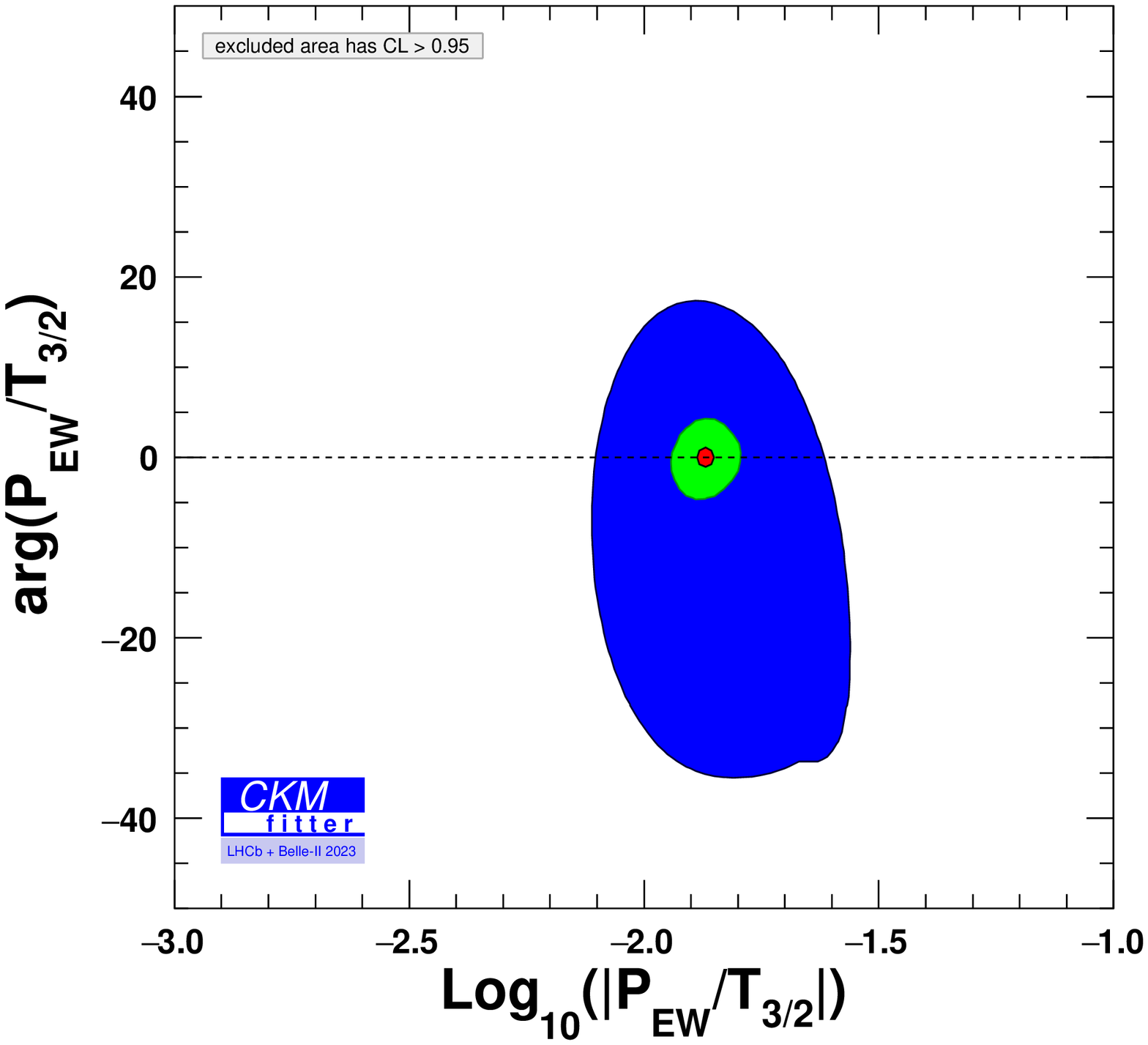}
\includegraphics[width=7cm]{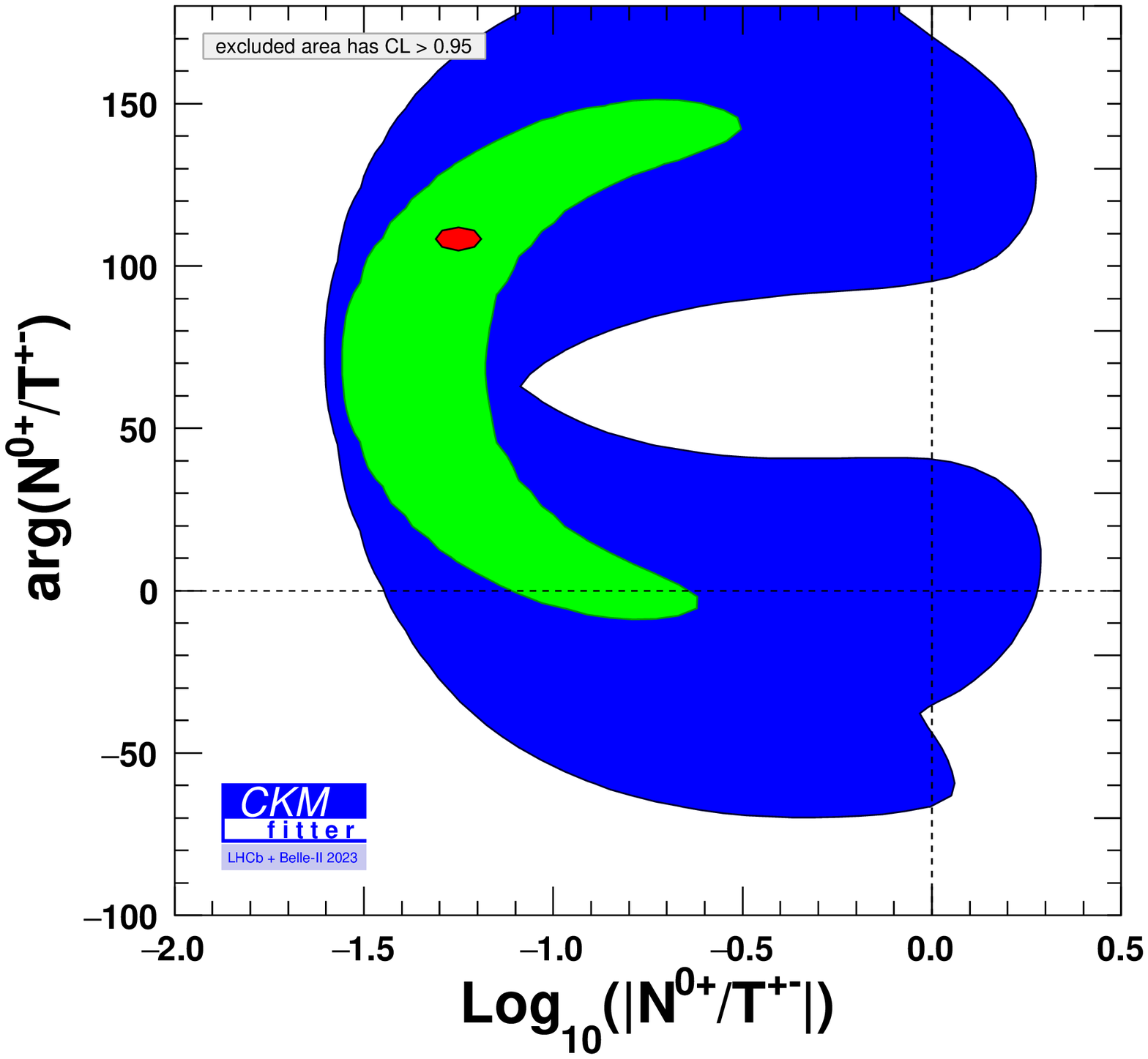}
\includegraphics[width=7cm]{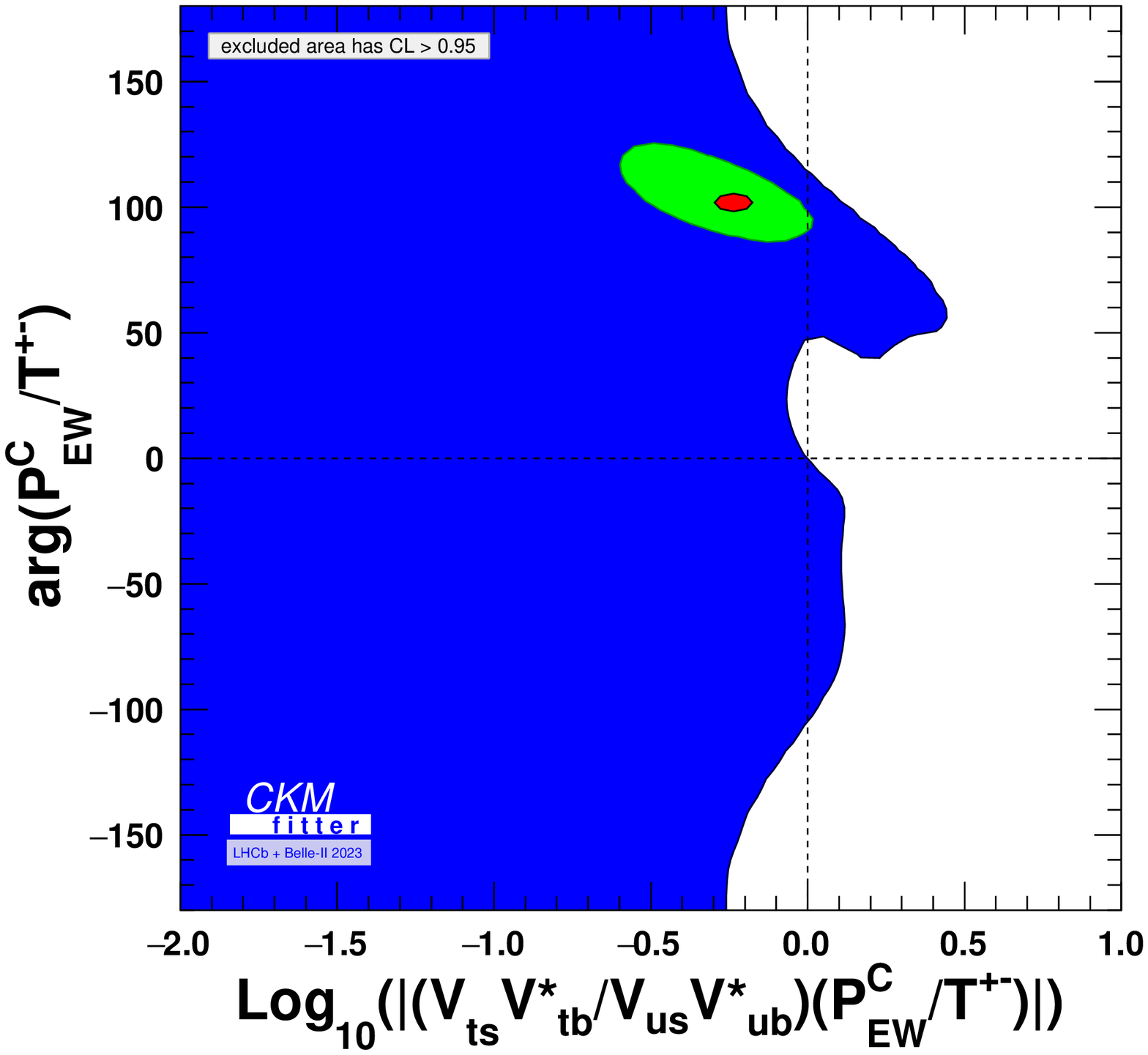}
\includegraphics[width=7cm]{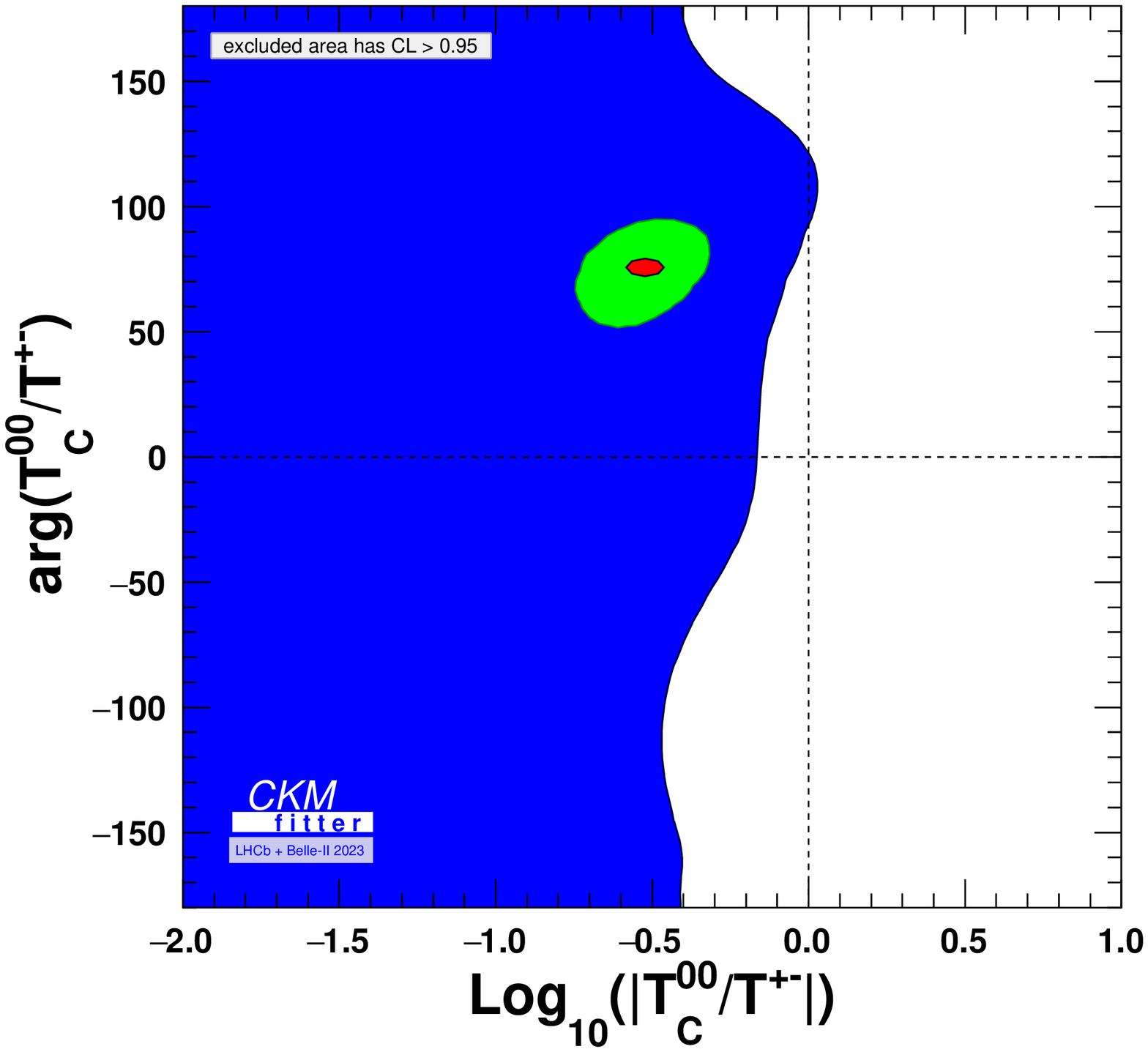}
\caption{
The expected two-dimensional constraints on the moduli and phases of various ratios
of hadronic parameters, using inputs from the first-step of the prospective study, based 
on results from the $B$-factories and expected sensitivities for LHCb Run1+Run2 (blue area); 
and using inputs from the second-step of the prospective study, based on the complete set 
of results from LHCb and Belle II (green area). The red spots in the figures represent the 
generation values obtained from Tab.~\ref{tab:IdealCase}. From top to bottom and left to 
right, the hadronic ratios are: 
$P^{+-}/T^{+-}$,
$P_{\rm EW}/T_{3/2}$,
$N^{0+}/T^{+-}$,
$P_{\rm EW}^{\rm C}/T^{+-}$,
and $T^{00}_{\rm C}/T^{+-}$, respectively.  
}
\label{fig:LHCbAndBelleII2023}
\end{center}
\end{figure*}

In this section, we study the impact of improved measurements of $K\pi\pi$ modes from the LHCb and Belle II experiments.
During the first run of the LHC, the LHCb experiment has collected large datasets of B-hadron decays, including charmless 
$B^0,B^+,B_s$ meson decays into tree-body modes. LHCb is currently collecting additional data in Run-2. In particular, due to the 
excellent performances of the LHCb detector for identifying charged long-lived mesons, the experiment has the potential for producing 
the most accurate charmless three-body results in the $B^+\rightarrow K^+\pi^-\pi^+$ mode, owing to high-purity event samples much larger 
than the ones collected by {\babar} and Belle. Using $3.0\ {\rm fb}^{-1}$ of data recorded during the LHC Run 1, first results on this 
mode are already available~\cite{Aaij:2014iva}, and a complete amplitude analysis is expected to be produced in the short-term future.
For the $B^0\rightarrow K^0_S\pi^+\pi^-$ mode, the event-collection efficiency is challenged by the combined requirements on reconstructing 
the $K^0_S\rightarrow \pi^+\pi^-$ decay and tagging the $B$ meson flavour, but nonetheless the  $B^0\rightarrow K^0_S\pi^+\pi^-$  data samples 
collected by LHCb are already larger than the ones from {\babar} and Belle. As it is more difficult to anticipate the reach of LHCb Dalitz-plot 
analyses for modes including $\pi^0$ mesons in the final state, the  $B^0\rightarrow K^+\pi^+\pi^0$, $B^+\rightarrow K^0_S\pi^+\pi^0$ 
$B^+\rightarrow K^+\pi^0\pi^0$ and $B^0\rightarrow K_S^0\pi^0\pi^0$ channels are not 
considered here. In addition, LHCb has also the potential for studying $B_s$ decay modes, and LHCb can reach $B\to KK\pi$ modes with branching ratios 
out of reach for $B$-factories.

The Belle II experiment~\cite{Urquijo:2015qsa}, currently in the stages of construction and commissioning, will operate in an experimental environment
very similar 
to the one of the {\babar} and Belle experiments. Therefore Belle II has the potential for studying all modes accessed by the $B$-factories,
with expected sensitivities that should scale in proportion to its expected total luminosity
(i.e., $50\ {\rm ab}^{-1}$). In addition, Belle II has the potential for accessing the $B^+\rightarrow K^+\pi^0\pi^0$ and $B^0\rightarrow K_S^0\pi^0\pi^0$
modes (for which the $B$-factories could not produce Dalitz-plot results) but these modes will  provide low-accuracy information,
redundant with some of the modes considered in this paper: therefore they are not included here.

Since both the LHCb and Belle II have the potential for studying large, high-quality samples of  $B^+\rightarrow K^+\pi^-\pi^+$, it 
is realistic to expect that the experiments will be able to extract a consistent, data-driven signal model to be used in all Dalitz-plot 
analysis, yielding systematic uncertainties significantly decreased with respect to the results from $B$-factories.

Finally for LHCb, since this experiment cannot perform $B$-meson counting as in a $B$-factory environment, the branching fractions need to be 
normalised with respect to measurements performed at {\babar} and Belle, until the advent of Belle II. This prospective study therefore is 
split into two periods: a first one based on the assumption of new results from LHCb Run1+Run2 only, and a second one using the complete set 
of LHCb and Belle II results. The corresponding inputs are gathered in App.~\ref{app:refprosp}. We use the reference scenario described in 
Tab.~\ref{tab:IdealCase} for the central values, so that we can guarantee the self-consistency of the inputs and we avoid reducing the 
uncertainties artificially because of  barely compatible measurements (which would occur if we used the central values of the current data 
and rescaled the uncertainties). The expected uncertainties, obtained from the extrapolations discussed previously, are described in 
Tab.~\ref{tab:LHCbAndBelleII}.

The blue area in Fig.~\ref{fig:LHCbAndBelleII2023} illustrates the potential for the first step of 
our prospective study ($B$-factories and LHCb Run1+Run2). For the input values used in the prospective, 
the modulus of the $P^{+-}/T^{+-}$ ratio will be constrained with a relative $10\%$ accuracy, 
and its complex phase will be constrained within $3$ degrees (we discuss 68\%~C.L. ranges in the following, whereas Fig.~\ref{fig:LHCbAndBelleII2023} shows 95\%~C.L. regions). Slightly tighter upper bounds 
on the $|T^{00}_{\rm C}/T^{+-}|$  and $|N^{0+}/T^{+-}|$ ratios may be set, albeit the relative phases of 
these rations will remain very poorly constrained. Assuming that the electroweak penguin is in 
agreement with the CPS/GPSZ prediction, its modulus will be constrained within $45\%$ and 
its phase within $14$ degrees. 

The addition of results from the Belle II experiment corresponds to the second step of this prospective study. As illustrated by the green area in Fig.~\ref{fig:LHCbAndBelleII2023}, 
the uncertainties on the modulus and phase of the $P^{+-}/T^{+-}$ ratio will decrease by factors of $1.4$ and $2.5$,
respectively. 
Owing to the addition of precision measurements by Belle II of 
the $B^0\to K^{*0}\pi^0$  Dalitz-plot parameters from the amplitude analysis of the $B^0\to K^+\pi^-\pi^0$ modes, 
the $T^{00}_{\rm C}/T^{+-}$ ratio can be constrained
 within 
a $22\%$ uncertainty for its modulus, and within $10$ degrees for its phase. Similarly, the uncertainties 
on the modulus and phase of the $P_{\rm EW}/T_{3/2}$ ratio  will decrease by factors $2.7$ and $2.9$, respectively. 
Concerning the colour-suppressed electroweak penguin, for which only a mild upper bound on its modulus was achievable 
within the first step of the prospective, can now be measured within a $22\%$ uncertainty for its modulus, and 
within $8$ degrees for its phase. Finally, the less stringent constraint will be achieved for the annihilation 
parameter. While its modulus can nevertheless be constrained between 0.3 and 1.5, the phase of 
this ratio may remain 
unconstrained in value, with just the sign of the phase being resolved. We add that one can also expect Belle II measurements for $B^+\to K^+\pi^0\pi^0$ and $B^0\to K_S\pi^0\pi^0$, however with larger uncertainties, so that we have not taken into account these decays.

In total, precise constraints on almost all hadronic parameters in the $B\rightarrow K^\star\pi$ system
will be achieved using the Dalitz-plot results from the LHCb and Belle II experiments, with a resolution of the current phase ambiguities. These constraints can be compared with various theoretical predictions, proving an important tool for testing models of hadronic contributions to charmless $B$ decays.

\section{Conclusion}

Non-leptonic B meson decays are very interesting processes both as probes of weak interaction and as tests of our understanding of QCD dynamics. 
They have been measured extensively at $B$-factories as well as at the LHCb experiment, but this wealth of data has not been fully exploited yet, 
especially for the pseudoscalar-vector modes which are accessible through Dalitz-Plot analyses of $B\to K\pi\pi$ modes. We have focused on the 
$B\to K^*\pi$ system which exhibits a large set of observables already measured. Isospin analysis allows us to express this decay in terms of CKM parameters and 6 complex hadronic amplitudes, but reparametrisation invariance 
prevents us from extracting simultaneously information on the weak phases and the hadronic amplitudes needed to describe these decays. We have 
followed two different approaches to exploit this data: either we extracted information on the CKM phase (after setting a condition on some of 
the hadronic amplitudes), or we determined of hadronic amplitudes (once we set the CKM parameters to their value from the CKM global fit~\cite{Charles:2004jd,Charles:2015gya,CKMfitterwebsite}).

In the first case, we considered two different strategies. We first reconsidered the CPS/GPSZ strategy proposed in Ref.~\cite{Ciuchini:2006kv,Gronau:2006qn}, 
amounting to setting a bound on the electroweak penguin in order to extract an $\alpha$-like constraint. We used a reference scenario inspired by the current data 
but with consistent central values and much smaller uncertainties in order to probe the robustness of the CPS/GPSZ method: it turns out that the method is easily biased if the 
bound on the electroweak penguin is not correct, even by a small amount. Unfortunately, this bound is not very precise from the theoretical point of view, 
which casts some doubt on the potential of this method to constrain $\alpha$. We have then considered a more promising alternative, consisting in setting a 
bound on the annihilation contribution. We observed that we could obtain an interesting stable $\beta$-like constraint and we discussed its potential to extract confidence intervals according 
to the accuracy of the bound used for the annihilation contribution.

In a second stage, we discussed how the data constrain the hadronic amplitudes, assuming the values of the CKM parameters. We performed an average of {\babar} and 
Belle data in order to extract constraints on various ratios of hadronic amplitudes, with the issue that some of these data contain several solutions to be combined 
in order to obtain a single set of inputs for the Dalitz-plot observables. The ratio $P^{+-}/T^{+-}$ is not very well constrained and exhibits two distinct preferred 
solutions, but it is not large and supports the expect penguin suppression. On the other hand, colour or electroweak suppression does not seem to hold, 
as illustrated by $|P_{\rm EW}/P^{+-}|$ (around 2),  $|P_{\rm EW}^{\rm C}/P_{\rm EW}|$ (around 1) or $|T^{00}_{\rm C}/T^{+-}|$ (mildly favouring values around 1). We however recall that some of these conclusions are 
very dependent on the {\babar} measurement on $\varphi^{00,+-}$ phase differences measured in $B^0\to K^+\pi^+\pi^0$: removing this input turns the ranges into mere upper 
bounds on these ratios of hadronic amplitudes.

For illustration purposes, we compared these results with typical theoretical expectations. 
We determined the hadronic amplitudes using an updated implementation of QCD 
factorisation. A good overall agreement between theory and experiment is found for most of the ratios of hadronic amplitudes, even though the experimental determinations remain often less accurate than the theoretical determinations in most instances. Nevertheless, two quantities still feature interesting properties. The ratio $P^{+-}/P_{\rm EW}$ could provide interesting constraints on the models used to describe power-suppressed contributions in QCD factorisation, keeping in mind the (precise) experimental determination of this ratio relies strongly on the $\varphi^{00,+-}$ phases measured by {\babar}, as discussed in the previous paragraph. The ratio $P_{\rm EW}^C/P_{\rm EW}$ is determined with similar accuracies  theoretically and experimentally, but the two determinations are not in good agreement, suggesting that this quantity could also be used to constrain QCD factorisation parameters.

Finally, we performed prospective studies, considering two successive stages based first on LHCb data from Run1 and Run2, then on the additional input from Belle II. 
Using our reference scenario and extrapolating the uncertainties of the measurements at both stages, we determined the confidence regions for the moduli and phases of 
the ratios of hadronic amplitudes. The first stage (LHCb only) would correspond to a significant improvement for $P^{+-}/T^{+-}$ and $P_{\rm EW}/T_{3/2}$, whereas the second 
stage (LHCb+Belle II) would yield tight constraints on $N^{0+}/T^{+-}$, $P_{\rm EW}^C/T^{+-}$ and $T^{00}_{\rm C}/T^{+-}$.

Non-leptonic $B$-meson decays remain an important theoretical challenge, and any contender should be able to explain not only the pseudoscalar-pseudoscalar modes 
but also the pseudoscalar-vector modes. Unfortunately, the current data do not permit such extensive tests, even though they hint at potential discrepancies with theoretical expectations concerning the hierarchies of hadronic amplitudes. However, our study suggests that a more thorough analysis of 
$B\to K\pi\pi$ Dalitz plots from LHCb and Belle II could allow for a precise determination of the hadronic amplitudes involved in $B\to K^*\pi$ decays thanks to the isobar approximation for three-body amplitudes. This will definitely 
shed some light on the complicated dynamics of weak and strong interaction at work in pseudo-scalar-vector modes, and it will provide important tests of our understanding of 
non-leptonic $B$-meson decays.

\section{Acknowledgments}

We would like to thank all our collaborators from the CKMfitter group for useful discussions, and Reina Camacho Toro for her collaboration on this project at an early stage. This project has received funding from the European Union Horizon 2020 research 
and innovation programme under the grant agreements No 690575. No 674896 and No. 692194. SDG acknowledges partial support from Contract FPA2014-61478-EXP.

\appendix

\section{Current experimental inputs}\label{App:Exp_inputs}

The full set real-valued physical observables, derived from the experimental inputs from {\babar} and Belle, is described in the following sections. The errors and correlation matrices  include both statistical and systematic uncertainties.

\begin{table*}[t]
\begin{center}
\begin{tabular}
{l|c|ccc}
 $B^0\rightarrow K^0_S\pi^+\pi^-$         & $~~~~~~~~~~$Global min$~~~~~~~~~~$  & ${\mathcal Re}\left[ \frac{q}{p} \frac{\overline{A}(K^{*-}\pi^+)}{A(K^{*+}\pi^-)} \right]$  
                                                                                                & ${\mathcal Im}\left[ \frac{q}{p} \frac{\overline{A}(K^{*-}\pi^+)}{A(K^{*+}\pi^-)} \right]$  
                                                                                                & ${\mathcal B}(K^{*+}\pi^-)$  \\
\hline
${\mathcal Re}\left[ \frac{q}{p} \frac{\overline{A}(K^{*-}\pi^+)}{A(K^{*+}\pi^-)} \right]$  & $0.428  \pm 0.473$    & 1.00 & 0.90 &  0.02 \\
${\mathcal Im}\left[ \frac{q}{p} \frac{\overline{A}(K^{*-}\pi^+)}{A(K^{*+}\pi^-)} \right]$  & $-0.690 \pm 0.302$    &      & 1.00 & -0.06 \\
${\mathcal B}(K^{*+}\pi^-)  (\times 10^{-6})$                                               & $8.290  \pm 1.189$    &      &      &  1.00 \\
\hline
\end{tabular}

\vspace{0.1cm}

\begin{tabular}
{l|c|ccc}
 $B^0\rightarrow K^0_S\pi^+\pi^-$         & Local min  ($\Delta {\rm NLL} = 0.16$)   & ${\mathcal Re}\left[ \frac{q}{p} \frac{\overline{A}(K^{*-}\pi^+)}{A(K^{*+}\pi^-)} \right]$  
                                                                                                & ${\mathcal Im}\left[ \frac{q}{p} \frac{\overline{A}(K^{*-}\pi^+)}{A(K^{*+}\pi^-)} \right]$  
                                                                                                & ${\mathcal B}(K^{*+}\pi^-)$  \\
\hline
${\mathcal Re}\left[ \frac{q}{p} \frac{\overline{A}(K^{*-}\pi^+)}{A(K^{*+}\pi^-)} \right]$  &  $-0.819 \pm 0.116$     & 1.00 & -0.19 & -0.15 \\
${\mathcal Im}\left[ \frac{q}{p} \frac{\overline{A}(K^{*-}\pi^+)}{A(K^{*+}\pi^-)} \right]$  &  $-0.049 \pm 0.494$     &      &  1.00 & -0.01 \\
${\mathcal B}(K^{*+}\pi^-)  (\times 10^{-6})$                                               &  $8.290  \pm 1.189$     &      &       &  1.00 \\
\end{tabular}
\caption{Central values and total (statistical and systematic) correlation matrix for the global (top) and local (bottom, $\Delta {\rm NLL} = 0.16$)  minimum solutions for the {\babar} $B^0\rightarrow K^0_S\pi^+\pi^-$ analysis.}
\label{tab:KSPiPi_babar}
\end{center}
\end{table*}

\begin{table}[t]
\begin{center}
\begin{tabular}
{l|c}
$B^+\rightarrow K^+\pi^-\pi^+$                                                                   & Value \\
\hline
$\left| \frac{\overline{A}(\overline{K}^{*0}\pi^-)}{A(K^{*0}\pi^+)} \right|$  & $1.033   \pm 0.047$   \\
${\mathcal B}(K^{*0}\pi^+) (\times 10^{-6})$                                  & $10.800  \pm 1.389$   \\
\end{tabular}
\caption{Central values of the observables for the {\babar} $B^+\rightarrow K^+\pi^-\pi^+$ analysis.}
\label{tab:KPiPi_babar}
\end{center}
\end{table}

\begin{table*}[t]
\begin{center}
{\begin{tabular}
{l|c|cccccc}
$B^0\rightarrow K^+\pi^-\pi^0$ &   Value                                                            & $\left| \frac{\overline{A}(K^{*-}\pi^+)}{A(K^{*+}\pi^-)} \right|$
                                                                                                      & ${\mathcal Re}\left[ \frac{A(K^{*0}\pi^0)}{A(K^{*+}\pi^-)} \right]$
                                                                                                      & ${\mathcal Im}\left[ \frac{A(K^{*0}\pi^0)}{A(K^{*+}\pi^-)} \right]$ 
                                                                                                      & ${\mathcal Re}\left[ \frac{\overline{A}(\overline{K}^{*0}\pi^0)}{\overline{A}(K^{*-}\pi^+)} \right]$
                                                                                                      & ${\mathcal Re}\left[ \frac{\overline{A}(\overline{K}^{*0}\pi^0)}{\overline{A}(K^{*-}\pi^+)} \right]$
                                                                                                      & ${\mathcal B}(K^{*0}\pi^0)$  \\
\hline
$\left| \frac{\overline{A}(K^{*-}\pi^+)}{A(K^{*+}\pi^-)} \right|$                                    & $0.742  \pm 0.091$  & 1.00 & 0.00 &  0.03 & -0.22 & -0.11 & -0.06 \\
${\mathcal Re}\left[ \frac{A(K^{*0}\pi^0)}{A(K^{*+}\pi^-)} \right]$                                  & $0.562  \pm 0.148$  &      & 1.00 &  0.68 &  0.33 & -0.01 &  0.44 \\
${\mathcal Im}\left[ \frac{A(K^{*0}\pi^0)}{A(K^{*+}\pi^-)} \right]$                                  & $-0.227 \pm 0.296$  &      &      &  1.00 & -0.07 &  0.00 & -0.13 \\
${\mathcal Re}\left[ \frac{\overline{A}(\overline{K}^{*0}\pi^0)}{\overline{A}(K^{*-}\pi^+)} \right]$ & $0.701  \pm 0.126$  &      &      &       &  1.00 &  0.25 &  0.55 \\
${\mathcal Im}\left[ \frac{\overline{A}(\overline{K}^{*0}\pi^0)}{\overline{A}(K^{*-}\pi^+)} \right]$ & $-0.049 \pm 0.376$  &      &      &       &       &  1.00 & -0.02 \\
${\mathcal B}(K^{*0}\pi^0)  (\times 10^{-6})$                                                        & $3.300  \pm 0.640$  &      &      &       &       &       &  1.00 \\
\end{tabular}}
\caption{Central values and total (statistical and systematic) correlation matrix for observables from the {\babar} $B^0\rightarrow K^+\pi^-\pi^0$ analysis.}
\label{tab:KPiPi0_babar}
\end{center}
\end{table*}

\begin{table*}[t]
\begin{center}
{\begin{tabular}
{l|c|cccccc}
$B^+\rightarrow K^0_S\pi^+\pi^0$ & Value                                                  & $\left| \frac{\overline{A}(K^{*-}\pi^0)}{A(K^{*+}\pi^0)} \right|$
                                                                                                      & ${\mathcal Re}\left[ \frac{A(K^{*+}\pi^0)}{A(K^{*0}\pi^+)} \right]$
                                                                                                      & ${\mathcal Im}\left[ \frac{A(K^{*+}\pi^0)}{A(K^{*0}\pi^+)} \right]$ 
                                                                                                      & ${\mathcal Re}\left[ \frac{\overline{A}(K^{*-}\pi^0)}{\overline{A}(\overline{K}^{*0}\pi^-)} \right]$
                                                                                                      & ${\mathcal Im}\left[ \frac{\overline{A}(K^{*-}\pi^0)}{\overline{A}(\overline{K}^{*0}\pi^-)} \right]$
                                                                                                      & ${\mathcal B}(K^{*+}\pi^0)$  \\
\hline
$\left| \frac{\overline{A}(K^{*-}\pi^0)}{A(K^{*+}\pi^0)} \right|$               & $0.533   \pm 1.403$                         & 1.00 & -0.26 &  0.01 & -0.70 & -0.22 & -0.16 \\
${\mathcal Re}\left[ \frac{A(K^{*+}\pi^0)}{A(K^{*0}\pi^+)} \right]$          & $1.415   \pm 6.952$                           &      & 1.00  & -0.23 &  0.12 & -0.51 &  0.90 \\
${\mathcal Im}\left[ \frac{A(K^{*+}\pi^0)}{A(K^{*0}\pi^+)} \right]$           & $-0.189  \pm 3.646$                         &      &       &  1.00 & -0.39 &  0.23 & -0.28 \\
${\mathcal Re}\left[ \frac{\overline{A}(K^{*-}\pi^0)}{\overline{A}(\overline{K}^{*0}\pi^-)} \right]$& $-0.106  \pm 2.687$  &      &       &       &  1.00 &  0.23 &  0.03 \\
${\mathcal Im}\left[ \frac{\overline{A}(K^{*-}\pi^0)}{\overline{A}(\overline{K}^{*0}\pi^-)} \right]$& $-0.851  \pm 4.278$  &      &       &       &       &  1.00 & -0.82 \\
${\mathcal B}(K^{*+}\pi^0)  (\times 10^{-6})$                                                                     & $9.200   \pm 1.480$       &      &       &       &       &       &  1.00 \\
\end{tabular}}
\caption{Central values and total (statistical and systematic) correlation matrix for observables from the {\babar} $B^+\rightarrow K^0_S\pi^+\pi^0$ analysis.}
\label{tab:K0PiPi0_babar}
\end{center}
\end{table*}

 \begin{table}[t]
\begin{center}
\begin{tabular}
{l|c}
$B^+\to K^{*+}\pi^0$ in $B^+\to K^+\pi^0\pi^0$                                                & value \\
\hline
${\mathcal B}(K^{*+}\pi^0)$   &  $(8.2 \pm 1.5 \pm 1.1)\times10^{-6}$   \\
$A_{CP}(K^{*+}\pi^0)$                                 & $ -0.06 \pm 0.24 \pm 0.04$   \\
\end{tabular}
\caption{Central values of the observables from the {\babar} analysis of $B^+\to K^{*+}(892)\pi^0$ quasi-two-body contribution to the $B^+\to K^+\pi^0\pi^0$.}
\label{tab:KstPi0}
\end{center}
\end{table}

\subsection{{\babar} results}\label{App:babar_inputs}

In this section, we describe the set of experimental inputs from the {\babar} experiment.

\begin{itemize}
\item $B^0\rightarrow K^0_S\pi^+\pi^-$~\cite{Aubert:2009me}. Two almost degenerate solutions were found differing only by $0.16$ 
 negative-log-likelihood ($\Delta {\rm NLL}$) units. The central values and correlation matrix of the measured observables for both solutions are shown in Tab.~\ref{tab:KSPiPi_babar}.
 
\item $B^+\rightarrow K^+\pi^-\pi^+$~\cite{Aubert:2008bj}. The central values of the observables for this analysis are shown in Tab.~\ref{tab:KPiPi_babar}. 
A linear correlation of $2\%$ was found between $\left| \frac{\overline{A}(\overline{K}^{*0}\pi^-)}{A(K^{*0}\pi^+)} \right|$ and ${\mathcal B}(K^{*0}\pi^+)$.

\item $B^0\rightarrow K^+\pi^-\pi^0$~\cite{BABAR:2011ae}. The central values and correlation matrix of the measured observables for this analysis are shown in 
Tab.~\ref{tab:KPiPi0_babar}.

\item $B^+\rightarrow K^0_S\pi^+\pi^0$~\cite{Lees:2015uun}. The central values and correlation matrix of the measured observables for this analysis are shown in 
Tab.~\ref{tab:K0PiPi0_babar}.

\item $B^+\to K^{*+}(892)\pi^0$ quasi-two-body contribution to the $B^+\to K^+\pi^0\pi^0$ final state~\cite{Lees:2011aaa}. 
The measured branching ratio and $CP$ asymmetry are shown in Tab.~\ref{tab:KstPi0} and they are used as uncorrelated inputs.

\end{itemize}
 
\begin{table*}[t]
\begin{center}
\begin{tabular}
{l|c|ccc}
 $B^0\rightarrow K^0_S\pi^+\pi^-$ & $~~~~~~~~~$Global min$~~~~~~~~~$  & ${\mathcal Re}\left[ \frac{q}{p} \frac{\overline{A}(K^{*-}\pi^+)}{A(K^{*+}\pi^-)} \right]$  
                                                                                                & ${\mathcal Im}\left[ \frac{q}{p} \frac{\overline{A}(K^{*-}\pi^+)}{A(K^{*+}\pi^-)} \right]$  
                                                                                                & ${\mathcal B}(K^{*+}\pi^-)$  \\
\hline
${\mathcal Re}\left[ \frac{q}{p} \frac{\overline{A}(K^{*-}\pi^+)}{A(K^{*+}\pi^-)} \right]$    & $0.790  \pm 0.145$    & 1.00 & 0.62 & -0.04 \\
${\mathcal Im}\left[ \frac{q}{p} \frac{\overline{A}(K^{*-}\pi^+)}{A(K^{*+}\pi^-)} \right]$    & $-0.206 \pm 0.398$    &      & 1.00 &  0.00 \\
${\mathcal B}(K^{*+}\pi^-)  (\times 10^{-6})$                                                 & $8.400  \pm 1.449$    &      &      &  1.00 \\
\hline
\end{tabular}

\begin{tabular}
{l|c|ccc}
 $B^0\rightarrow K^0_S\pi^+\pi^-$ &   Local min   ($\Delta {\rm NLL} = 7.5$)        & ${\mathcal Re}\left[ \frac{q}{p} \frac{\overline{A}(K^{*-}\pi^+)}{A(K^{*+}\pi^-)} \right]$  
                                                                                                & ${\mathcal Im}\left[ \frac{q}{p} \frac{\overline{A}(K^{*-}\pi^+)}{A(K^{*+}\pi^-)} \right]$  
                                                                                                & ${\mathcal B}(K^{*+}\pi^-)$  \\
\hline
${\mathcal Re}\left[ \frac{q}{p} \frac{\overline{A}(K^{*-}\pi^+)}{A(K^{*+}\pi^-)} \right]$ &  $0.808  \pm 0.110$      & 1.00 &  0.01 & -0.06 \\
${\mathcal Im}\left[ \frac{q}{p} \frac{\overline{A}(K^{*-}\pi^+)}{A(K^{*+}\pi^-)} \right]$ &  $0.010  \pm 0.439$      &      &  1.00 &  0.00 \\
${\mathcal B}(K^{*+}\pi^-)  (\times 10^{-6})$                                              &  $8.400  \pm 1.449$      &      &       &  1.00 \\
\end{tabular}
\caption{Central values  and total (statistical and systematic) correlation matrix for the global (top) and local solution (bottom, $\Delta {\rm NLL} = 7.5$) minimum solutions of the observables from the Belle $B^0\rightarrow K^0_S\pi^+\pi^-$ analysis.}
\label{tab:KSPiPi_belle}
\end{center}
\end{table*}

 \begin{table}[t]
\begin{center}
\begin{tabular}{l|c}
$B^+\rightarrow K^+\pi^-\pi^+$                                                 & value \\
\hline
$\left| \frac{\overline{A}(\overline{K}^{*0}\pi^-)}{A(K^{*0}\pi^+)} \right|$   & $0.861   \pm 0.059$   \\
${\mathcal B}(K^{*0}\pi^+)  (\times 10^{-6})$                                  & $9.670   \pm 1.061$   \\
\end{tabular}
\caption{Central values of the observables from the Belle $B^+\rightarrow K^+\pi^-\pi^+$ analysis.}
\label{tab:KPiPi_belle}
\end{center}
\end{table}

\subsection{Belle results}\label{App:belle_inputs}

In this section, we describe the set of experimental inputs from the Belle experiment.

\begin{itemize}

\item $B^0\rightarrow K^0_S\pi^+\pi^-$~\cite{Dalseno:2008wwa}. Two solutions were found differing by $7.5$ 
 $\Delta {\rm NLL}$. The central values and correlation matrix of the measured observables for both solutions are shown in 
 Tab.~\ref{tab:KSPiPi_belle}.
  
\item $B^+\rightarrow K^+\pi^-\pi^+$~\cite{Garmash:2006bj}. The central values of the observables for this analysis are shown in Tab.~\ref{tab:KPiPi_belle}.
 A nearly vanishing correlation was found between $\left| \frac{\overline{A}(\overline{K}^{*0}\pi^-)}{A(K^{*0}\pi^+)} \right|$ and ${\mathcal B}(K^{*0}\pi^+)$.

\end{itemize}

\subsection{Combined {\babar} and Belle results}\label{App:comb_inputs}

\begin{figure}[t]
\begin{center}
\includegraphics[width=8.5cm, angle=0]{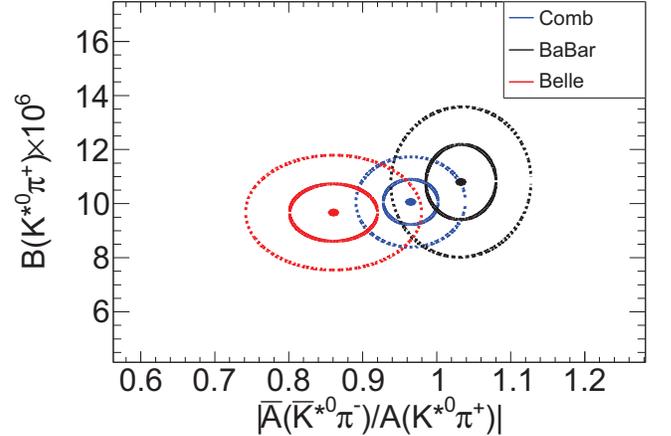}
\caption{Contours at 1 (solid) and 2 (dotted) $\sigma$ in the $\left| \frac{\overline{A}(\overline{K}^{*0}\pi^-)}{A(K^{*0}\pi^+)} \right|$ vs ${\mathcal B}(K^{*0}\pi^+)$ plane 
for the {\babar} (black) and Belle (red) results, as well as the combination (blue).}
\label{fig:Combination_babarbelle_KPiPi}
\end{center}
\end{figure}

The {\babar} and Belle results for the $B^0\rightarrow K^0_S\pi^+\pi^-$ and $B^+\rightarrow K^+\pi^-\pi^+$ analyses shown previously have been combined in the usual 
way for sets of independent measurements. The combination for the $B^+\rightarrow K^+\pi^-\pi^+$ mode is straightforward as the results exhibit only one solution, 
as shown in Fig.~\ref{fig:Combination_babarbelle_KPiPi}. The resulting central values are shown in Tab.~\ref{tab:KPiPiKSPiPi_babarbelle}. 
A vanishing linear correlation is found between $\left| \frac{\overline{A}(\overline{K}^{*0}\pi^-)}{A(K^{*0}\pi^+)} \right|$ and ${\mathcal B}(K^{*0}\pi^+)$.

The combination of the {\babar} and Belle measurements for the $B^0\rightarrow K^0_S\pi^+\pi^-$ mode is more complicated as the results feature several solutions 
which are relatively close in units of $\Delta {\rm NLL}$. In order to combine this measurements we proceed as follows:

\begin{itemize}
 \item We combine each solution of the {\babar} analysis with each one of the Belle results.

 \item In the goodness of fit of the combination ($\chi^2_{\rm min}$), we add the $\Delta {\rm NLL}$ of each {\babar} and Belle solution. In the case of the global  minimum the corresponding $\Delta {\rm NLL}$ is zero.
 
 \item Finally, we take the envelope of the four combinations as the final result.
\end{itemize}

\begin{figure*}[t]
\begin{center}
\includegraphics[width=5.9cm, angle=0]{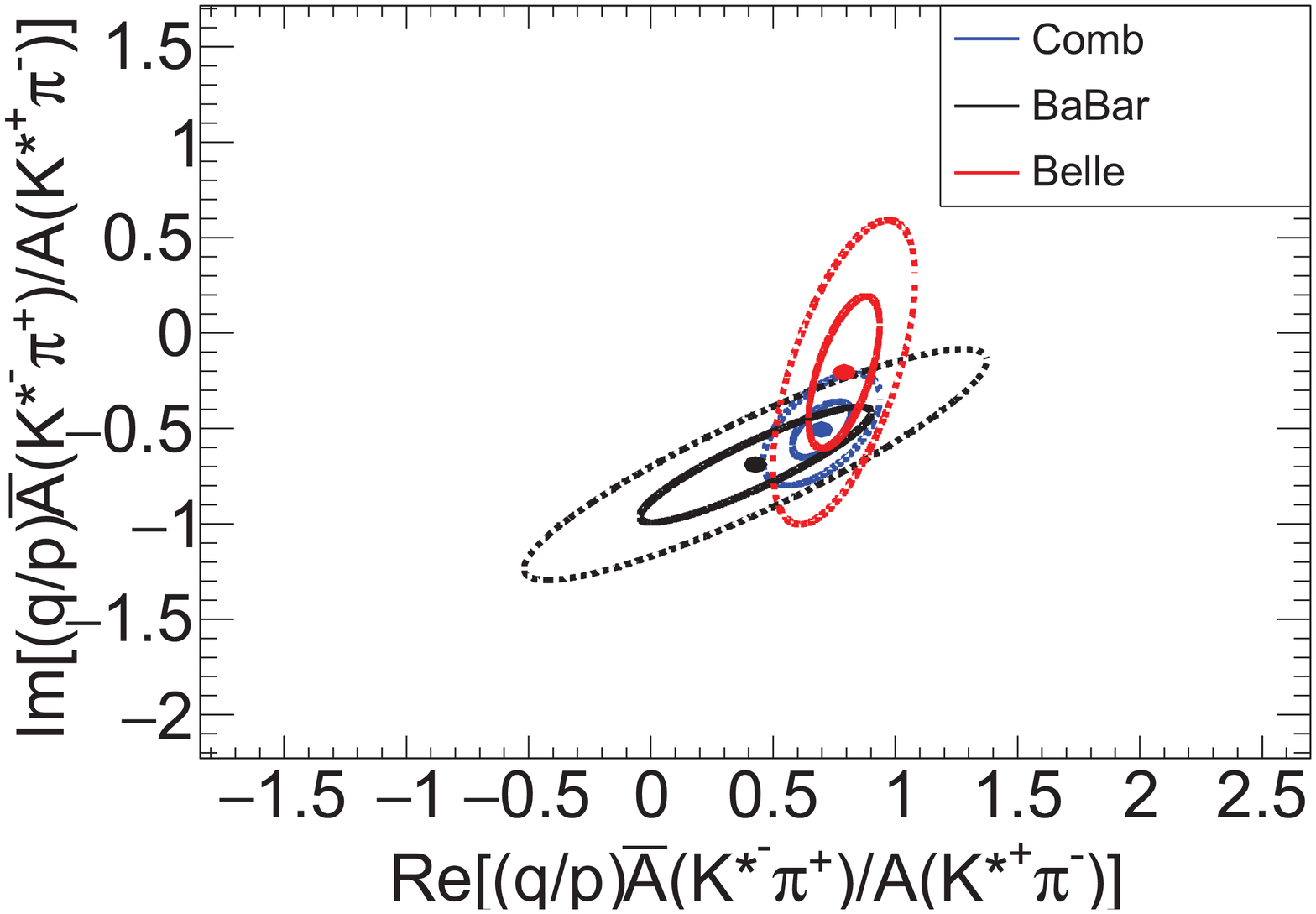}
\includegraphics[width=5.9cm, angle=0]{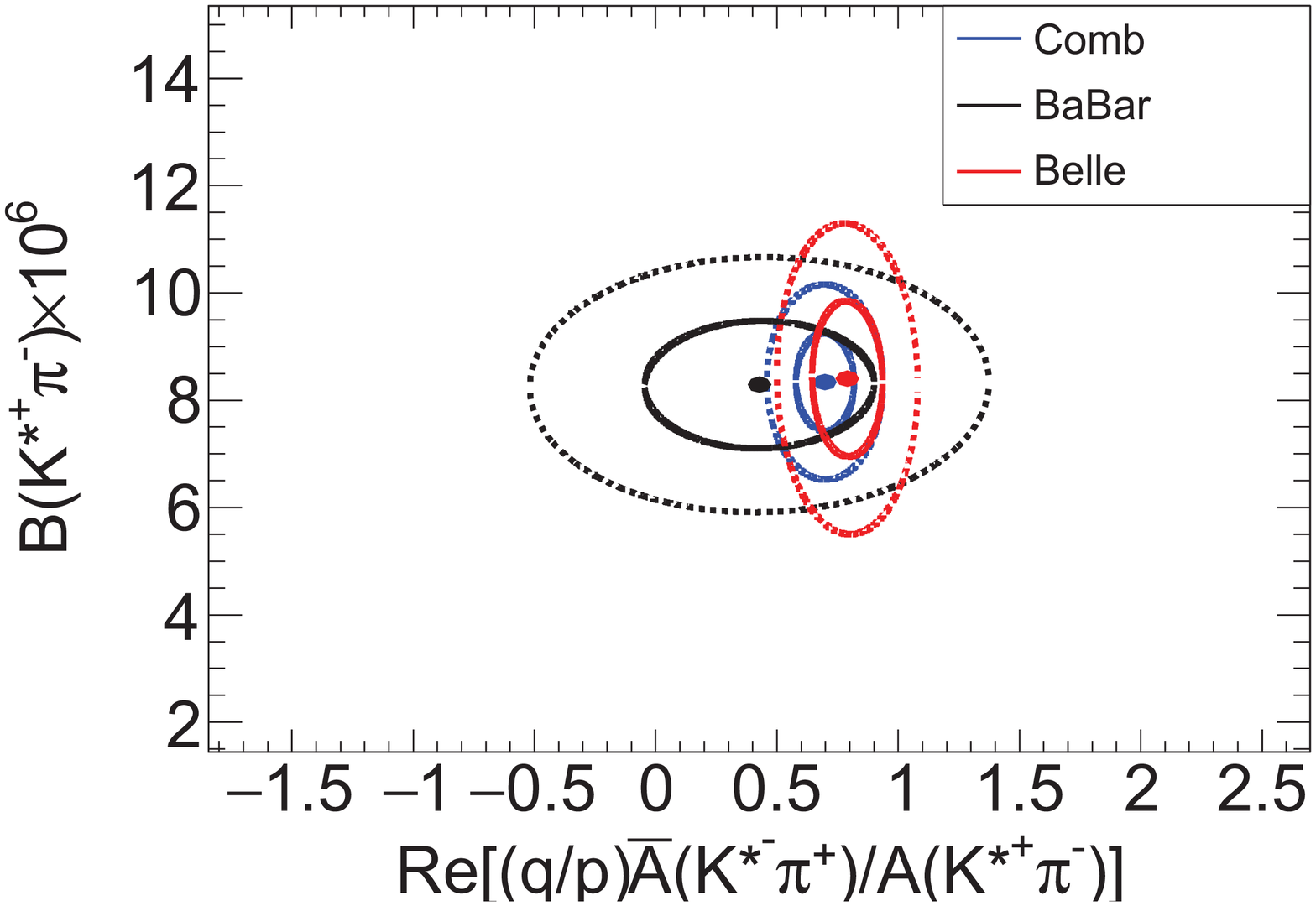}
\includegraphics[width=5.9cm, angle=0]{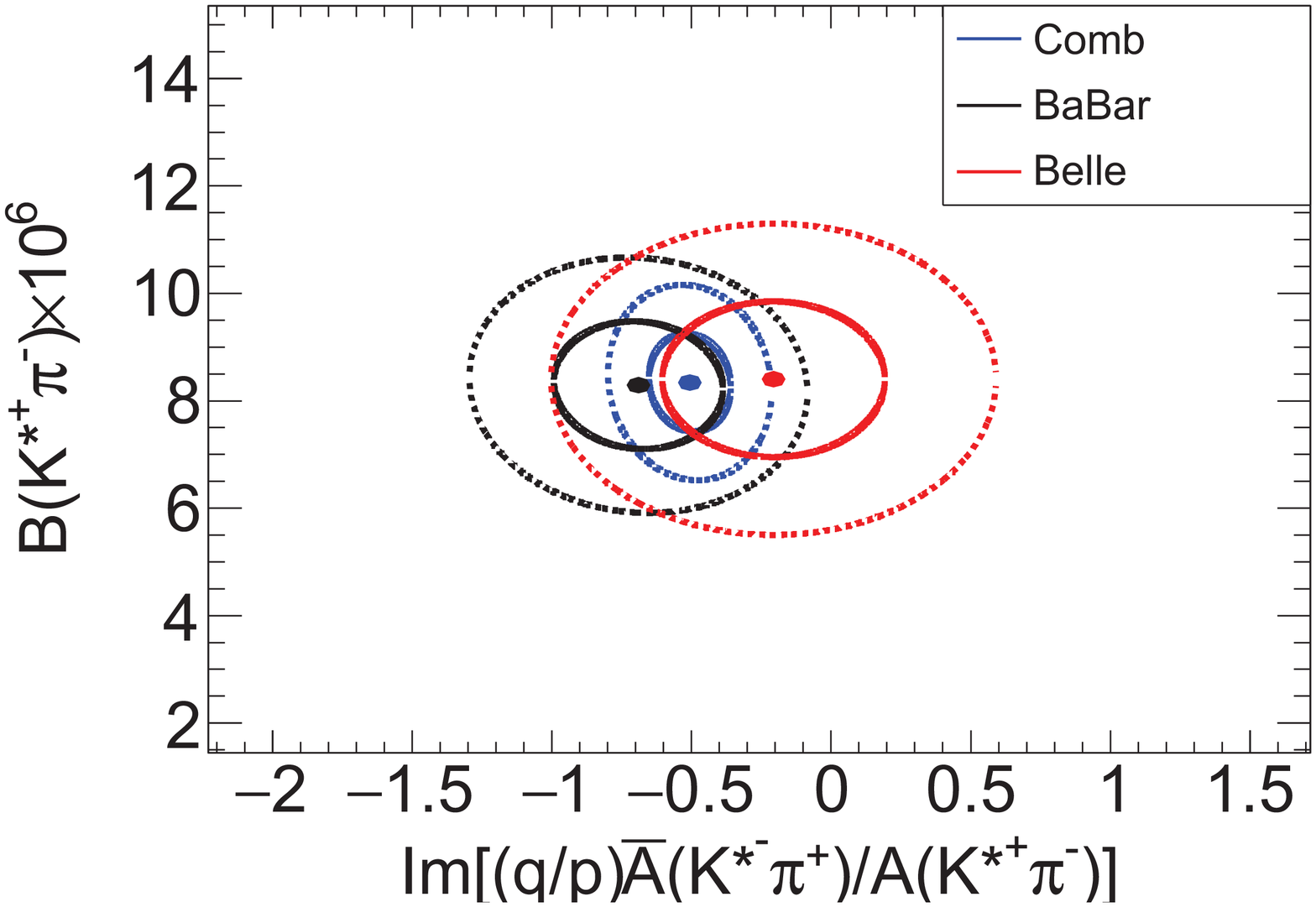}
\caption{Contours at 1 (solid) and 2 (dotted) $\sigma$ in the ${\mathcal Re}\left[ \frac{q}{p} \frac{\overline{A}(K^{*-}\pi^+)}{A(K^{*+}\pi^-)} \right]$ vs 
${\mathcal Im}\left[ \frac{q}{p} \frac{\overline{A}(K^{*-}\pi^+)}{A(K^{*+}\pi^-)} \right]$ (left), ${\mathcal Re}\left[ \frac{q}{p} \frac{\overline{A}(K^{*-}\pi^+)}{A(K^{*+}\pi^-)} \right]$ 
vs ${\mathcal B}(K^{*+}\pi^-)$ (middle) and ${\mathcal Im}\left[ \frac{q}{p} \frac{\overline{A}(K^{*-}\pi^+)}{A(K^{*+}\pi^-)} \right]$ vs ${\mathcal B}(K^{*+}\pi^-)$ (right) planes for the 
{\babar} (black) and Belle (red) results, as well as the combination (blue).}
\label{fig:Combination_babarbelle_KSPiPi}
\end{center}
\end{figure*}

We find the following $\chi^2_{\rm min}$ for the four combinations: 1.1, 8.7, 9.5 and 98.3. As the closest combination from the global minimum differs 
by 7.6 units in $\chi^2_{\rm min}$, we have decided to focus on the global minimum for the phenomenological analysis. The combination for this global minimum is shown in 
Fig.~\ref{fig:Combination_babarbelle_KSPiPi}. The resulting central values and covariance matrix are shown in Tab.~\ref{tab:KPiPiKSPiPi_babarbelle}.

\begin{table*}[t]
\begin{center}
\begin{tabular}
{l|c}
$B^+\rightarrow K^+\pi^-\pi^+$                                                                & Value \\
\hline
$\left| \frac{\overline{A}(\overline{K}^{*0}\pi^-)}{A(K^{*0}\pi^+)} \right|$   & $0.965   \pm 0.037$   \\
${\mathcal B}(K^{*0}\pi^+)  (\times 10^{-6})$                                  & $10.062  \pm 0.835$   \\
\end{tabular}

\vspace{0.2cm}

\begin{tabular} {l|c|ccc}
$B^0\rightarrow K^0_S\pi^+\pi^-$  & Value                    & ${\mathcal Re}\left[ \frac{q}{p} \frac{\overline{A}(K^{*-}\pi^+)}{A(K^{*+}\pi^-)} \right]$  
                                                                                                & ${\mathcal Im}\left[ \frac{q}{p} \frac{\overline{A}(K^{*-}\pi^+)}{A(K^{*+}\pi^-)} \right]$  
                                                                                                & ${\mathcal B}(K^{*+}\pi^-)$  \\
\hline
${\mathcal Re}\left[ \frac{q}{p} \frac{\overline{A}(K^{*-}\pi^+)}{A(K^{*+}\pi^-)} \right]$    & $0.698  \pm 0.120$   & 1.00 & 0.58 & -0.01 \\
${\mathcal Im}\left[ \frac{q}{p} \frac{\overline{A}(K^{*-}\pi^+)}{A(K^{*+}\pi^-)} \right]$    & $-0.506 \pm 0.146$   &      & 1.00 & -0.09 \\
${\mathcal B}(K^{*+}\pi^-)  (\times 10^{-6})$                                                 & $8.340  \pm 0.910$   &      &      &  1.00 \\
\end{tabular}
\caption{Central values of the observables from the $B^+\rightarrow K^+\pi^-\pi^+$ (top) and $B^0\rightarrow K^0_S\pi^+\pi^-$ (bottom) analysis obtained by combining {\babar} and Belle results.}
\label{tab:KPiPiKSPiPi_babarbelle}
\end{center}
\end{table*}

These combined results for the $B^0\rightarrow K^0_S\pi^+\pi^-$ and $B^+\rightarrow K^+\pi^-\pi^-$ modes are used with the {\babar} results for the 
$B^0\rightarrow K^+\pi^-\pi^0$ and $B^+\rightarrow K^0_S \pi^+\pi^0$ as inputs for the phenomenological analysis using the current experimental 
measurements.

\section{Two-body non leptonic amplitudes in QCD factorisation} \label{app:QCDFinputs}

We compute the $B\to K^*\pi$ amplitudes in the framework of QCD factorisation, using the results of Ref.~\cite{Beneke:2003zv}. We take the semileptonic $B\to\pi$ and $B\to K\pi$ form factors from computations based on Light-Cone Sum Rules~\cite{Ball:2004rg,Straub:2015ica}. The parameters for the light-meson distribution amplitudes that enter hard-scattering contributions are consistently taken from the last two references. On the other hand the first inverse moment of the $B$-meson distribution amplitude $\lambda_B$ is taken from Ref.~\cite{Braun:2003wx}. Quark masses are taken from review by the FLAG group~\cite{Aoki:2016frl}. Our updated inputs are summarised in Table~\ref{tab:QCDFinputs}.
\begin{table}
\begin{tabular}{ll|ll}
Input & Value & Input & Value \\\hline
$\alpha_1(K^*)$ & $0.06\pm 0\pm 0.04$ & $\alpha_1(K^*,\perp)$ & $0.04\pm 0\pm 0.03$ \\
$\alpha_2(K^*)$ & $0.16\pm 0\pm 0.09$ & $\alpha_2(K^*,\perp)$ & $0.10\pm 0\pm 0.08$ \\
$f_\perp(K^*)$ & $0.159\pm 0\pm 0.006$ & $A_0[B\to K^*](0)$ & $0.356\pm 0\pm 0.046$ \\
$\alpha_2(\pi)$ & $0.062\pm 0\pm 0.054$ & $F_0[B\to\pi](0)$ & $0.258\pm 0\pm 0.031$ \\
$\lambda_B$ &$0.460\pm 0\pm 0.110$ & $\bar m_b$ & $4.17$\\
$\bar m_s$ & $0.0939\pm 0\pm 0.0011$ & $m_q/m_s$ & $\sim 0$
\end{tabular}
\caption{Input values for the hadronic parameters that enter QCD factorisation predictions: moments of the distribution amplitudes for mesons, decay constants, form factors and quark masses. Dimensionful quantities are in GeV. The $\pm 0$ in second position means that all uncertainties are considered as coming from a theoretical origin and they are treated according to the Rfit approach. See the text for references.}
\label{tab:QCDFinputs}
\end{table}

We stress that the calculations  of Ref.~\cite{Beneke:2003zv} correspond to Next-to-Leading Order (NLO). Since then, some NNLO contributions have been computed~\cite{BHWS,Bell:2007tv,Bell:2009nk,Beneke:2009ek,Bell:2015koa}, that we neglect in view of the sizeable uncertainties on the input parameters: this is sufficient for our illustrative purposes (see Section~\ref{QCDFcomparison}).

\section{Reference scenario and prospective studies}\label{app:refprosp}

Some of the experimental results collected in App.~\ref{App:Exp_inputs} are affected by large uncertainties, and the central values are not 
always fully consistent with SM expectations. 
This is not a problem when we want to extract values of the hadronic parameters from the data, but it makes rather unclear 
the discussion of the accuracy of specific models (say, for the extraction of weak angles) or the prospective studies assuming 
improved experimental measurements, see Secs.~\ref{sec:CKM} and \ref{sec:prospect}. 

For this reason, we design a reference scenario described in Tab.~\ref{tab:IdealCase}. The values on hadronic parameters are 
chosen to reproduce the current best averages of branching fractions and $CP$ asymmetries  in $B\rightarrow K^*\pi$ roughly. As 
most observable phase differences among these modes are poorly constrained by the results currently available, we do no attempt 
at reproducing their central values and we use the values resulting from the hadronic parameters. The hadronic amplitudes are 
constrained to respect the naive assumptions: $|P_{\rm EW}/T_{3/2}| \simeq 1.35\%$, $|P^C_{\rm EW}| < |P_{\rm EW}|$ and $|T^{00}_{\rm C}| < |T^{+-}|$. 
The best values of the hadronic parameters yield the values of branching ratios and $CP$ asymmetries gathered in Tab.~\ref{tab:IdealCase}. As can be seen, the overall agreement is fair, but it is not good for all observables. Indeed, as discussed in Sec.~\ref{sec:Hadronic}, the current data do not favour all the hadronic hierarchies that we have imposed to obtain our reference scenario in Tab.~\ref{tab:IdealCase}.

For the studies of different methods to extract CKM parameters described in Sec.~\ref{sec:CKM}, we fit the values
of hadronic parameters by assigning  small, arbitrary, uncertainties to the physical observables:
$\pm 5\%$ for branching ratios, $\pm 0.5\%$ for $CP$ asymmetries, and $\pm 5^\circ$ for interference phases.

For the prospective studies described in Sec.~\ref{sec:prospect}, we estimate future experimental uncertainties at two different stages. 
We first consider a list of expected measurements from LHCb, using the combined Run1 and Run2 data. 
We then reassess the expected results including Belle II measurements. Our method to project
uncertainties in the two stages is based on the statistical scaling of data samples ($1/\sqrt{N_{\rm evts}}$), corrected for 
additional factors due 
to particular detector performances and analysis technique features, as described below.

LHCb Run1 and Run2 data will significantly increase the statistics mainly for the fully charged final states
$B^0\rightarrow K^0_S(\rightarrow \pi^+\pi^-)\pi^+\pi^-$ and $B^+\rightarrow K^+\pi^-\pi^+$, with an expected increase of 
about $3$ and $40$, respectively~\cite{B0toKspipi_LHCb:2012iva,BtoKpipi_LHCb:2013iva}.
For these modes, we assume a signal-to-background ratio similar to the
ones measured at $B$ factories (this may represent an underestimation of the potential sensitivity of LHCb data, but 
this assumption has a very minor impact on the results of our prospective study). The statistical scaling 
factor thus defined can be applied as such to direct $CP$ asymmetries, but some additional aspects must be considered 
in the scaling of uncertainties for other observables. For time-dependent $CP$ asymmetries, the difference in flavour-tagging performances (the effective tagging efficiency $Q$) should be taken into account. In the $B$-factory environment, 
a quality factor  $Q_{\rm B-factories} \sim 30$~\cite{FavourTagging_BaBar:2009iva,FavourTagging_Belle:2012iva} was achieved,
while for  LHCb a smaller value is used ($Q_{\rm LHCb} \sim 3$~\cite{FavourTagging_LHCb:2012iva}), 
which entails an additional factor $(Q_{\rm B-factories}/Q_{\rm LHCb})^{1/2} \sim 3.2$ in the scaling of uncertainties. 
For  branching 
ratios, LHCb is not able to directly count the number of $B$ mesons produced, and it is necessary to resort to a normalisation using final states for which  the branching ratio has been measured elsewhere (mainly at $B$-factories). This 
additional source of uncertainty is taken into account in the projection of the error. Finally, in our prospective studies, we 
adopt the pessimistic view of neglecting potential measurements from LHCb for modes with $\pi^0$ mesons in the final state 
(e.g., $B^0\rightarrow K^+\pi^-\pi^0$ and $B^+\rightarrow K^0_S\pi^+\pi^0$), as it is difficult to anticipate the evolution in the performances for $\pi^0$ reconstruction and phase space resolution.

Belle II~\cite{Urquijo:2015qsa} expects to surpass by a factor of $\sim 50$ the total
statistics collected by the $B$-factories. As the experimental environments will be very similar,
we just scale the current uncertainties by this statistical factor.

Starting from the statistical uncertainties from Babar and scaling them according to the above procedure, we obtain our projections of uncertainties on physical observables, 
shown in Tab.~\ref{tab:LHCbAndBelleII}, where 
the current uncertainties are compared with the projected ones for the first ($B$-factories combined with LHCb Run1 and Run2) and 
second (adding Belle II) stages described previously.

\begin{table*}[t]
\begin{center}
\begin{tabular}
{lcc|lcc}
Hadronic amplitudes                                                 & Magnitude  & Phase ($^\circ$) & Observable                                   & Measurement                & Value
\\
\hline
$T^{+-}$                                                            & 2.540      &    0.0     & ${\mathcal B}(B^0\rightarrow K^{*+}\pi^-)$         &   $8.4 \pm 0.8$          & $7.1$
\\
$T^{00}_{\rm C}$                                                    & 0.762      &   75.8     & ${\mathcal B}(B^0\rightarrow K^{*0}\pi^0)$         &   $3.3 \pm 0.6$          & $1.6$
\\
$N^{0+}$                                                            & 0.143      &  108.4     & ${\mathcal B}(B^+\rightarrow K^{*+}\pi^0)$         &   $8.2 \pm 1.8$          & $8.5$
\\
$P^{+-}$                                                            & 0.091      &   -6.5     & ${\mathcal B}(B^+\rightarrow K^{*0}\pi^+)$         &  $10.1^{+0.8}_{-0.9}$    & $10.9$
\\
$P_{\rm EW}$                                                        & 0.038      &   15.2     & $A_{CP}(B^0\rightarrow K^{*+}\pi^-)$               &  $-0.23 \pm 0.06$             & $-0.129$
\\
$P_{\rm EW}^{\rm C}$                                                & 0.029      &  101.9     & $A_{CP}(B^0\rightarrow K^{*0}\pi^0)$               &  $-0.15 \pm 0.13$            & $+0.465$
\\
$\left|\frac{V_{ts}V_{tb}^*P^{+-}}{V_{us}V_{ub}^*T^{+-}}\right|$    & 1.809      &            & $A_{CP}(B^+\rightarrow K^{*+}\pi^0)$               &  $-0.39 \pm 0.12$            & $-0.355$
\\
$\left|T^{00}_{\rm C}/T^{+-}\right|$                                & 0.300      &            & $A_{CP}(B^+\rightarrow K^{*0}\pi^+)$               &  $+0.038 \pm 0.042$          & $+0.039$
\\
$\left|N^{0+}/T^{00}_{\rm C}\right|$                                & 0.187      &            &                                                    &                          &
\\
$\left|P_{\rm EW}/P^{+-}\right|$                                    & 0.421      &            &                                                    &                          &
\\
$\left|P_{\rm EW}/(T^{+-} + T^{00}_{\rm C})\right|/R$               & 1.009      &            &                                                    &                          &
\\
$\left|P_{\rm EW}^{\rm C}/P_{\rm EW}\right|$                        & 0.762      &            &                                                    &                          &
\\
\end{tabular}
\caption{Values chosen for our reference scenario. The values on hadronic parameters (left columns) are chosen to roughly reproduce the reference values of branching fractions (in units of $10^{-6}$) and 
$CP$ asymmetries in $B\rightarrow K^*\pi$  (right columns). The reference input values come from the current HFLAV averages~\cite{Amhis:2016xyh}, except for
$A_{CP}(B^+\rightarrow K^{*+}\pi^0)$, where the value is taken from Ref.~\cite{Lees:2015uun}. The values of the hadronic parameters yield the branching ratios and $CP$ asymmetries of the last column.}
\label{tab:IdealCase}
\end{center}
\end{table*}

\begin{table*}[t]
\begin{center}
\begin{tabular}{c|c|c|c|c}
Observable	& Analysis				& Current uncertainty	& LHCb (Run1+Run2)	& LHCb+Belle II
\\ \hline
${\mathcal Re}\left[ \frac{q}{p} \frac{\overline{A}(K^{*-}\pi^+)}{A(K^{*+}\pi^-)} \right]$	
		& $B^0\rightarrow K^0_S\pi^+\pi^-$	& $0.11$ 		& $0.04$		& $0.01$	
\\
${\mathcal Im}\left[ \frac{q}{p} \frac{\overline{A}(K^{*-}\pi^+)}{A(K^{*+}\pi^-)} \right]$	
		& $B^0\rightarrow K^0_S\pi^+\pi^-$ 	& $0.16$ 		& $0.11$		& $0.02$	
\\
${\mathcal B}(K^{*+}\pi^-)$
		& $B^0\rightarrow K^0_S\pi^+\pi^-$	& $0.69$		& $0.32$		& $0.09$
\\
$\left| \frac{\overline{A}(K^{*-}\pi^+)}{A(K^{*+}\pi^-)} \right|$
		& $B^0\rightarrow K^+\pi^-\pi^0$	& $0.06$		& $0.06$		& $0.01$
\\
${\mathcal Re}\left[ \frac{A(K^{*0}\pi^0)}{A(K^{*+}\pi^-)} \right]$
		& $B^0\rightarrow K^+\pi^-\pi^0$	& $0.11$		& $0.11$		& $0.02$
\\
${\mathcal Im}\left[ \frac{A(K^{*0}\pi^0)}{A(K^{*+}\pi^-)} \right]$
		& $B^0\rightarrow K^+\pi^-\pi^0$	& $0.23$		& $0.23$		& $0.03$
\\
${\mathcal Re}\left[ \frac{\overline{A}(\overline{K}^{*0}\pi^0)}{\overline{A}(K^{*-}\pi^+)} \right]$
		& $B^0\rightarrow K^+\pi^-\pi^0$	& $0.10$		& $0.10$		& $0.01$ 
\\
${\mathcal Im}\left[ \frac{\overline{A}(\overline{K}^{*0}\pi^0)}{\overline{A}(K^{*-}\pi^+)} \right]$
		& $B^0\rightarrow K^+\pi^-\pi^0$	& $0.30$		& $0.30$		& $0.04$ 
\\
${\mathcal B}(K^{*0}\pi^0) $
		& $B^0\rightarrow K^+\pi^-\pi^0$	& $0.35$		& $0.35$		& $0.05$
\\
$\left| \frac{\overline{A}(\overline{K}^{*0}\pi^-)}{A(K^{*0}\pi^+)} \right|$
		& $B^+\rightarrow K^+\pi^-\pi^+$	& $0.04$		& $0.005$		& $0.004$
\\
${\mathcal B}(K^{*0}\pi^+)$
		& $B^+\rightarrow K^+\pi^-\pi^+$	& $0.81$		& $0.50$		& $0.11$
\\
$\left| \frac{\overline{A}(K^{*-}\pi^0)}{A(K^{*+}\pi^0)} \right|$
		& $B^+\rightarrow K^0_S\pi^+\pi^0$	& $0.15$		& $0.15$		& $0.02$
\\
${\mathcal Re}\left[ \frac{A(K^{*+}\pi^0)}{A(K^{*0}\pi^+)} \right]$
		& $B^+\rightarrow K^0_S\pi^+\pi^0$	& $0.16$		& $0.16$		& $0.02$	
\\ 
${\mathcal Im}\left[ \frac{A(K^{*+}\pi^0)}{A(K^{*0}\pi^+)} \right]$
		& $B^+\rightarrow K^0_S\pi^+\pi^0$	& $0.30$		& $0.30$		& $0.04$	
\\ 
${\mathcal Re}\left[ \frac{\overline{A}(K^{*-}\pi^0)}{\overline{A}(\overline{K}^{*0}\pi^-)} \right]$
		& $B^+\rightarrow K^0_S\pi^+\pi^0$	& $0.21$		& $0.21$		& $0.03$
\\ 
${\mathcal Im}\left[ \frac{\overline{A}(K^{*-}\pi^0)}{\overline{A}(\overline{K}^{*0}\pi^-)} \right]$
		& $B^+\rightarrow K^0_S\pi^+\pi^0$	& $0.13$		& $0.13$		& $0.02$
\\ 
${\mathcal B}(K^{*+}\pi^0)$
		& $B^+\rightarrow K^0_S\pi^+\pi^0$	& $0.92$		& $0.92$		& $0.13$
\\
\end{tabular}
\caption{Prospective scenarios for statistical uncertainties on the $B\rightarrow K^{\star}\pi$ observables. The extrapolations
are based on the current statistical uncertainties from {\babar} results. The uncertainties on the branching fractions are given in units of $10^{-6}$.}
\label{tab:LHCbAndBelleII}
\end{center}
\end{table*}

\end{document}